\documentclass[traditabstract,onecolumn]{aa}

\usepackage{graphicx}
\usepackage{psfig}
\usepackage{natbib}
\usepackage{savesym}
\usepackage{amsmath}
\savesymbol{iint}
\usepackage{txfonts}
\restoresymbol{TXF}{iint}

\bibpunct{(}{)}{;}{a}{}{,}

\newcommand{\ssi}{_i}
\newcommand{\ssj}{_j}
\newcommand{\ssij}{_{ij}}

\begin{document}

\title{SEREN - A new SPH code for star and planet formation simulations}

\subtitle{Algorithms and tests}

\author{D. A. Hubber \inst{1,2,3,4}, 
        C. P. Batty \inst{2},
        A. McLeod \inst{2,5} \and
        A. P. Whitworth \inst{2} }

\institute{Department of Physics and Astronomy, University of Sheffield, 
           Hicks Building, Hounsfield Road, Sheffield S3 7RH, UK \and
           School of Physics and Astronomy, Cardiff University, 
           Queens Buildings, The Parade, Cardiff, CF24 3AA, Wales, UK \and
           Institute for Theoretical Astrophysics, University of Oslo, 
           Pb 1029 Blindern, 0315 Oslo, Norway \and
	   Centre of Mathematics for Applications, University of Oslo, 
           Pb 1053 Blindern, 0316 Oslo, Norway \and
           Astronomical Institute, Academy of Sciences of the Czech Republic, 
           Bo\v{c}n\'{i} II 1401, 141 31 Praha 4, Czech Republic}
           
\offprints{D.Hubber@sheffield.ac.uk}

\authorrunning{Hubber, Batty, McLeod \& Whitworth}

\titlerunning{SEREN}

\date{February 3rd, 2011}

\abstract {We present SEREN, a new hybrid Smoothed Particle Hydrodynamics and N-body code designed to simulate astrophysical processes such as star and planet formation.  It is written in Fortran 95/2003 and has been parallelised using OpenMP.  SEREN is designed in a flexible, modular style, thereby allowing a large number of options to be selected or disabled easily and without compromising performance.   SEREN uses the conservative `grad-h' formulation of SPH, but can easily be configured to use traditional SPH or Godunov SPH.  Thermal physics is treated either with a barotropic equation of state, or by solving the energy equation and modelling the transport of cooling radiation.  A Barnes-Hut tree is used to obtain neighbour lists and compute gravitational accelerations efficiently, and an hierarchical time-stepping scheme is used to reduce the number of computations per timestep.  Dense gravitationally bound objects are replaced by sink particles, to allow the simulation to be evolved longer, and to facilitate the identification of protostars and the compilation of stellar and binary properties. At the termination of a hydrodynamical simulation, SEREN has the option of switching to a pure N-body simulation, using a 4th-order Hermite integrator, and following the ballistic evolution of the sink particles (e.g. to determine the final binary statistics once a star cluster has relaxed). We describe in detail all the algorithms implemented in SEREN and we present the results of a suite of tests designed to demonstrate the fidelity of SEREN and its performance and scalability.  Further information and additional tests of SEREN can be found at the web-page {\rm http://www.astro.group.shef.ac.uk/seren}.}

  % {Code}
  % {Code}
  % {Code}
 
\keywords{Hydrodynamics - Methods: numerical - Stars: formation}

\maketitle

%%%%%%%%%%%%%%%%%%%%%%%%%%%%%%%%%%%%%%%
\section{Introduction} \label{S:INTRO}%
%%%%%%%%%%%%%%%%%%%%%%%%%%%%%%%%%%%%%%%

Star formation problems are amongst the most demanding in computational astrophysics, requiring a large number of physical processes to be be modeled (e.g. hydrodynamics, self-gravity, optically thick radiative cooling, gas chemistry, ionization, gas-ion coupling, magneto-hydrodynamics, radiative and mechanical feedback) over a very large range of physical conditions (i.e. gas densities from $\sim\! 10^{-20}\,{\rm g}\,{\rm cm}^{-3}$ to $\sim\! 10^{+1}\,{\rm g}\,{\rm cm}^{-3}$, and gas temperatures from $\sim\! 10{\rm K}$ to $\sim\! 10^7{\rm K}$). It is non-trivial to include all of the above physics in a single code which works over such a wide range of physical conditions and produces accurate results in an efficient manner. There are also often multiple methods available to model these processes, and one must choose the most appropriate and/or optimal method to study a given problem. The principal choice is whether to use an Eulerian grid-based code to simulate the fluid dynamics, or a Lagrangian particle-based code. Grid-based schemes are capable of modelling incompressible fluid dynamics accurately and efficiently, but for highly compressible fluids, where the density can take a large range of values, expensive adaptive-mesh-refinement techniques are required. Particle-based schemes, such as Smoothed Particle Hydrodynamics, do not model hydrodynamical processes as well as grid-based schemes (Agertz et al. 2007), but they can model highly compressible fluids through a large range of scales with ease.  This makes particle codes well suited to modelling self-gravitating fluids such as those involved in star formation.  A number of publicly available codes using either static or adaptive-mesh-refinement grids (e.g. ZEUS, Stone \& Norman 1992; { FLASH, Fryxell et al. 2000; RAMSES, Teyssier 2002; ENZO, Abel, Bryan \& Norman 2002}) or particles (e.g. GADGET, Springel, Yoshida \& White 2001; GADGET2, Springel 2005; VINE, Wetzstein et al. 2009, Nelson et al. 2009) are available, and have been applied to a variety of different phenomena in interstellar gas dynamics, star and galaxy formation, and cosmology.

Here we present SEREN, a new multi-dimensional self-gravitating hydrodynamics and N-body code.  SEREN uses the Smoothed Particle Hydrodynamics (SPH) algorithm to model fluid dynamics, in combination with tree-gravity and hierarchical block-timestepping routines.  It also includes a variety of specialist routines designed to tackle star and planet formation problems, such as sink particles (Bate, Bonnell \& Price 1995), and a 4th order Hermite N-body integrator (Makino \& Aarseth 1992) to follow the ballistic evolution of a star cluster once its gas has been accreted or dispersed.

The purposes of this paper are (i) to describe the algorithms implemented in SEREN, and (ii) to demonstrate the fidelity of SEREN -- i.e. that the algorithms are coded correctly and reproduce known results in tests, so that future publications presenting simulations performed with SEREN can refer to this paper for a full description of the code. In Section \ref{S:OVERVIEW}, we give a brief overview of SEREN and all of its main features, { and compare these features with those available in other available astrophysical SPH codes}. In Section \ref{S:SPH}, we describe in detail the Smoothed Particle Hydrodynamics algorithms used. In Section \ref{S:SPHGRAVITY}, we describe the implementation of self-gravity in SPH. In Section \ref{S:THERMAL}, we briefly discuss the available thermal physics modules, including the transport of heating, cooling and ionizing radiation. In Section \ref{S:TIMEINT}, we discuss the integration schemes and time-stepping criteria. In Section \ref{S:SINKS}, we discuss the implementation of sink particles. In Section \ref{S:NBODY}, we discuss the 4th order Hermite N-body integrator and the additional features contained within it (e.g. binary identification). In Section \ref{S:TREE}, we discuss the implementation of the Barnes-Hut tree, and how it is used to determine neighbour lists and calculate gravitational accelerations. In Section \ref{S:TESTS}, we present a large suite of tests, to demonstrate that SEREN simulates correctly the physical processes it is intended to capture. In Section \ref{S:OPTIMISATION}, we discuss the memory optimisations used. In Section \ref{S:PARALLELISATION}, we discuss the techniques used to parallelise SEREN using OpenMP, and we demonstrate how SEREN scales on shared-memory machines. In Section \ref{S:FUTUREPLANS}, we outline the major features that are still to be implemented.

%%%%%%%%%%%%%%%%%%%%%%%%%%%%%%%%%%%%%%%%%%%%%%%%%%%%%%%%%%%%%%%
\section{Overview of SEREN and other codes} \label{S:OVERVIEW}%
%%%%%%%%%%%%%%%%%%%%%%%%%%%%%%%%%%%%%%%%%%%%%%%%%%%%%%%%%%%%%%%

SEREN is a multi-dimensional self-gravitating SPH and N-body code. It has been designed for star and planet formation problems, but it can easily be adapted to simulate other astrophysical phenomena.  SEREN is written in Fortran 95 (with some Fortran 2003 features) and is parallelised using OpenMP.  It is written in a highly modular style, with a large number of features that can be switched on or off using Makefile options and conditional compilation flags.  It can be compiled for one, two or three-dimensions, although it is most optimal in three-dimensional mode.   We list here the main algorithms and features included in SEREN: 
\begin{itemize}
\item Standard SPH (e.g. Monaghan 1992), `grad-h' SPH (Springel \& Hernquist 2002; Price \& Monaghan 2004b), and Godunov SPH (Inutsuka 2002; Cha \& Whitworth 2003)
\item Kernel-softened self-gravity (Price \& Monaghan 2007)
\item Artificial dissipation (Lattanzio \& Monaghan 1985, Balsara 1995, Monaghan 1997, Morris \& Monaghan 1997, Price 2008)
\item 2nd-order Runge-Kutta, 2nd-order Predictor-Corrector and 2nd-order kick-drift-kick and drift-kick-drift Leapfrog integration schemes 
\item Hierarchical block time-stepping (e.g. Hernquist \& Katz 1989)
\item Periodic boundary conditions, including periodic gravity (Hernquist, Bouchet \& Suto 1991; Klessen 1997)
\item Several particle types: self-gravitating gas particles, non-gravitating inter-cloud particles, static or non-static boundary particles
\item Octal-spatial trees for neighbour-searching and gravity (Barnes \& Hut 1986, Pfalzner \& Gibbon 1996)
\item Simple isothermal, polytropic or barotropic equations of state, solution of the energy equation with associated radiation transport (Stamatellos et al. 2007), and propagation of ionizing radiation using HEALPix rays (Bisbas et al. 2009)
\item Sink particles (Bate, Bonnell \& Price 1995)
\item 4th order Hermite N-body integrator (Makino \& Aarseth 1992)
\item Identification of binaries and calculation of binary properties (e.g. Aarseth 2003)
\end{itemize}
We control which algorithms are used in SEREN using Makefile options, so only the employed subroutines are compiled and included in the executable file.  The parameters which determine how the selected algorithms function are set in a separate parameters file.  In Sections \ref{S:SPH} to \ref{S:TREE}, we describe in more detail the implementation of these algorithms in SEREN.

Several other SPH codes are available to the astrophysics community for performing simulations of self-gravitating hydrodynamics.  While these codes share a common set of basic features, most contain specialised algorithms to model certain astrophysical processes, or are optimised to perform a particular class of simulation.   We briefly discuss the algorithms and features in other available astrophysical SPH codes, in order to highlight to potential users the relative merits of each code for solving particular problems { and how they contrast with the features implemented in SEREN}.  We only discuss here those codes that have a refereed or archived publication containing details of the implementation and tests.

%%%%%%%%%%%%%%%%%%%%%%%%%%%%%%%%%%%%%%%%
\subsection{GADGET \& GADGET 2} \label{SS:GADGET2}
GADGET (Springel, Yoshida \& White 2001) and GADGET 2 (Springel 2005) are written in C and parallelised using MPI.  While the original GADGET code was designed to investigate galaxy formation problems, GADGET 2 was designed to investigate large-scale cosmological problems such as galaxy cluster formation and the formation of structure in the Universe (e.g. Springel et al. 2005). 
MPI can be used very efficiently when the work distributed to all CPUS is automatically load-balanced.  Therefore, the approximately uniform (large-scale) density distribution used in cosmological simulations is a problem that an MPI code like GADGET 2 can handle efficiently on very large clusters with 1,000s of CPUs (e.g. Springel 2005).  
GADGET 2 uses a Peano-Hilbert space-filling curve in order to determine how to distribute the particles amongst the available processors. This improves the scalability, by reducing communication overheads.
GADGET 2 uses a conservative SPH formulation combined with solving the entropy equation for the thermal physics (Springel \& Hernquist 2002).  Particle properties can be integrated using either a Leapfrog-KDK or Leapfrog-DKD integration scheme, in combination with a hierarchical block-timestep scheme.  The calculation of gravitational forces is split into short and long-range computations;  short range forces are computed using a Barnes-Hut tree (which is efficient for clumpy density distributions), and long-range forces are computed using a particle-mesh scheme (which is efficient for smoother density distributions).
GADGET 2 contains the ability to model several different particle types relevant to galaxy and cosmology simulations, namely gas, cold-dark matter and star particles.  Star particles usually represent a whole cluster of stars, in comparison to sink particles in SEREN which represent individual stars, or unresolved small, multiple systems.

%%%%%%%%%%%%%%%%%%%%%%%%%%%%%%%%%
\subsection{GASOLINE} \label{SS:GASOLINE}
GASOLINE (Wadsley, Stadel \& Quinn; arXiv:astro-ph/0303521v1) is written in Fortran and is parallelised for shared-memory machines.  
GASOLINE uses the standard formulation of SPH (e.g. Monaghan 1992) with $(\alpha,\beta)$ viscosity (Monaghan \& Gingold 1983) and a Balsara switch (Balsara 1995) for reducing unwanted shear viscosity.
GASOLINE can use two separate trees;  a K-D tree for neighbour searching and a Barnes-Hut octal tree (Barnes \& Hut 1986) for calculating gravitational forces.  The code computes multipole moments up to hexadecapole-order to compute gravitational forces efficiently, but only uses the geometrical MAC for evaluating the cell-interaction list.  Ewald summation is also available for simulating periodic boxes. GASOLINE contains a number of options for treating thermal physics, including an implicit integrator for solving the energy equation.  A number of cooling and ionisation processes can be selected, as well as a simple heating-feedback prescription due to star formation.  GASOLINE uses a Leapfrog-KDK integration scheme for advancing particle positions and velocities, along with a standard hierarchical block-timestep scheme.

%%%%%%%%%%%%%%%%%%%%%%%%%%%%%%%%%
\subsection{VINE} \label{SS:VINE}
VINE (Wetzstein et al. 2009; Nelson et al. 2009) is written in Fortran 95 and parallelised using OpenMP.  As with GADGET and GADGET 2, VINE has been designed to investigate galaxy and cosmological problems.  
VINE has also been parallelised using OpenMP.  It has been tested on up to 128 CPUs and scales well provided the problem size is large enough.  
VINE uses a nearest-neighbour binary tree (e.g. Benz et al. 1990) to compute gravitational forces and to search for neighbours efficiently.  VINE also has the facility to use GRAPE boards (e.g. Makino et al. 2003) and thus can significantly speed up the calculation of the gravitational forces for particular problems. VINE does not use a conservative form of SPH, but rather uses the traditional form of SPH (Monaghan 1992).  VINE contains a variety of particle types similar to GADGET 2, such as gas, cold-dark matter and star particles.

%%%%%%%%%%%%%%%%%%%%%%%%%%%%%%%%%
\subsection{MAGMA} \label{SS:MAGMA}
MAGMA (Rosswog \& Price 2007) is an Smoothed Particle Magneto-hydrodynamics (SPMHD) code which is parallelised using OpenMP.  MAGMA has been designed to model compact objects, such as binary-neutron stars.  MAGMA uses the conservative `grad-h' SPH scheme for computing hydro and gravitational forces, and `Euler potentials' for solving the ideal MHD equations; this enforces ${\rm div}\,{\bf B} = 0$ by design.  The code includes dissipative artificial viscosity, conductivity and resistivity terms, with switches such as time-dependent viscosity (Morris \& Monaghan 1997) for reducing dissipation.  Thermal physics includes a relativistic equation-of-state for modelling the interiors of neutron stars, and a method for modelling neutrino emission.  No additional particle types (e.g. sink particles) are included. Gravitational forces are computed with a binary tree (Benz et al. 1990) and particle-particle interactions are computed with kernel-softened gravity (Price \& Monaghan 2007).  Particle positions and velocities are integrated with a second-order predictor-corrector scheme, using an individual timestep scheme.

%%%%%%%%%%%%%%%%%%%%%%%%%%%%%%%%%
\subsection{EvoL} \label{SS:EvoL}
EvoL (Merlin et al. 2010) is written in Fortran 95 and is parallelised using MPI.  EvoL was designed to investigate cosmological structure, galaxy-cluster and galaxy formation problems, similar to GADGET 2 and VINE.  
As with other galaxy/cosmological codes, EvoL can model self-gravitating gas, cold dark matter and star particles.  EvoL models the gas-dynamics using a modified `grad-h' SPH formulation, and also contains terms that correct for unequal-mass particles.  Gravity is calculated using a Barnes-Hut tree, and neighbouring particle gravitational forces are computed with a conservative scheme similar to Price \& Monaghan (2007), but using the number density instead of the mass density, which again is beneficial when using unequal mass particles.  EvoL uses a Leapfrog-KDK scheme for integrating particle positions and velocities.  A standard hierarchical block-timestep scheme is employed, along with the instantaneous timestep-reduction procedure (Saitoh \& Makino 2009) to ensure the timesteps used for neighbouring particles are not greatly different.  EvoL also contains the ability to evolve the particle positions using the X-SPH method (e.g. Monaghan 1992) which can prevent particle interpenetration.

%%%%%%%%%%%%%%%%%%%%%%%%%%%%%%%%%%%%%%%%%%%%%%%%%%%%%%%%
\section{Smoothed Particle Hydrodynamics} \label{S:SPH}%
%%%%%%%%%%%%%%%%%%%%%%%%%%%%%%%%%%%%%%%%%%%%%%%%%%%%%%%%

SPH is a Lagrangian hydrodynamics scheme which uses particles to represent the fluid (Gingold \& Monaghan 1977; Lucy 1977). SEREN contains three different variants of SPH: the standard implementation (Monaghan 1992), the conservative `grad-h' implementation (Springel \& Hernquist 2002, Price \& Monaghan 2004b) and the Godunov implementation (Cha \& Whitworth 2003; Eqns (9) \& (10)).  The `grad-h' implementation is the favoured, default implementation in SEREN.

In SPH, particle properties are smoothed over a length scale, $h$, called the {\it smoothing length}, using a weighting function, $W({\bf r},h)$, called the {\it kernel function}. The fluid is thus still a continuous medium despite being represented by a finite number of discrete particles. The volume over which a particle is smoothed is called its {\it smoothing kernel}. Particle $i$ interacts hydrodynamically with all other SPH particles, $j$, that lie inside the smoothing kernel of $i$ ({\it gather}), and/or whose own smoothing kernels overlap $i$ ({\it scatter}). These particles are referred to as the {\it neighbours} of $i$. The smoothing length determines the spatial resolution and can in principle be set to any value. The simplest choice is to keep $h$ uniform in space and constant in time, throughout the simulation. However, to take advantage of the Lagrangian nature of SPH, it is often desirable to set the smoothing length of an SPH particle to be of order the local mean particle separation. The resolution then automatically adapts to the local conditions, providing an adaptability that is much more difficult to achieve with grid codes. SEREN contains two choices for the kernel function, both of which have finite extent, $r_{_{\rm MAX}}={\cal R}h:\;$ the M4 cubic spline kernel (Monaghan \& Lattanzio 1985) with ${\cal R}=2$, and the quintic spline kernel (Morris 1996) with ${\cal R}=3$. Detailed properties of these kernels are given in Appendix \ref{A:KERNEL}.

Since `grad-h' is the default implementation of SPH in SEREN, we briefly describe its main features here. In order to guarantee conservation of momentum, angular momentum and energy, the SPH fluid equations are derived from the Euler-Lagrange equations. This requires that the smoothing length of a particle be either constant, a function of the particle's co-ordinates, or a function of some property that is itself a function of the particle's co-ordinates. We follow Springel \& Hernquist (2002) and Price \& Monaghan (2004b) in making the smoothing length a function of the density. Specifically, for particle $i$ we put 
\begin{eqnarray} \label{EQN:HRHO}
h\ssi  &=& \eta_{_{\rm SPH}} \left( \frac{m\ssi }{\rho\ssi } \right)^{\frac{1}{D}}\,,
\end{eqnarray}
where $m\ssi$ is the mass of particle $i$, $\rho\ssi$ is the SPH density at the position of particle $i$, $D$ is the spatial dimensionality, and $\eta_{_{\rm SPH}}$ is a parameter that controls the mean number of neighbours, $\;\bar{N}_{_{\rm NEIB}} \simeq 2\,{\cal R}\,\eta_{_{\rm SPH}} \;,\;\; \pi\,({\cal R}\eta_{_{\rm SPH}})^2 \;,\;\;(4\pi /3)({\cal R}\eta_{_{\rm SPH}})^3\,\,$ in one, two and three dimensions respectively. $\;\rho\ssi$ is calculated using 
\begin{eqnarray} \label{EQN:SPHRHO}
\rho\ssi  &=& \sum \limits_{j=1}^{N}  m\ssj W({\bf r}\ssij,h\ssi)\,,
\end{eqnarray}
where ${\bf r}\ssij \equiv {\bf r}\ssi - {\bf r}\ssj$, and the summation includes particle $i$ itself. Since the smoothing length is needed in order to calculate the density in Eqn. (\ref{EQN:SPHRHO}) and vice-versa in Eqn. (\ref{EQN:HRHO}), $h\ssi$ and $\rho\ssi$ are obtained by iteration.

Once $h$ and $\rho$ are evaluated for all particles, the terms in the SPH fluid equations can be computed. The momentum equation is 
\begin{eqnarray} \label{EQN:GRADHMOMEQN}
\frac{d{\bf v}\ssi }{dt} &=& -
\sum \limits_{j=1}^{N}  m\ssj  \left( 
\frac{P\ssi}{\Omega\ssi \rho\ssi^2} \nabla\ssi W({\bf r}\ssij ,h\ssi) + 
\frac{P\ssj}{\Omega\ssj \rho\ssj^2} \nabla\ssi W({\bf r}\ssij ,h\ssj) \right)
\,,
\end{eqnarray}
where $P_i$ is the pressure of particle $i$, $\nabla\ssi W$ is the gradient of the kernel function at the position of particle $i$, and 
\begin{eqnarray} \label{EQN:OMEGA}
\Omega\ssi  &=& 1 - \frac{\partial h\ssi }{\partial \rho\ssi } 
\sum \limits_{j=1}^{N}  m\ssj  \frac{\partial W}{\partial h}
({\bf r}\ssij , h\ssi )\,.
\end{eqnarray}
$\Omega\ssi$ is a dimensionless quantity that corrects for the spatial variability of $h$. $\partial h\ssi / \partial \rho_i$ is obtained explicitly from Eqn. (\ref{EQN:HRHO}). $\partial W / \partial h$ is obtained from the kernel function (see Appendix \ref{A:KERNEL}). The SPH energy equation is 
\begin{eqnarray} \label{EQN:GRADHENEQN}
\frac{du\ssi }{dt} &=& 
\frac{P\ssi }{\Omega\ssi  \rho\ssi ^2} \sum \limits_{j=1}^{N}
m\ssj  {\bf v}\ssij \cdot \nabla W\ssij ({\bf r}\ssij , h\ssi )\,,
\end{eqnarray}
where ${\bf v}\ssij \equiv {\bf v}\ssi - {\bf v}\ssj$. Since the mass of each particle is constant, and the density is computed using Eqn. (\ref{EQN:SPHRHO}), there is no need to solve the SPH continuity equation.

The summations in Eqns. (\ref{EQN:SPHRHO}), (\ref{EQN:GRADHMOMEQN}), (\ref{EQN:OMEGA}) and (\ref{EQN:GRADHENEQN}) are formally over all particles in the simulation. However, since the kernels used in SEREN both have finite extent, the summations are actually only over the neighbours of particle $i$. SEREN uses a Barnes-Hut tree (Barnes \& Hut 1986) to obtain neighbour lists. The procedures for constructing and walking the tree are described in Section \ref{S:TREE}.

%%%%%%%%%%%%%%%%%%%%%%%%%%%%%%%%%%%%%%%%%%%%%%%%%%%%%%%%%%%%%%%%%%%%%%
\subsection{Artificial viscosity and conductivity} \label{SS:ARTDISS}%
%%%%%%%%%%%%%%%%%%%%%%%%%%%%%%%%%%%%%%%%%%%%%%%%%%%%%%%%%%%%%%%%%%%%%%

In most formulations of SPH, artificial viscosity terms are needed to ensure that shocks are captured, i.e. that converging particle streams do not interpenetrate, but rather form a contact discontinuity, and that kinetic energy is converted into thermal energy at the shock, thereby generating entropy. SEREN includes two different forms of artificial viscosity: the standard $(\alpha,\beta)$ formulation (Monaghan \& Gingold 1983), and the formulation based on Riemann solvers (Monaghan 1997).  The Monaghan-Riemann formulation is the default in SEREN, and involves adding the following extra terms to the momentum and energy equations, 
\begin{eqnarray}\label{EQN:MON97ARTVISC}
\left(\frac{d{\bf v}\ssi}{dt}\right)_{_{\rm DISS}}&=&\sum\limits_{j=1}^{N}\,\frac{m\ssj}{\overline{\rho}\ssij}\,\left\{\alpha v_{_{\rm SIG}} {\bf v}\ssij\cdot\hat{\bf r}\ssij\right\}\,\overline{\nabla\ssi W}\left({\bf r}\ssij ,h\ssi ,h\ssj\right)\,,\\\label{EQN:MON97ENERGYDISS}
\left(\frac{du\ssi}{dt}\right)_{_{\rm DISS}}&=&-\,\sum\limits_{j=1}^{N}\,\frac{m\ssj}{\overline{\rho}\ssij}\,\left\{\frac{\alpha v_{_{\rm SIG}}({\bf v}\ssij\cdot\hat{\bf r}\ssij)^2}{2}+\alpha'v_{_{\rm SIG}}'( u\ssi -u\ssj)\right\}\,\hat{\bf r}\ssij\cdot\overline{\nabla\ssi W}\left({\bf r}\ssij ,h\ssi ,h\ssj\right)\,,
\end{eqnarray}
where $\alpha$ and $\alpha'$ are user specified coefficients of order unity, $v_{_{\rm SIG}}$ and $v_{_{\rm SIG}}'$ are signal speeds, $\hat{\bf r}_{ij}={\bf r}_{ij}/|{\bf r}_{ij}|$, and 
\begin{eqnarray}
\overline{\nabla\ssi W}({\bf r}\ssij , h\ssi , h\ssj) &=&
\frac{\nabla\ssi W({\bf r}\ssij , h\ssi) + 
\nabla\ssi W({\bf r}\ssij , h\ssj)}{2}\,.
\end{eqnarray}
This form of artificial dissipation is chosen as the default because (a) it has a physically informed motivation, and (b) it can be generalised to model dissipation in other quantities while giving just as good results as the standard $(\alpha,\beta)$ viscosity when modelling shocks. The dissipation term on the right-hand side of Eqn. (\ref{EQN:MON97ARTVISC}) and the first term on the right-hand side of Eqn. (\ref{EQN:MON97ENERGYDISS}) represent artificial viscosity -- i.e. exchange of momentum between neighbouring particles which are approaching or receding from one another, and conversion of the kinetic energy lost into thermal energy -- and they are moderated by the signal speed $v_{_{\rm SIG}} = c\ssi + c\ssj - {\bf v}\ssij \cdot \hat{\bf r}\ssij$. The second term on the right-hand side of Eqn. (\ref{EQN:MON97ENERGYDISS}) represents artificial conductivity, and acts to smooth out gradients in the specific internal energy. For purely hydrodynamic simulations, Price (2008) advocates that the artificial conductivity be moderated by the signal speed
\begin{eqnarray} \label{EQN:PRICE2008COND}
v_{_{\rm SIG}}'&=&\sqrt{\frac{|P\ssi -P\ssj|}{\overline{\rho}\ssij}}\,.
\end{eqnarray}
However, in self-gravitating simulations this can drain thermal energy from dense condensations, thereby artificially accelerating gravitational contraction. Wadsley et al. (2006) have proposed the alternative signal speed 
\begin{eqnarray} \label{EQN:SIGNALVEL3}
v_{_{\rm SIG}}'&=&|{\bf v}\ssij\cdot\hat{\bf r}\ssij |
\end{eqnarray}
for artificial conductivity. Both Eqn. (\ref{EQN:PRICE2008COND})  and Eqn. (\ref{EQN:SIGNALVEL3}) are included as options in SEREN.  We note that when the Godunov-SPH formulation is selected, we can disable the artificial {\it viscosity} since the Riemann solver should allow us to capture shocks accurately.  We may need to retain the artificial {\it conductivity} since { our simple implementation of a Riemann solver into SPH does not address that problem}.

%%%%%%%%%%%%%%%%%%%%%%%%%%%%%%%%%%%%%%%%%%%%%%%%%%%%%%%%%%%%%%%%%%%%%
\subsubsection{Artificial viscosity switches} \label{EQN:VISCSWITCH}%
%%%%%%%%%%%%%%%%%%%%%%%%%%%%%%%%%%%%%%%%%%%%%%%%%%%%%%%%%%%%%%%%%%%%%

Artificial viscosity can have undesirable side effects. In the absence of shocks it can lead to kinetic energy being dissipated at an unacceptably high rate, i.e. much faster than would happen with physical viscosity; this is an important consideration in simulations of interstellar turbulence. It can also deliver an unacceptably high shear viscosity, and thereby corrupt shear flows; this is an important consideration in simulations of the long-term evolution of accretion discs. A number of devices has been proposed to reduce the artificial viscosity in regions where it is not needed. Three such viscosity limiters are included in SEREN. The first is the switch proposed by Balsara (1995) in which $\alpha$ is multiplied by the dimensionless quantity $\frac{1}{2}(f\ssi + f\ssj)$, where 
\begin{eqnarray} \label{EQN:FBALSARA}
f\ssi &=&\frac{|\nabla\cdot{\bf v}|\ssi }{|\nabla\cdot{\bf v}|\ssi\;+\;|\nabla\times{\bf v}|\ssi\; +\;0.001\,c\ssi /h\ssi}\,.
\end{eqnarray}
In regions of strong compression (i.e. shocks), the $|\nabla \cdot {\bf v}|$ terms tend to dominate over the $|\nabla \times {\bf v}|$ term, so $f\ssi  \rightarrow 1$. In regions where vorticity dominates (i.e. shear flows), the $|\nabla \times {\bf v}|$ term dominates, so $f\ssi  \rightarrow 0$.  

The second device (which can be used in conjunction with the first) is time-dependent viscosity (Morris \& Monaghan 1997). In time-dependent viscosity, each particle $i$ has its own value of $\alpha_i$, which evolves according to the equation
\begin{eqnarray} \label{EQN:TDVISC}
\frac{d\alpha\ssi }{dt}&=& \frac{\alpha_{_{\rm MIN}}-\alpha\ssi}{\tau\ssi}+S\ssi\,.
\end{eqnarray}
Here $\alpha_{_{\rm MIN}}$ is the default value of $\alpha_i$, and $\tau\ssi$ is the e-folding time on which $\alpha_i$ relaxes to $\alpha_{_{\rm MIN}}$, if the source term,
\begin{eqnarray} \label{EQN:TDSOURCE}
S\ssi&=&{\rm MAX}\left\{-\left(\nabla\cdot{\bf v}\right)\ssi,0\right\}\,\left(\alpha_{_{\rm MAX}}-\alpha\ssi\right)\,,
\end{eqnarray}
vanishes. Reasonable results are obtained with $\alpha_{_{\rm MIN}}=0.1$, since a small residual artificial viscosity is needed to suppress high-frequency particle noise. The e-folding time is given by $\tau\ssi = {\cal C}\,h\ssi/c\ssi$ with ${\cal C}\sim 5$ (i.e. roughly a sound-crossing time for the smoothing kernel). The source term ensures that if particle $i$ enters a shock, $\alpha_i$ quickly increases towards $\alpha_{_{\rm MAX}}\sim 1$, but as soon as the shock is passed it decays back to $\alpha_{_{\rm MIN}}$. If we use $(\alpha,\beta)$ viscosity, then we set $\beta\ssi = 2\alpha\ssi$.

{ The third device is the pattern-matching switch described by Cartwright \& Stamatellos (2010; arXiv:1004.3694). This switch is very effective in pure Keplerian discs, i.e. non-self-gravitating equilibrium discs modelled in the frame of reference of the central star, but has not yet been adapted to work in more general situations.}

%%%%%%%%%%%%%%%%%%%%%%%%%%%%%%%%%%%%%%%%%%%%
\section{Self-gravity} \label{S:SPHGRAVITY}%
%%%%%%%%%%%%%%%%%%%%%%%%%%%%%%%%%%%%%%%%%%%%

Although the calculation of gravitational accelerations resembles an N-body problem, with forces between point masses, one should -- for consistency with the calculation of SPH accelerations -- take proper account of fact that the underlying density field, given by Eqn. (\ref{EQN:SPHRHO}), is actually continuous, and the gravitational potential is related to this continuous density field by Poisson's equation, $\nabla^2\Phi =4\,\pi\,G\,\rho$. Price \& Monaghan (2007) derive the equations of self-gravitating SPH by including the gravitational potential in the Lagrangian, and then proceeding as in Price \& Monaghan (2004). It is then necessary to introduce two additional kernel functions, the gravitational acceleration kernel ($\phi'$) and the gravitational potential kernel ($\phi$, called the softening kernel by Price \& Monaghan 2007),
\begin{eqnarray}\label{EQN:GRAVFORCEKERNEL}
\phi'({\bf r},h)&=&\frac{4\,\pi}{r^2}\int\limits_0^r{W({\bf r}',h)\,r'^2\,dr'}\,,\\\label{EQN:GRAVPOTKERNEL}
\phi({\bf r},h)&=&4\,\pi\left(-\frac{1}{r}\int\limits_0^r{W({\bf r}',h)\,r'^2\,dr'}+\int\limits_0^r{W({\bf r}',h)\,r'\,dr'}-\int\limits_0^{{\cal R}h}{W({\bf r}',h)\,r'\,dr'}\right)\,.
\end{eqnarray}
As with the basic kernel function, $W$, and its other derivatives, these new gravitational kernels are computed in advance, on a grid, and stored, so that subsequently values can be obtained efficiently by interpolation. The forms of both of these kernels are discussed in Appendix \ref{A:KERNEL}. Using these kernels, Price \& Monaghan (2007) show that the gravitational acceleration of particle $i$ is 
\begin{eqnarray}
\left(\frac{d{\bf v}\ssi}{dt}\right)_{_{\rm GRAV}}&=&-\,G\sum\limits_{j=1}^{N}{m\ssj\,\overline{\phi'}({\bf r}\ssij,h\ssi,h\ssj)\,\hat{\bf r}\ssij}\,-\,\frac{G}{2}\sum\limits_{j=1}^{N}\left\{\frac{\zeta_i}{\Omega_i}\nabla W\ssi({\bf r}\ssij,h\ssi)+\frac{\zeta_j}{\Omega_j}\nabla W\ssi({\bf r}\ssij,h\ssj)\right\}\,,
\label{EQN:GRADHGRAV1}
\end{eqnarray}
where
\begin{eqnarray} 
\overline{\phi'}({\bf r}\ssij , h\ssi , h\ssj ) &=&
\frac{\phi'({\bf r}\ssij , h\ssi ) + \phi'({\bf r}\ssij , h\ssj )}{2}\,,
\label{EQN:PHIPBAR}
\end{eqnarray}
\begin{eqnarray}
\zeta_i = \frac{\partial h\ssi}{\partial h\ssj} 
\sum \limits_{j=1}^{N} m_j \frac{\partial \phi}{\partial h}
({\bf r}\ssij, h\ssi)\,,
\label{EQN:GRADHZETA}
\end{eqnarray}
and $\Omega\ssi$ is given by Eqn. (\ref{EQN:OMEGA}). The two summation terms in Eqn. (\ref{EQN:GRADHGRAV1}) are, respectively, the kernel-softened gravitational acceleration, and the `grad-h' corrections that account for adaptive smoothing lengths. The $\zeta\ssi$ term is calculated and stored when other SPH quantities are calculated (i.e. $\rho\ssi$, $(\nabla\cdot{\bf v})\ssi$, $\Omega\ssi$, etc.). To compute $\zeta\ssi$ requires $\partial \phi /\partial h$, which can be calculated and stored, once the form of $W$ has been specified (see Appendix \ref{A:KERNEL}). The gravitational potential at the position of particle $i$ due to all other particles is
\begin{eqnarray} \label{EQN:KSGRAVPOT}
\Phi\ssi &=& G\, \sum \limits_{j=1}^{N} m\ssj \,
\overline{\phi}({\bf r}\ssij , h\ssi, h\ssj) \,,
\end{eqnarray}
where
\begin{eqnarray}\label{EQN:PHIBAR}
\overline{\phi}({\bf r}\ssij , h\ssi , h\ssj) &=&
\frac{\phi({\bf r}\ssij , h\ssi) + \phi({\bf r}\ssij , h\ssj)}{2}\,.
\end{eqnarray}
If we choose standard SPH or Godunov SPH, the second summation in Eqn. (\ref{EQN:GRADHGRAV1}) is omitted, and the total energy is not as well conserved (see Price \& Monaghan 2007).

To compute gravitational accelerations exactly, using Eqns. (\ref{EQN:GRADHGRAV1}), (\ref{EQN:PHIPBAR}) \& (\ref{EQN:GRADHZETA}), requires a summation over all particle pairs and is therefore an ${\cal O}(N^2)$ process. To speed up the computation of gravitational accelerations, SEREN uses a Barnes-Hut tree (Barnes \& Hut 1986).  The resulting gravitational accelerations are not exact, but the resulting small fractional errors are considered acceptable, since there are other comparable or larger sources of error. The implementation of the gravity tree is described in Section \ref{S:TREE}.

%%%%%%%%%%%%%%%%%%%%%%%%%%%%%%
\subsection{Periodic gravity}%
%%%%%%%%%%%%%%%%%%%%%%%%%%%%%%

Cosmological simulations (e.g. Springel et al. 2005) and simulations of turbulent molecular clouds (e.g. Klessen, Heitsch \& Mac Low 2000) often set out to model a representative piece of an infinite (or much more extended) medium, by assuming that the infinite medium consists of an infinite number of replicas of the main computational domain, extending periodically in all directions, and then employing periodic boundary conditions. For purely hydrodynamic simulations, periodic wrapping is sufficient to give acceptable boundary conditions. When self-gravity is invoked, we must include a contribution to the acceleration from all the replicas of the computational domain, extending to infinity. SEREN does this using the Ewald method (Hernquist, Bouchet \& Suto 1991; Klessen 1997). If the computational domain is a cube of side-length $L$, the total gravitational acceleration exerted on particle $i$ by all of the infinite replicas of particle $j$ (but not directly by the particle $j$ itself) is 
\begin{eqnarray}
\left( \frac{d{\bf v}\ssi}{dt} \right)_{_{\rm EWALD,j}} &=& G\,m\ssj 
\left( {\bf f}({\bf r}\ssij) + \frac{{\bf r}\ssij}{|{\bf r}\ssij|^3}
\right)\,, \label{EQN:EWALD1}
\end{eqnarray}
where
\begin{eqnarray}
{\bf f}({\bf r}) &=& 
 - \sum \limits_{\bf n} \frac{{\bf r} - {\bf n}L}
{|{\bf r} - {\bf n}L|^3} \left\{ {\rm erfc}\,(\alpha |{\bf r} - 
{\bf n}L|)  + \frac{2 \alpha}{\sqrt{\pi}} |{\bf r} - {\bf n}L| 
\exp\left(- \alpha^2 |{\bf r} - {\bf n}L|^2 \right) \right\} 
 - \frac{1}{L^3} \sum \limits_{\bf k} 
\frac{4\pi{\bf k}}{k^2} \exp \left( - \frac{k^2}{4\alpha^2} \right) 
\sin ({\bf k \cdot r}) \label{EQN:EWALD2}
\end{eqnarray}
and $\alpha = 2/L$.  The first summation in Eqn. (\ref{EQN:EWALD2}) is over all replicas in all directions (i.e. all ${\bf n}$-space) and the second summation is over all phase-space (i.e. all ${\bf k}$-space). The summations converge rapidly and can be truncated for $|{\bf r}-{\bf n}L|<3.6L$ and $k^2<40\pi^2/L^2$. SEREN computes the dimensionless correction forces for a wide range of separations and tabulates the values in a look-up table.  

%%%%%%%%%%%%%%%%%%%%%%%%%%%%%%%%%%%%%%%%%%%%
\section{Thermal physics} \label{S:THERMAL}%
%%%%%%%%%%%%%%%%%%%%%%%%%%%%%%%%%%%%%%%%%%%%

SEREN contains several equation-of-state (EOS) algorithms which can be selected using Makefile options. In all cases we assume that the gas is ideal, and so the pressure and specific internal energy are related by
\begin{eqnarray}\label{EQN:IDEALGAS}
P&=&\frac{\rho k_{_{\rm B}}T}{\overline{m}}\;\,=\;\,(\gamma-1)\rho u\,,
\end{eqnarray}
where $k_{_{\rm B}}$ is Boltzmann's constant, $\overline{m}$ is the mean gas-particle mass, and $\gamma$ is the ratio of specific heats. With Options 1 to 3 below there is no need to solve the SPH energy equation, whereas with Options 4 and 5 there is.

\vspace{0.2cm}

\noindent {\sc 1. Isothermal equation of state.} If the gas is isothermal at temperature $T_{_{\rm O}}$, 
\begin{eqnarray}
P&=&c_{_{\rm O}}^2\,\rho\,, \label{EQN:ISOTHERMALEOS}
\end{eqnarray}
with constant isothermal sound speed,  $c_{_{\rm O}}=\left(k_{_{\rm B}}T_{_{\rm O}}/\overline{m}\right)^{1/2}$.

\vspace{0.2cm}

\noindent {\sc 2. Polytropic equation of state.} The polytropic EOS has the form
\begin{eqnarray} \label{EQN:POLYTROPIC}
P &=& K\,\rho^{\eta}
\end{eqnarray}
where $K$ is the polytropic constant and $\eta$ is the polytropic exponent; the polytropic index is $n=(\eta-1)^{-1}$.

\vspace{0.2cm}

\noindent {\sc 3. Barotropic equation of state.} SEREN includes a barotropic equation of state of the form 
\begin{eqnarray}
T&=&T_{_{\rm O}}\,\rho\left\{1+\left(\frac{\rho}{\rho_{_{\rm CRIT}}}\right)^{\gamma-1}\right\}\,. 
\label{EQN:BAROTROPIC}
\end{eqnarray}
This mimics the behaviour of interstellar gas in molecular clouds, where the gas is optically thin to its cooling radiation and approximately isothermal (at $T_{_{\rm O}}\sim 10\,{\rm K}$) when the density is low ($\rho<\rho_{_{\rm CRIT}}\sim 10^{-13}\,{\rm g}\,{\rm cm}^{-3}$), and optically thick to its own cooling radiation and approximately adiabatic (e.g. with $\gamma\simeq 5/3$) at higher densities ($\rho>\rho_{_{\rm CRIT}}$).

\vspace{0.2cm}

\noindent {\sc 4. Adiabatic equation of state.} We integrate the internal energy equation explicitly (using Eqn. \ref{EQN:GRADHENEQN}) and then calculate the thermal pressure from Eqn. (\ref{EQN:IDEALGAS}). Changes in the specific internal energy are solely due to compressional and/or viscous heating.

\vspace{0.2cm}

\noindent {\sc 5. Radiative cooling.} The method of Stamatellos et al. (2007) is used to capture realistically the main effects of radiative heating and cooling (in the optically thin, thick and intermediate regimes), but without the expense of a full radiative transfer calculation. This algorithm uses local functions of state (namely the density, temperature and gravitational potential) to compute an approximate optical depth to infinity, and hence to obtain an approximate cooling rate. This cooling rate is then used to solve the energy equation implicitly, and hence to determine the thermal evolution of the gas.

\vspace{0.2cm}

\noindent {\sc 6. Ionising radiation.} SEREN also includes the option to model a single discrete source of ionising radiation (i.e. an OB star or tight cluster of OB stars) using the algorithm of Bisbas et al. (2009). This algorithm generates an isotropic distribution of HEALPix rays, which are split into smaller child rays wherever finer resolution is needed. The rays propagate until they reach the ionisation front, where they are terminated. Particles well inside the H{\sc ii} region are given a high temperature ($\sim 10,000\,{\rm K}$) and particles well outside the H{\sc ii} region are treated with one of the EOS algorithms listed above. There is a region with thickness of order the local smoothing length in which the temperature variation is smoothed, so as to avoid problems associated with abrupt temperature discontinuities.

%%%%%%%%%%%%%%%%%%%%%%%%%%%%%%%%%%%%%%%%%%%%%
\section{Time integration} \label{S:TIMEINT}%
%%%%%%%%%%%%%%%%%%%%%%%%%%%%%%%%%%%%%%%%%%%%%

%%%%%%%%%%%%%%%%%%%%%%%%%%%%%%%%%%%%%%%%%%%%%%%%%%%%%%%
\subsection{Integration schemes} \label{SS:INTSCHEMES}%
%%%%%%%%%%%%%%%%%%%%%%%%%%%%%%%%%%%%%%%%%%%%%%%%%%%%%%%

SEREN offers a choice of four integration schemes: 2nd-order Runge-Kutta, 2nd-order Leapfrog (kick-drift-kick {\it and} drift-kick-drift) and 2nd-order Predictor-Corrector. The default choice is { the 2nd-order Leapfrog drift-kick-drift}:
\begin{eqnarray}
{\bf r}^{n+1/2}\ssi &=& {\bf r}^n\ssi + {\bf v}^n\ssi\,\frac{\Delta t}{2}\,, \label{EQN:2LV1} \\
{\bf v}^{n+1/2}\ssi &=& {\bf v}^n\ssi + {\bf a}^{n-1/2}\ssi\,\frac{\Delta t}{2}\,, \label{EQN:2LV2} \\
u^{n+1/2}\ssi &=& u^{n}\ssi + \dot{u}^{n-1/2}\ssi\,\frac{\Delta t}{2}\,, \label{EQN:2LV3} \\
{\bf v}^{n+1}\ssi &=& {\bf v}^n\ssi + {\bf a}^{n+1/2}\ssi\,\Delta t\,, \label{EQN:2LV4} \\
{\bf r}^{n+1}\ssi &=& {\bf r}^n\ssi + \frac{1}{2}({\bf v}^{n}\ssi + {\bf v}^{n+1}\ssi)\,\Delta t\,, \label{EQN:2LV5} \\
u^{n+1}\ssi &=& u^{n}\ssi + \dot{u}^{n+1/2}\ssi\,\Delta t\,. \label{EQN:2LV6} 
\end{eqnarray}
The main advantage of this scheme is that it only requires one acceleration calculation per timestep, as opposed to two in the 2nd-order Runge-Kutta scheme.  Leapfrog schemes (both the Leapfrog kick-drift-kick and drift-kick-drift) are {\it symplectic} (i.e. they conserve phase-space) and so they are more stable for orbital integration (for example, in disc simulations). They are also, in principle, time-reversible for constant, global timesteps.  The use of block time-stepping breaks exact time-reversibility (see Section \ref{SS:BLOCKTIMESTEPS}), and also breaks exact momentum and angular momentum conservation. The other integration schemes are included because some perform better than the Leapfrog scheme in non--self-gravitating problems, and to allow comparison with other codes that use different integrators.

%%%%%%%%%%%%%%%%%%%%%%%%%%%%%%%%%%%%%%%%%%%%%%%%%%%
\subsection{Optimal timesteps} \label{SS:TIMESTEP}%
%%%%%%%%%%%%%%%%%%%%%%%%%%%%%%%%%%%%%%%%%%%%%%%%%%%

SEREN calculates (but does not explicitly use) the {\it optimal} timestep for particle $i$, $\Delta t_i$, by determining the minimum value of three separate timesteps. The first is based on a modified Courant condition of the form 
\begin{eqnarray} \label{EQN:DTCOURANT}
\Delta t_{_{\rm COUR}} &=& \gamma_{_{\rm COUR}}
\frac{h_i}{(1 + 1.2 \alpha) c_i + (1 + 1.2 \beta) h_i |\nabla \cdot {\bf v}|_i}\,.
\end{eqnarray}
The denominator contains the $h_i |\nabla \cdot {\bf v}|_i$ term (which is frame-independent) instead of the absolute speed, $|{\bf v}|_i$ (which is normally used in the Courant condition). The terms involving $\alpha$ and $\beta$ in the denominator account for particles that are in the vicinity of shocks. The second timestep condition is an acceleration condition similar to those used in some N-body codes, i.e. 
\begin{eqnarray} \label{EQN:DTACCEL}
\Delta t_{_{\rm ACCEL}} &=& \gamma_{_{\rm ACCEL}}
\sqrt{\frac{h_i}{|{\bf a}|_i+\eta_a}}
\end{eqnarray}
where $\eta_a$ is a small positive acceleration to ensure the denominator does not at any time fall to zero. The third timestep condition is the heating condition, which limits the fractional change in the internal energy per timestep, 
\begin{eqnarray} \label{EQN:DTENERGY}
\Delta t_{_{\rm ENERGY}} &=& \gamma_{_{\rm ENERGY}} 
\frac{u_i}{|du/dt|_i+\eta_{\dot{u}}}\,,
\end{eqnarray}
where $\eta_{\dot{u}}$ is a small positive heating rate to ensure the denominator does not fall to zero. This timestep criterion is only used when the SPH energy equation (Eqn. \ref{EQN:GRADHENEQN}) is solved explicitly. If we solve the energy equation implicitly (e.g. Stamatellos et al. 2007), we only use the Courant and acceleration timesteps, Eqns. (\ref{EQN:DTCOURANT}) and (\ref{EQN:DTACCEL}), to compute the optimal timestep for particle $i$, $\Delta t_i$. 

%%%%%%%%%%%%%%%%%%%%%%%%%%%%%%%%%%%%%%%%%%%%%%%%%%%%%%%%%%%%%%%%%%%%
\subsection{Hierarchical block timesteps} \label{SS:BLOCKTIMESTEPS}%
%%%%%%%%%%%%%%%%%%%%%%%%%%%%%%%%%%%%%%%%%%%%%%%%%%%%%%%%%%%%%%%%%%%%

SEREN uses hierarchical block time-stepping (e.g. Aarseth 2003) to reduce the run-time of a simulation. In a typical star formation simulation, only a small fraction of the particles might require very small timesteps, for example those passing through a shock or those near the centre of a condensation. If a global timestep is used, accelerations are recalculated for all particles, irrespective of whether the recalculation is really needed. Instead, we allow each particle to have its own timestep, chosen from a binary hierarchy of possible values, $\Delta t_n = 2^n\,\Delta t_{_{\rm MIN}}$, where $n = 0,1,2, ... , n_{_{\rm MAX}}$. Particle $i$ is then allocated the largest value of $\Delta t_n$ from this hierarchy that is smaller than its optimal timestep, $\Delta t_i$ (based on Eqns. \ref{EQN:DTCOURANT}, \ref{EQN:DTACCEL} and \ref{EQN:DTENERGY}). By restricting the ratio of timesteps to integer powers of $2$, we ensure that the particles are always synchronised at the end of the largest timestep, $\Delta t_{_{\rm MAX}} = 2^{n_{_{\rm MAX}}}\,\Delta t_{_{\rm MIN}}$.  

The acceleration of a particle is then recalculated with a frequency determined by its allocated timestep, $\Delta t_n$. The most expensive parts of this recalculation are those associated with walking the trees. At any time, the positions, velocities and thermodynamic properties of particles whose accelerations do not yet need to be recalculated are simply estimated by extrapolation. 

The timestep for a particle is recalculated at the end of its current timestep, using Eqns. (\ref{EQN:DTCOURANT}) to (\ref{EQN:DTENERGY}).  When the allocated timestep of a particle decreases (i.e. it moves to lower $n$ in the hierarchy), there is no problem, because any lower timestep in the hierarchy is automatically synchronised with the higher one from which the particle is descending. On the other hand, this is not necessarily the case when a particle's allocated timestep increases (i.e. it moves to higher $n$ in the hierarchy).  In this situation, we have to check that the lower timestep is correctly synchronised with the higher one before we can move the particle up (i.e. increase its allocated timestep). In addition, we only allow a particle to increase its allocated timestep one level at a time.

As shown by Saitoh \& Makino (2009), SPH can perform poorly when neighbouring particles have very different timesteps. For example, in a high Mach-number shock, the particles may interpenetrate because particles from the low-density pre-shock gas have much longer timesteps than those in the high-density post-shock gas, and therefore in a single timestep they advance deep into the shocked region. SEREN mitigates this effect by broadcasting each particle's allocated timestep to all it neighbours. If one of the neighbours $j$ of particle $i$ has an allocated timestep which is more than two levels higher in the hierarchy { (i.e. more than a factor 4 longer; $t_j > 4\,t_i$), the neighbour's timestep is automatically reduced to $t_j = 4\,t_i$ as soon as the timestep hierarchy is correctly synchronised.}

%%%%%%%%%%%%%%%%%%%%%%%%%%%%%%%%%%%%%%%%%
\section{Sink particles} \label{S:SINKS}%
%%%%%%%%%%%%%%%%%%%%%%%%%%%%%%%%%%%%%%%%%

Sink particles are used in SPH to allow simulations of star formation to be followed longer (Bate, Bonnell \& Price 1995, hereafter BBP95).  Gravitational collapse inevitably leads to high densities, short smoothing lengths, high accelerations, and therefore short timesteps. Under these circumstances, even the use of block time-stepping (Section \ref{SS:BLOCKTIMESTEPS}) cannot prevent run-times from becoming impractically long. To circumvent this problem, we replace dense condensations with sink particles. A sink particle possesses the collective properties of the condensation it represents (i.e. mass, centre-of-mass position and net momentum) but does not retain any information about the internal structure and evolution of the condensation. Thus SPH particles that would otherwise have continued evolving inexorably towards higher density (thereby using up ever increasing amounts of CPU-time) are instead excised from the simulation. This means that the dynamics of the remaining more diffuse gas, and the formation of additional condensations, can be followed in an acceptable run-time. The assumption is made that -- in the absence of feedback from the resulting protostar -- the only important effect that the material inside a sink particle will have on its surroundings is due to its gravitational field. Thus sink particles interact gravitationally, but not hydrodynamically, with other sink and SPH particles.

A sink particle is created when an SPH particle satisfies all the stipulated sink-creation criteria. These criteria are divided into default criteria and optional criteria. The SPH particle which triggers the formation of a sink is referred to as the seed particle. The default criteria for sink creation are then (i) that the SPH density of the seed particle is greater than $\rho_{_{\rm SINK}}$, and (ii) that there is no other sink particle within $2r_{_{\rm SINK}}$ of the seed particle (i.e. a sink particle should not be formed overlapping a pre-existing sink particle). In principle, $\rho_{_{\rm SINK}}$ and $r_{_{\rm SINK}}$ can be chosen independently. However, the results are only realistic if the material going initially into a sink particle is resolved. The default procedure in SEREN is to set $r_{_{\rm SINK}}={\cal R}h_s$, where $h_s$ is the smoothing length of the seed particle. This means that different sink particles have slightly different radii. The option exists in SEREN to prescribe a universal $r_{_{\rm SINK}}$.

The four optional sink-creation criteria are (iii) that the mean density of the seed particle and all its neighbours exceeds $\rho_{_{\rm SINK}}$ (this ensures that a stochastic density fluctuation does not result in the formation of a sink); (iv) that the SPH velocity divergence of the seed particle is negative, $\left(\nabla\cdot{\bf v}\right)_s<0$ (this ensures that the particles going into the sink are condensing, and not being sheared apart); (v) that the SPH acceleration divergence of the seed particle is negative, $\left(\nabla\cdot{\bf a}\right)_s<0$ (this ensures that the condensation is not being torn apart by tidal forces); (vi) that the total mechanical energy of the seed particle and its neighbours (kinetic plus gravitational potential energy in the centre-of-mass frame) is negative.

Only one sink particle can be created in any one timestep; otherwise the possibility would exist to generate multiple overlapping sinks. At each timestep, SEREN loops over all the SPH particles, and finds those whose SPH density (or, if required, mean density) exceeds $\rho_{_{\rm SINK}}$. These candidate seed particles are then ordered in a list of decreasing SPH density (or mean density), and SEREN runs through this list until it finds a seed particle that satisfies all the creation criteria, and creates a sink particle out of this seed particle and all its neighbours.

An SPH particle $i$ is accreted by an existing sink particle $s$ if (a) the SPH particle lies inside the sink-particle's radius, $|{\bf r}_i-{\bf r}_s|\leq r_{_{\rm SINK}}$, and (b) the kinetic plus gravitational energy of the two-body system comprising the sink-particle and the SPH-particle is negative. The SPH particle's mass, linear and angular momentum are then assimilated by the sink particle, and the SPH particle itself is removed from the simulation. When determining which SPH particles are accreted by which sink particles, we first compile a list of all the SPH particles which are to be accreted by each sink, and only when these lists are complete do we update the sink properties (mass, position, momentum) to account for the SPH particles it has just assimilated. This is necessary because otherwise the accretion process would depend on the order in which the SPH particles were interrogated.

%%%%%%%%%%%%%%%%%%%%%%%%%%%%%%%%%%%%%%%%%%%%
\section{N-body integrator} \label{S:NBODY}%
%%%%%%%%%%%%%%%%%%%%%%%%%%%%%%%%%%%%%%%%%%%%

Simulations of star formation very often result in the formation of multiple stellar systems. Such simulations are modelled with hydrodynamical codes until most of the gas has been accreted by protostars, or dispersed by feedback; this is referred to as the {\it accretion phase}. In the absence of magnetic fields and feedback, the accretion phase is driven entirely by the competition between thermal pressure, viscosity, and gravity. The system then enters the {\it ballistic phase}, in which N-body dynamics modify the final clustering and binary properties, typically over a period of several tens of crossing times (Van Albada 1968). SPH simulations are often terminated after the accretion phase and not evolved through the ballistic phase.

SEREN includes an N-body integrator, so that it can follow both the accretion phase and the ballistic phase, in a single simulation.  SEREN switches from an SPH simulation of the accretion phase, to an N-body simulation of the ballistic phase, if one of two conditions are met: either the simulation has reached the end-time stipulated in the parameters file, or a critical fraction of the original gas mass has been accreted by sink particles.  At the switch-over, SEREN identifies any SPH particles which are strongly bound to a particular sink. On the assumption that these SPH particles are either about to be accreted by that sink, or will form a tightly bound disc around it (and eventually be accreted or form a planetary system), they are instantaneously accreted by the sink to which they are bound. This ensures that their contribution to the overall gravitational potential is not suddenly lost at the switch-over.  

One problem that corrupts N-body codes is inaccuracies resulting from close interactions.  These can build up over the course of a simulation, or materialise quickly in near head-on interactions, causing large energy errors. A variety of techniques has been employed to alleviate this problem, such as using very short timesteps (e.g. Portegies Zwart et al. 2001), gravity-softening (Aarseth 2003) or transformation of the equations of motion (e.g. KS regularization; Stiefel \& Scheifele 1971).  In SEREN, the N-body code retains the kernel-softened gravity used in the SPH code, in order to ensure that the gravitational accelerations are computed consistently between the two parts of a simulation.  This has the advantage of preventing large energy errors due to close interactions, but has the disadvantage of preventing the formation of close binaries (separations less than $r_{_{\rm SINK}}$).

%%%%%%%%%%%%%%%%%%%%%%%%%%%%%%%%%%%%%%%%%%%%%%%%%%%
\subsection{Hermite integrator} \label{SS:HERMITE}%
%%%%%%%%%%%%%%%%%%%%%%%%%%%%%%%%%%%%%%%%%%%%%%%%%%%

The N-body integrator of choice is a fourth-order Hermite integrator (Makino 1991; Makino \& Aarseth 1992). The Hermite integrator has been presented in two different forms in the literature, either as a fourth-order leapfrog scheme or as a fourth-order predictor-corrector scheme (Aarseth 2003).  SEREN uses the predictor-corrector version of the Hermite integrator. Both forms are considered superior to other 4th-order N-body integrators, in the sense of giving better energy conservation and allowing longer timesteps (Makino 1991; Aarseth 2003).  The leapfrog version of the Hermite scheme also maintains many of the properties of a traditional 2nd order leapfrog integrator (for example, it is symplectic), but it is of higher order by virtue of using both the acceleration and its first time derivative.  

The N-body code uses a global timestep informed by the Aarseth (2001) criterion,
\begin{eqnarray}
\Delta t\ssi &=& \gamma \, \sqrt{\frac{|{\bf a}\ssi| |\ddot{\bf a}\ssi| + 
|\dot{\bf a}\ssi|^2}{|\dot{\bf a}\ssi| |\dddot{\bf a}\ssi| + 
|\ddot{\bf a}\ssi|^2}}\,. \label{EQN:DTAARSETH}
\end{eqnarray}
Here $\dot{\bf a}$, $\ddot{\bf a}$ and $\dddot{\bf a}$ are, respectively, the 1st, 2nd and 3rd time derivatives of the acceleration, calculated at the end of the previous timestep; $\gamma$ is an accuracy factor of order $\sim 0.1$ (Makino \& Aarseth 1992). Next we calculate the acceleration and its time-derivative (sometimes called the {\it jerk}) at the beginning of the step. The acceleration is given by 
\begin{eqnarray}
{\bf a}^n\ssi &=& -\,G\, \sum \limits_{j=1}^{N}{m\ssj\,\overline{\phi'}({\bf r}\ssij,h\ssi,h\ssj)\,\hat{\bf r}\ssij}\,,
\end{eqnarray}
where $\overline{\phi'}$ is the same gravitational-acceleration kernel as used in calculating kernel-softened gravitational accelerations in SPH (Eqn. \ref{EQN:GRAVFORCEKERNEL}). The kernel-softening means we must account for the rate of change of the kernel function and include extra terms in the expression for the jerk (Makino \& Aarseth 1992). Using the same notation as in Section \ref{S:SPHGRAVITY}, the expression for the jerk becomes 
\begin{eqnarray}
\dot{\bf a}^n\ssi &=& -\,G\,\sum \limits\ssj 
{\frac{m\ssj\,\overline{\phi'}({\bf r}\ssij , h\ssi , h\ssj)}{|{\bf r}\ssij |}{\bf v}\ssij} \; + \;3\,G\, \sum \limits\ssj^{N}
{\frac{m\ssj\,({\bf r}\ssij  \cdot {\bf v}\ssij)\,\overline{\phi'}({\bf r}\ssij , h\ssi , h\ssj)}{|{\bf r}\ssij |^3} {\bf r}\ssij } \; 
- \;4\,\pi\,G\, \sum \limits\ssj^{N} {\frac{m\ssj\,\,({\bf r}\ssij  \cdot {\bf v}\ssij)\, 
\overline{W}({\bf r}\ssij , h\ssi , h\ssj)}{|{\bf r}\ssij |^2}{\bf r}\ssij}\,.
\label{EQN:JERK}
\end{eqnarray}
The particle positions and velocities are then advanced to the end of the timestep, 
\begin{eqnarray}
{\bf r}^{n+1}\ssi &=& {\bf r}\ssi^{n} + {\bf v}\ssi^{n} \Delta t + \frac{1}{2}{\bf a}\ssi^{n} \Delta t^2 + \frac{1}{6} \dot{\bf a}\ssi^{n} \Delta t^3\,, \label{EQN:HERMPREDICT1} \\
{\bf v}^{n+1}\ssi &=& {\bf v}\ssi^{n} + {\bf a}\ssi^{n} \Delta t + \frac{1}{2}\dot{\bf a}\ssi^{n} \Delta t^2\,. \label{EQN:HERMPREDICT2}
\end{eqnarray}
We calculate the acceleration and jerk again using the new positions and velocities.  We can thus calculate the second and third time derivatives at the beginning of the step (Makino \& Aarseth 1992),
\begin{eqnarray}
\ddot{\bf a}\ssi^{n} &=& \frac{2 \left( -3({\bf a}\ssi^{n} - {\bf a}\ssi^{n+1}) - (2\dot{\bf a}\ssi^{n} + \dot{\bf a}\ssi^{n+1})\Delta t \right)}{\Delta t^2}\,,  \label{EQN:A2} \\
\dddot{\bf a}\ssi^{n} &=& \frac{6 \left( 2({\bf a}\ssi^{n} - {\bf a}\ssi^{n+1}) + (\dot{\bf a}\ssi^{n} + \dot{\bf a}\ssi^{n+1})\Delta t \right)}{\Delta t^3}\,. \label{EQN:A3}
\end{eqnarray}
Finally, we add the higher order terms to the position and velocity vectors, 
\begin{eqnarray}
{\bf r}\ssi^{n+1} &=& {\bf r}\ssi^{n+1} + \frac{1}{24}\ddot{\bf a}\ssi^{n} \Delta t^4 + \frac{1}{120} \dddot{\bf a}\ssi^{n} \Delta t^5\,, \label{EQN:HERMCORRECT1} \\
{\bf v}\ssi^{n+1} &=& {\bf v}\ssi^{n+1} + \frac{1}{6}\ddot{\bf a}\ssi^{n} \Delta t^3 +\frac{1}{24}\dddot{\bf a}\ssi^{n} \Delta t^4\,. \label{EQN:HERMCORRECT2}
\end{eqnarray} 
The values of ${\bf a}$, $\dot{\bf a}$, $\ddot{\bf a}$ and $\dddot{\bf a}$ computed at the end of the timestep allow the code to calculate the next time step using Eqn. (\ref{EQN:DTAARSETH}). This is not possible on the very first timestep, and there we use explicit equations to calculate $\ddot{\bf a}$ and $\dddot{\bf a}$ (e.g. Aarseth 2001; his Eqns. 6 and 7); all subsequent timesteps are determined using $\ddot{\bf a}$ and $\dddot{\bf a}$ from Eqns. (\ref{EQN:A2}) and (\ref{EQN:A3}).  

%%%%%%%%%%%%%%%%%%%%%%%%%%%%%%%%%%%%%%%%%%%%%%%%%%%%%%%%%%%%%%%%%%%%
\subsection{Identification of multiple systems} \label{SS:BINARYID}%
%%%%%%%%%%%%%%%%%%%%%%%%%%%%%%%%%%%%%%%%%%%%%%%%%%%%%%%%%%%%%%%%%%%%

During the N-body simulation, SEREN automatically searches for binaries and hierarchical triples and quadruples. There is no single robust method for identifying a bound, stable multiple system that contains an arbitrary number of components. We use a simple two-stage procedure. The first stage is to identify all binary systems present at the current time. This involves calculating the two-body energies of all star-pairs in the simulation. If (a) stars 1 and 2 are found to be mutually most-bound (i.e. the two-body energy of stars 1 and 2 is a minimum and negative for both stars), and (b) stars 1 and 2 are not bound to any other stars (i.e. the two-body energies of 1 and 2 with all other stars are positive), then they are identified as a bound binary system. If the primary and secondary masses are $m_1$ and $m_2$ respectively, the instantaneous relative displacement is ${\bf r}_{12} \equiv {\bf r}_1 - {\bf r}_2$, and the instantaneous relative velocity is ${\bf v}_{12} = {\bf v}_1 - {\bf v}_2$, then the two-body energy and angular momentum are\begin{eqnarray}\label{EQN:2BODYENERGY}
E_{b} &=& \frac{1}{2}\mu\,|{\bf v}_{12}|^2\, + \,G\,m_1\,m_2\,
\overline{\phi}({\bf r}_{12},h_1,h_2)\,,\\\label{EQN:2BODYANGMOM}
{\bf L} &=& \mu \, {\bf r}_{12} \times {\bf v}_{12} \,, \label{EQN:2BODYANGMOM}
\end{eqnarray}
where $\mu=m_1m_2/(m_1+m_2)$. The orbital binary parameters are then given by
\begin{eqnarray} 
q &=& \frac{m_2}{m_1}\,, \label{EQN:MASSRATIO} \\
a &=& - \frac{G\,m_1\,m_2}{2E_{b}}\,, \label{EQN:SMA} \\
e &=& \left( 1 - \frac{|{\bf L}|^2}{G\,a\,(m_1 + m_2)\,\mu^2} \right)^{1/2}\,. 
\label{EQN:ECCENT} 
\end{eqnarray}

The next stage is to search for hierarchical systems. In order to facilitate this search, each binary found in the previous step is replaced by a single ghost-binary particle. We then repeat the procedure performed in the first stage, searching for any mutually most-bound pairs, but now using the ghost-binaries and the remaining unattached stars. If a ghost-binary is found to be most-bound to a single star and vice-versa, they are identified as an hierarchical triple, and the orbit of the system is calculated as above and recorded.  If two ghost-binaries are found to be most-bound to each other, then they are recorded as an hierarchical quadruple.

%%%%%%%%%%%%%%%%%%%%%%%%%%%%%%
\section{Tree} \label{S:TREE}%
%%%%%%%%%%%%%%%%%%%%%%%%%%%%%%

SEREN uses an implementation of the Barnes-Hut tree (Barnes \& Hut 1986, Pfalzner \& Gibbon 1996) to rapidly obtain neighbour lists for SPH interactions, and to efficiently calculate gravitational accelerations. The Barnes-Hut tree is an octal-spatial decomposition tree that splits the volume of the tree cells at each level into eight equal-volume cubic sub-cells (or four equal-area square sub-cells in 2D) -- recursively until only a few, or zero, particles remain in each sub-cell. The cells at which a branch of the tree terminates are called {\it leaf cells}. SEREN decomposes the particles as an ensemble, in a similar manner to the algorithm described by Pfalzner \& Gibbon (1996) (as distinct from the original Barnes-Hut method, which considers one particle at a time as the tree structure is built). The Pfalzner \& Gibbon algorithm makes it easier to parallelise the tree-build routine using OpenMP.

We construct two separate trees, one for particles that experience hydrodynamic accelerations and one for particles that experience gravitational accelerations. This is advantageous because these accelerations are computed using different cell properties. In the case where all SPH particles are self-gravitating, we can build the tree structure once, copy this structure to the second tree, but then stock the two trees (i.e. calculate the properties of the tree cells) separately.  Since the timestep criteria restrict how far particles can move in any one timestep, the tree structure will not change appreciably from one timestep to the next.  Therefore we only build the tree structure every $\,\sim\! 10$ timesteps, but restock it every timestep.

%%%%%%%%%%%%%%%%%%%%%%%%%%%%%%%%%
\subsection{Neighbour searching}%
%%%%%%%%%%%%%%%%%%%%%%%%%%%%%%%%%

The Barnes-Hut {\it neighbour} tree is constructed using all SPH particles. For each cell, we record (i) the position of centre of the bounding box containing all particles in the cell, (ii) the maximum distance of all particles from the bounding box centre, (iii) the maximum smoothing length of all particles in the cell, and, if it is a leaf cell, (iv) the identifiers of all particles contained in the cell.  Storing these quantities enables us to find neighbours efficiently, either by gather (i.e. all particles for which $|{\bf r}\ssij|^2 \leq {\cal R}^2h\ssi^2$), or by scatter (i.e. all particles for which $|{\bf r}\ssij|^2 \leq {\cal R}^2h\ssj^2$), or both. When we perform a tree-search of this type, we obtain a {\it potential neighbour list}, which is guaranteed to contain all of the true neighbours but normally also contains non-neighbours. There is no need to cull this list, because all non-neighbours which are passed to the SPH routines have no effect, since the kernel and its derivatives are zero for non-neighbours.

%%%%%%%%%%%%%%%%%%%%%%%%%%%%%%%%%%%%%%%%%%%%%%
\subsection{Tree gravity} \label{SS:TREEGRAV}%
%%%%%%%%%%%%%%%%%%%%%%%%%%%%%%%%%%%%%%%%%%%%%%

The Barnes-Hut {\it gravity} tree is built using only self-gravitating SPH particles. For each cell, we record by default (i) the total mass of the cell, (ii) the position of the centre of mass, and, if it is a leaf cell, (iii) the IDs of all SPH particles contained in the cell. Additionally, we can compute and store higher-order multipole terms, in order to calculate the gravity of a cell to greater accuracy. A multipole expansion can in principle be made up to any order, although it is usually optimal to truncate after only a few terms. The monopole term is simply the centre of mass term for each cell and the dipole term is always zero if calculated with respect to the centre of mass of the cell. In SEREN, we provide the option to include either the quadrupole moment terms, or the quadrupole and octupole moment terms.  The equations for the quadrupole and octupole moment tensors of a cell are given in Appendix \ref{A:MULTIPOLE}. The quadrupole moment tensor is a traceless symmetric matrix, $Q$, meaning there are 5 independent terms to be stored for each cell. The octupole moment tensor is a more complicated rank-3 tensor, $S$, whose symmetries result in 10 independent terms which must be stored for each cell. The gravitational potential at the position of particle $i$ due to cell $c$, up to octupole order, is
\begin{eqnarray}
\phi_{_{\rm GRAV}} &=& - \frac{GM_c}{|{\bf r}|} - \frac{GQ_{ab,c} r_a r_b}{2\,|{\bf r}|^5} - \frac{GS_{ab,c}r_a^2 r_b + GS_{123,c}r_1r_2r_3}{2\,|{\bf r}|^7}\, 
\label{EQN:TREEGRAVPOT}
\end{eqnarray}
where ${\bf r}={\bf r}_i-{\bf r}_c$ is the position of particle $i$ relative to cell $c$, $(r_1,r_2,r_3)$ are the Cartesian components of ${\bf r}$, and we employ the Einstein summation convention (i.e. we sum over repeated indices). If we define $\hat{\bf e}_{a}$ to be the unit vector in the $a^{\rm th}$ Cartesian direction, the gravitational acceleration of particle $i$, due to cell $c$, up to octupole order, is 
\begin{eqnarray}
\left( \frac{d{\bf v}}{dt} \right)_{_{\rm GRAV}} &=& 
- \frac{GM_c}{|{\bf r}|^3}{\bf r}
+ \frac{GQ_{ab,c} r_a}{2\,|{\bf r}|^5}\hat{\bf e}_b 
- \frac{5}{2}\frac{GQ_{ab,c} r_a r_b}{2\,|{\bf r}|^7}{\bf r}\nonumber\\&& 
+ \frac{GS_{ab,c} r_a r_b}{|{\bf r}|^7}\hat{\bf e}_a
+ \frac{GS_{ab,c} r_a^2}{|{\bf r}|^7}\hat{\bf e}_b 
- \frac{7GS_{ab,c} r_a^2 r_b}{|{\bf r}|^9}{\bf r}  
- \frac{7GS_{123,c} r_1 r_2 r_3}{2\,|{\bf r}|^9}{\bf r}
+ \frac{GS_{123,c}}{|{\bf r}|^7} 
\left(r_2 r_3\hat{\bf e}_1 + r_3 r_1\hat{\bf e}_2 + r_1 r_2\hat{\bf e}_3 \right) 
\,.
\label{EQN:TREEGRAVFORCE}
\end{eqnarray}
When walking the gravity tree, the code must interrogate cells to decide whether to use the multipole expansion or to open up the cell and interrogate its child cells.  This decision is determined by the {\it multipole-acceptance criterion} (MAC). SEREN includes a simple Geometric MAC and a GADGET-style MAC; it also includes a new Eigenvalue MAC, which uses the eigenvalues of the quadrupole moment terms to determine whether to open a cell.

%%%%%%%%%%%%%%%%%%%%%%%%%%%%%%%%%%%%%%%%%%%%%%%%
\subsubsection{Geometric MAC} \label{SS:GEOMAC}%
%%%%%%%%%%%%%%%%%%%%%%%%%%%%%%%%%%%%%%%%%%%%%%%%

The Geometric MAC uses the size of the cell, $\ell_c$ (i.e. its longest corner-to-corner length), and its distance from the particle, $|{\bf r}_i-{\bf r}_c|$, to calculate the angle the cell subtends at the particle, $\theta_{ci}=\ell_c/|{\bf r}_i-{\bf r}_c|$. If $\theta_c$ is smaller than some pre-defined tolerance, $\theta_{_{\rm MAC}}$, the gravitational acceleration due to the cell is given by the multipole expansion, Eqn. (\ref{EQN:TREEGRAVFORCE}). If this criterion is not satisfied, the cell is opened and the sub-cells on the next level are interrogated in the same way. If a leaf cell is opened, we store the identifiers of all the particles contained in it, and compute their contribution to the net gravitational acceleration directly (using Eqns. \ref{EQN:GRADHGRAV1}, \ref{EQN:PHIPBAR} \& \ref{EQN:GRADHZETA}). For computational efficiency, the code calculates and stores for each cell the quantity ${\cal S}_c=(\ell_c/\theta_{_{\rm MAC}})^2$. An unnecessary square-root operation is then avoided by applying the geometric MAC in the form
\begin{eqnarray} \label{EQN:GEOMAC2}
|{\bf r}_i-{\bf r}_c|^2&\geq&{\cal S}_c\hspace{1.0cm}(\mbox{\sc cell\;does\;not\;need\;to\;be\;opened})\,.
\end{eqnarray}

%%%%%%%%%%%%%%%%%%%%%%%%%%%%%%%%%%%%%%%%%%%%%%%%%%%%%%
\subsubsection{GADGET-style MAC} \label{SS:GADGETMAC}%
%%%%%%%%%%%%%%%%%%%%%%%%%%%%%%%%%%%%%%%%%%%%%%%%%%%%%%

Springel, Yoshida \& White (2001) have formulated another type of MAC for the SPH code GADGET. This MAC uses an approximation to the leading error term in the multipole expansion to calculate for each cell the smallest distance from the cell at which the multipole expansion can be used. GADGET includes quadrupole moment corrections, and so the leading error term is the octupole term.  However, Springel, Yoshida \& White suggest that the octupole moment term is small in a homogeneous density field, in which case the hexadecapole term is the largest error term. For a cell $c$ of total mass $M_c$ and linear size $\ell_c$, an approximation to the magnitude of the acceleration of particle $i$ due to the hexadecipole term is $\;a_{_{\rm HEX}}\sim GM_c\ell_c^4/|{\bf r}_i-{\bf r}_c|^6$. If $a_{_{\rm HEX}}$ is less than some user-defined fraction of the total gravitational acceleration of particle $i$, i.e. $a_{_{\rm HEX}} < \alpha_{_{\rm MAC}}\,|{\bf a}_{_{\rm GRAV}}|$, the multipole expansion is used; otherwise cell $c$ must be opened to the next level. Since the current acceleration of particle $i$ is not yet available, the code uses the acceleration from the previous timestep as an approximation. The code therefore calculates and stores for each cell the quantity $\chi_c=(GM_c\ell_c^4/\alpha_{_{\rm MAC}})^{1/3}$, and then applies the GADGET-style MAC in the form
\begin{eqnarray} \label{EQN:GADGETMAC}
|{\bf r}_i-{\bf r}_c|^2&\geq&\chi_c\,|{\bf a}_{_{\rm GRAV}}|^{-1/3}\hspace{1.0cm}(\mbox{\sc cell\;does\;not\;need\;to\;be\;opened}).
\end{eqnarray}
When the quadrupole and octupole moment terms are not used, the leading error term is the quadrupole term. Therefore an approximation to the acceleration of the quadrupole term, $a_{_{\rm QUAD}}=G\,M_c\ell_c^2/|{\bf r}_i-{\bf r}_c|^4$, is used instead. In this case the code calculates and stores for each cell the quantity $\chi_c'=(GM_c\ell_c^2/\alpha_{_{\rm MAC}})^{1/2}$, and then applies the GADGET-style MAC in the form
\begin{eqnarray} \label{EQN:GADGET2MAC}
|{\bf r}_i-{\bf r}_c|^2&\geq&\chi_c'\,|{\bf a}_{_{\rm GRAV}}|^{-1/2}\hspace{1.0cm}(\mbox{\sc cell\;does\;not\;need\;to\;be\;opened}).
\end{eqnarray}
{ Since we do not have a value of ${\bf a}_{_{\rm GRAV}}$ on the very first timestep, we use the Geometric MAC with $\theta_{_{\rm MAC}} = 1.0$ to obtain an initial estimate and then revert to Eqns. (\ref{EQN:GADGETMAC}) \&  (\ref{EQN:GADGET2MAC}). Eqns. (\ref{EQN:GADGETMAC}) \& (\ref{EQN:GADGET2MAC}) do not guarantee a maximum fractional force error, but rather attempt to set an upper limit on the error contribution from each cell. It is therefore possible that the error is larger than desired.  Therefore we use the Geometric MAC with $\theta_{_{\rm MAC}} = 1.0$, alongside the GADGET-style MAC, as a safety measure for the rare cases where Eqns. (\ref{EQN:GADGETMAC}) \& (\ref{EQN:GADGET2MAC}) are inadequate.}

%%%%%%%%%%%%%%%%%%%%%%%%%%%%%%%%%%%%%%%%%%%%%%%%%%%
\subsubsection{Eigenvalue MAC} \label{SS:EIGENMAC}%
%%%%%%%%%%%%%%%%%%%%%%%%%%%%%%%%%%%%%%%%%%%%%%%%%%%

We introduce here a new Eigenvalue MAC, based on the quadrupole moment terms of a cell. Salmon \& Warren (1994) originally suggested using higher-order multipole moments directly to formulate a MAC, but they only used upper limits to the multipole moment terms to constrain the leading error term of the multipole expansion. This resulted in a more conservative MAC than was actually required to achieve the desired accuracy, and hence more expensive tree walks. The Eigenvalue MAC is formulated by determining the maximum values of the gravitational potential (or acceleration) due to the quadrupole moment terms of a cell. The quadrupole moment tensor is a real, symmetric and traceless matrix. It therefore has three real eigenvalues, $\lambda_1, \lambda_2, \lambda_3$. From Eqn. (\ref{EQN:TREEGRAVPOT}), the gravitational potential due to the quadrupole moment term is 
\begin{eqnarray}
\phi_{_{\rm QUAD}}&=&-\,\frac{G\,Q_{ab,c}r_ar_b}{2\,|{\bf r}|^5}\,.
\label{EQN:QUADGRAVPOT}
\end{eqnarray}
The term in the numerator, $Q_{ab,c} r_a r_b$, is the {\it quadratic form} between the quadrupole matrix $Q$ and the vector ${\bf r}$. It can be shown (e.g. Riley, Hobson \& Bence 1997) that the quadratic form has a maximum absolute value given by $|\lambda_{_{\rm MAX}}|\,|{\bf r}|^2$, where $\lambda_{_{\rm MAX}}$ is the largest { in magnitude} of the three eigenvalues. We therefore solve the eigenvalue equation,
\begin{eqnarray}\label{EQN:DEPRESSEDCUBIC}
{\rm det}\left[Q-\lambda\,I\right]&=&A\,+\,B\lambda\,+\,C\lambda^2\,-\,\lambda^3\;\,=\;\,0\,,\\
A&=&-\,Q_{33}Q_{12}^2\,-\,Q_{22}Q_{13}^2\,-\,Q_{11}Q_{23}^2\,+\,2Q_{12}Q_{13}Q_{23}\,+\,Q_{11}Q_{22}Q_{33}\hspace{1.0cm}\equiv\;\,{\rm det}[Q]\,,\\
B&=&Q_{12}^2\,+\,Q_{13}^2\,+\,Q_{23}^2\,-\,Q_{11}Q_{22}\,-\,Q_{11}Q_{33}\,-\,Q_{22}Q_{33}\,,\\
C&=&Q_{11}\,+\,Q_{22}\,+\,Q_{33}\hspace{7.2cm}\equiv\;\,{\rm Tr}[Q]\,.
\end{eqnarray}
Since, by design, $C={\rm Tr}[Q]=0$, Eqn. (\ref{EQN:DEPRESSEDCUBIC}) is a {\it depressed cubic equation}, i.e. a cubic equation with no quadratic term. Since also $Q$ is real and symmetric, the eigenvalues are real, and Eqn. (\ref{EQN:DEPRESSEDCUBIC}) can be solved by the method of Vieta (e.g. Martin 1998). In particular, the largest eigenvalue is
\begin{eqnarray}
\lambda_{_{\rm MAX}}&=&\sqrt{\frac{4B}{3}}\;\,=\;\,2\,\sqrt{\frac{Q_{12}^2\,+\,Q_{13}^2\,+\,Q_{23}^2\,-\,Q_{11}Q_{22}\,-\,Q_{11}Q_{33}\,-\,Q_{22}Q_{33}}{3}}\,.
\end{eqnarray}
We therefore require that the magnitude of the quadrupole moment potential, $|\phi_{_{\rm QUAD}}|=G\lambda_{_{\rm MAX}}/2|{\bf r}_i-{\bf r}_c|^3$, be less than some user-defined fraction of the total potential, $|\phi_{_{\rm QUAD}}|<\alpha_{_{\rm MAC}}|\phi_{_{\rm GRAV}}|$. The code approximates $\phi_{_{\rm GRAV}}$ with the value from the previous timestep, and calculates and stores for each cell the quantity $\xi_c=\{G^2(Q_{12}^2+Q_{13}^2+Q_{23}^2-Q_{11}Q_{22}-Q_{11}Q_{33}-Q_{22}Q_{33})/3\alpha_{_{\rm MAC}}^2\}^{1/3}$. The Eigenvalue MAC is then applied in the form
\begin{eqnarray} \label{EQN:EIGENMAC2}
|{\bf r}_i-{\bf r}_c|^2&\geq&\xi_c\,|\phi_{_{\rm GRAV}}|^{-2/3}\hspace{1.0cm}(\mbox{\sc cell\;does\;not\;need\;to\;be\;opened}).
\end{eqnarray}
{ Eqn. (\ref{EQN:EIGENMAC2}) does not guarantee a maximum fractional error, but attempts to limit the error contribution from each cell.  Therefore we also use the Geometric MAC with $\theta_{_{\rm MAC}} = 1.0$, alongside the Eigenvalue MAC, as an extra safety measure.}

%%%%%%%%%%%%%%%%%%%%%%%%%%%%%%%%%%%%%%%%%%%%%%%%%%%%%%%%%%%%%%%%%%%%%%%%%%%
\subsubsection{SPH-neighbour cell-opening criterion} \label{SS:SPHNEIBMAC}%
%%%%%%%%%%%%%%%%%%%%%%%%%%%%%%%%%%%%%%%%%%%%%%%%%%%%%%%%%%%%%%%%%%%%%%%%%%%

The multipole expansions used in SEREN assume that each SPH particle in a cell is a point-mass. In contrast, the derivation of the equation of motion takes account of the finite extent of an SPH particle (i.e. kernel-softened gravity; see Section \ref{S:SPHGRAVITY}). The SPH-neighbour criterion therefore requires that any cell that might contain neighbours of particle $i$ be opened. The code calculates and stores for each cell the quantity $d_c^2={\rm MAX}_j\{(|{\bf r}_j-{\bf r}_c|+{\cal R}h_j)\}$, where the maximum is over all the particles $j$ in the cell; $d_c$ is the maximum extent of the smoothing kernels of the particles in cell $c$. The overall cell-opening criterion then takes the form
 \begin{eqnarray} \label{EQN:SPHMAC}
|{\bf r}_i-{\bf r}_c|^2&\geq&{\rm MAX}\left\{({\cal R}h_i)^2;\;\;d_c^2;\;\,{\cal S}_c\;\,\mbox{\sc or}\;\,\chi_c|{\bf a}_{_{\rm GRAV}}|^{-1/3}\;\,\mbox{\sc or}\;\,\chi_c'|{\bf a}_{_{\rm GRAV}}|^{-1/2}\;\,\mbox{\sc or}\;\,\xi_c|\phi_{_{\rm GRAV}}|^{-2/3}\right\}\hspace{0.5cm}(\mbox{\sc cell\;does\;not\;need\;to\;be\;opened}).
\end{eqnarray}
This additional criterion adds an extra overhead to calculating the properties of the cells, and also to the gravity walk, since there are now two cell-opening criteria to check. However, in highly clustered geometries such as those found in gravitational collapse problems, this extra check brings significant accuracy and speed benefits.

%%%%%%%%%%%%%%%%
\begin{figure*} 
\centerline{\psfig{figure=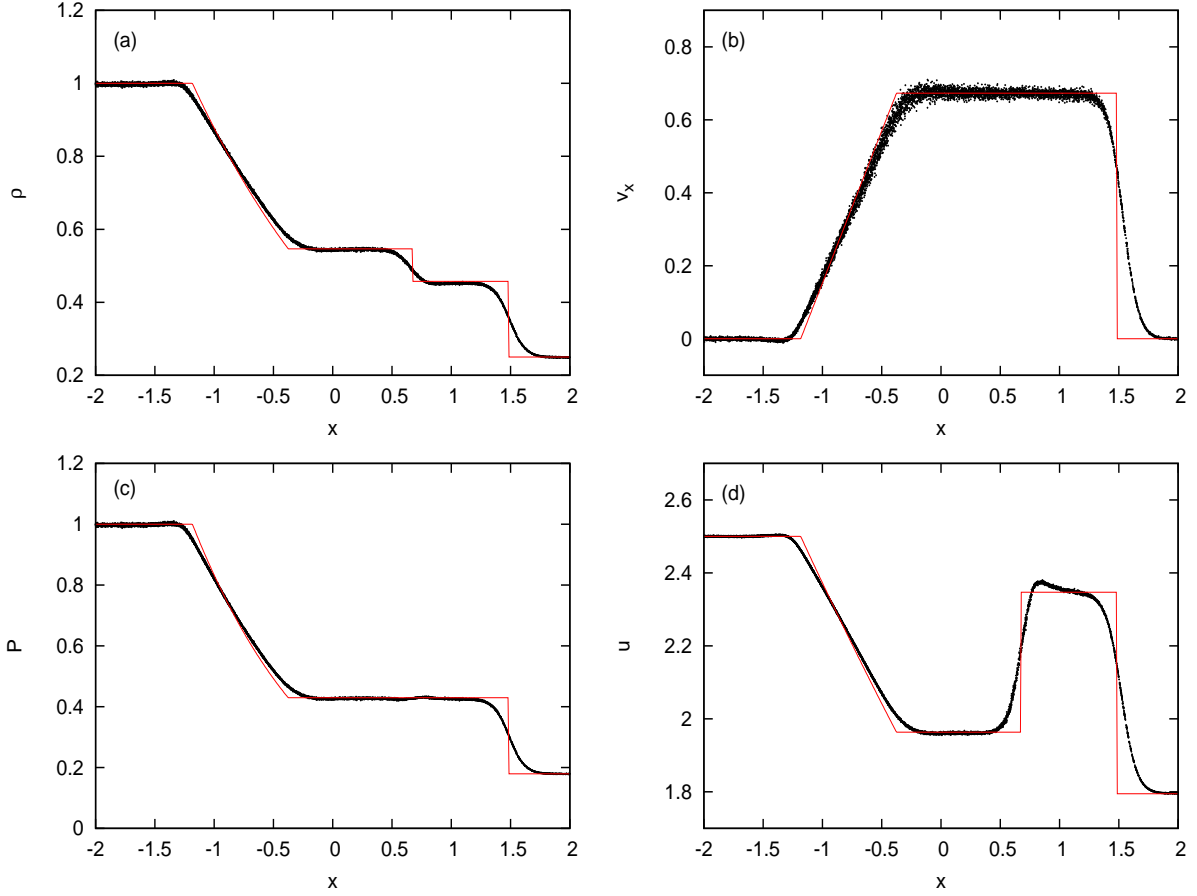,height=12.0cm,width=16cm,angle=270}}
\caption{Results of the adiabatic shock test using the `grad-h' SPH formulation (Price \& Monaghan 2004) showing (a) the density, (b) the x-velocity, (c) the thermal pressure, and (d) the specific internal energy after a time $t = 1.0$. The black dots represent the results from the SPH simulation and the red lines show the semi-analytic solution obtained using a Riemann solver.}
\label{FIG:ADSOD}
\end{figure*}
%%%%%%%%%%%%%%%%

%%%%%%%%%%%%%%%%%%%%%%%%%%%%%%%
\section{Tests} \label{S:TESTS}

We have performed a large number of standard and non-standard tests to demonstrate that the algorithms in SEREN have been implemented correctly and perform well. It is not practical to test all possible combinations of the options available in SEREN, and we have therefore chosen tests which demonstrate the performance of particular algorithms.  Where possible, we compare the test results with known analytic or semi-analytic solutions. Where an algorithm has been developed in another SPH code and the subroutine then imported into SEREN (e.g. the radiative cooling module of Stamatellos et al. 2007), or has been written directly into SEREN as an independent module (e.g. the HEALPix module for treating ionising radiation; Bisbas et al. 2009), the testing is not described here, and the interested reader is referred to the original paper.

%%%%%%%%%%%%%%%%%%%%%%%%%%%%%%%%%%%%%%%%%%%%%%%%%%%%%%%%%%%
{
\subsection{Generation of initial conditions} \label{SS:IC}
The generation of initial conditions in SPH often needs careful consideration, since particle noise and edge effects can impact negatively on test simulations such as those described here.  For example, random initial conditions suffer from Poisson-noise in the particle distribution which leads to high-frequency noise in the density and particle accelerations.  A safer approach is to generate a so-called `glass' distribution of particles.  A glass is a semi-regular structure in which all the particles are roughly equidistant from each other.

In order to generate a glass, we initially place equal-mass particles randomly in a periodic box.  The particles are evolved using SEREN, with artificial viscosity to dissipate the kinetic energy, until the particles have settled into an equilibrium structure.  We use an isothermal EOS and a Courant factor of $\gamma_{_{\rm COUR}} = 0.2$.  Once settled, the particle boxes can be replicated and joined together to create larger settled particle distributions, and uniform-density spheres can be cut from a box. All of the glass distributions used in this paper are set-up using this method.  We note that glass-structures can be set up with different methods (e.g. `repulsive' gravity in GADGET2; Springel 2005).
}

%%%%%%%%%%%%%%%%
\begin{table}[b]
\centering
\begin{tabular}{crccrcc}\hline\hline
        & \hspace{1.2cm}{\sc adiabatic sod test} & $x < 0$ & $x > 0$ & \hspace{1.5cm}{\sc isothermal colliding flows} & $x < 0$ & $x > 0$ \\ \hline
$\rho$  &                                        & 1.0     & 0.25    &                                                & 1.0     & 1.0     \\
P       &                                        & 1.0     & 0.1795  &                                                & 1.0     & 1.0     \\
$v_{x}$ &                                        & 0.0     & 0.0     &                                                & 4.0     & -4.0    \\
\end{tabular}
\caption{Initial conditions for the adiabatic Sod test (columns 2 and 3), and for the colliding flows test (columns 4 and 5).}
\label{TAB:ADSOD}
\end{table}
%%%%%%%%%%%

%%%%%%%%%%%%%%%
\begin{figure*} 
\centerline{\psfig{figure=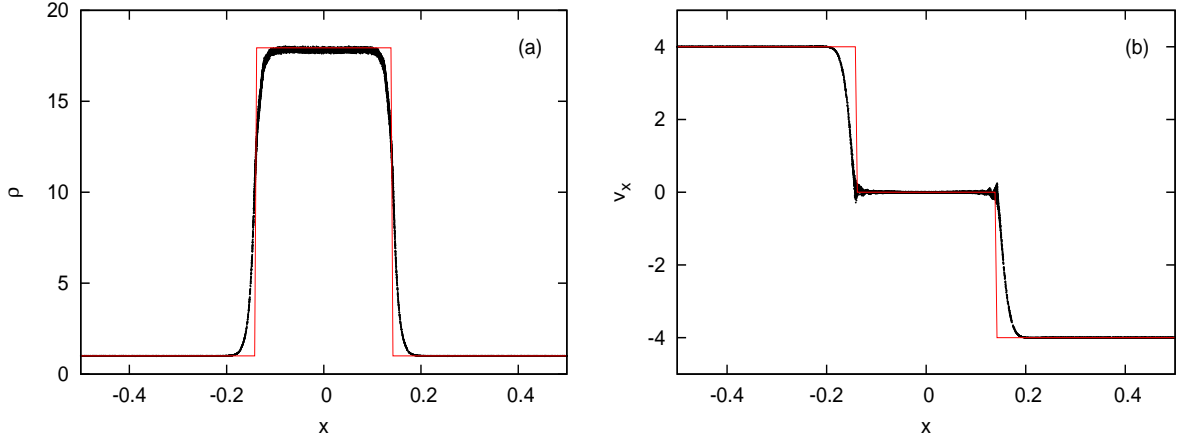,height=6.0cm,width=16cm,angle=270}}
\caption{Results of the colliding flows test using the standard SPH equations with the Monaghan (1997) artificial viscosity showing (a) the density and (b) the x-velocity, after a time $t = 0.6$. The black dots represent the results from the SPH simulation and the red lines show the analytic solution.}
\label{FIG:COL-3D-MON}
\end{figure*}
%%%%%%%%%%%%%

%%%%%%%%%%%%%%%
\begin{figure*} 
\centerline{\psfig{figure=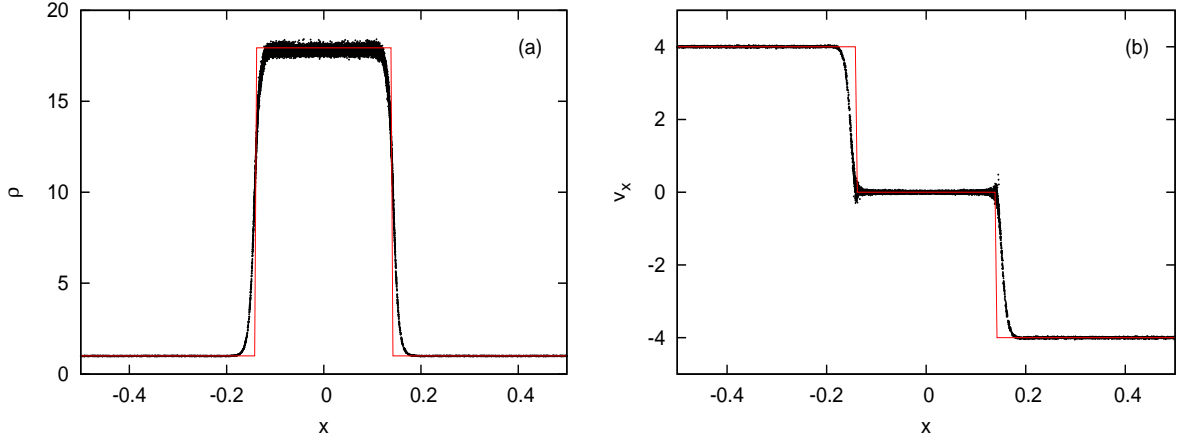,height=6.0cm,width=16cm,angle=270}}
\caption{Results of the colliding flows test using the standard SPH equations and time-dependent $(\alpha,\beta)$ viscosity (Morris \& Monaghan 1997) showing (a) the density and (b) the x-velocity, after a time $t = 0.6$. The black dots represent the results from the SPH simulation and the red lines show the analytic solution.}
\label{FIG:COL-3D-MON-TD}
\end{figure*}
%%%%%%%%%%%%%

%%%%%%%%%%%%%%%%%%%%%%%%%%%%%%%%%%%%%%%%%%%%%%%%%%%%%%%%%%%
\subsection{Adiabatic Sod test (Sod 1978)} \label{SS:ADSOD}

The initial conditions for this test are summarised in Table \ref{TAB:ADSOD} (left side). The computational domain is $-4\leq x\leq +4,\;0\leq y\leq 1,\;0\leq z\leq 1$, and periodic wrapping is invoked in all three dimensions. Initially (at $t=0$), the left-hand half of the domain ($x<0$) contains a high-density, high-pressure gas, represented by 64,000 particles, and the right-hand half ($x>0$) contains a low-density, low-pressure gas, represented by 16,000 particles.  The particles have been relaxed to a glass, and are at rest; they have equal mass.  The gas evolves adiabatically, with adiabatic exponent $\gamma = 1.4$. We therefore solve the momentum and energy equations, using both artificial viscosity and artificial conductivity, to moderate the discontinuities in velocity and temperature, respectively. We perform this test in 3D (since this is the dimensionality of star-formation simulations) using the default `grad-h' SPH method, the Monaghan (1997) artificial viscosity and the Price (2008) artificial conductivity.

Fig. \ref{FIG:ADSOD} shows the density, x-velocity, thermal pressure and specific internal energy profiles (black dots), and the accurate 1-D solution obtained using a Riemann solver (red lines), at the end of the simulation ($t=1$) in the interval $|x|<2$. A rarefaction wave is propagating into the high-density gas on the left (its head is at $x\sim -1.3$), and a shock wave is propagating into the low-density gas on the right (it has reached $x\sim1.5$). There is also a contact discontinuity (at $x\sim0.6$), since the gas from the right has higher specific entropy than that from the left. The SPH results reproduce the gross features of the accurate solution well, but the discontinuities are inevitably spread over a few smoothing lengths.

%%%%%%%%%%%%%%%%%%%%%%%%%%%%%%%%%%%%%%%%%%%%%%%%
\subsection{Colliding flows test} \label{SS:COL}

The initial conditions for this test are summarised in Table \ref{TAB:ADSOD} (right side). The computational domain is $-4\leq x\leq +4,\;0\leq y\leq 0.2,\;0\leq z\leq 0.2$, and periodic wrapping is invoked in the $y$ and $z$ dimensions, but not in the $x$ dimension. Initially, the density is uniform, but the gas in the left-hand half of the computational domain $(x < 0)$ has velocity $v_x = +4$, and the gas in the right-hand half $(x > 0)$ has velocity $v_x = -4$; the gas is represented by 128,000 equal-mass particles which have been relaxed to a glass.  { The velocities are smoothed near $x = 0$ instead of having an unresolved x-velocity discontinuity}. { Therefore the discontinuous velocity profile, ${\bf v}'({\bf r})$, is replaced by the smoothed velocity, }
\begin{eqnarray} \label{EQN:SMOOTHEDVEL}
{\bf v}\ssi = \sum \limits_{j=1}^{N} 
{ \frac{m\ssj}{\rho\ssj}{\bf v}'\ssj\,W({\bf r}\ssij,h\ssi) }\,.
\end{eqnarray}
The gas is isothermal, with dimensionless sound speed $c_{_{\rm S}}=1$, so the code does not need to solve the energy equation, nor does it need to invoke artificial conductivity.  This test demonstrates how well artificial viscosity enables the code to suppress particle interpenetration and capture shocks.  { We perform the test in 3D, using standard SPH with Monaghan (1997) with and without time-dependent artificial viscosity  (Morris \& Monaghan 1997).  For the time-dependent viscosity simulation, we set $\alpha_{\rm MAX} = 2$ (and $\beta_{\rm MAX} = 4$) with $\alpha_{_{\rm MIN}} = 0.1$. We adopt a global timestep for both simulations.

Fig. \ref{FIG:COL-3D-MON} compares the SPH density and x-velocity as a function of $x$ (black dots) with the analytic solution (red line), for the standard SPH run, and Fig. \ref{FIG:COL-3D-MON-TD} makes the same comparison for the time-dependent viscosity run, both at $t = 0.6$. The peak density and the width of the shock are in agreement with the analytic solution for both runs, but the discontinuities in density and velocity are smeared out over a few smoothing lengths. This smearing is an inherent feature of SPH simulations. The time-dependent viscosity performs almost as well as the standard Monaghan (1997) viscosity, with a little more scatter in the post-shock density.  The scatter in the density is partly a result of increased particle disorder at the shock front due to a small amount of particle penetration when the shock forms (Figs. \ref{FIG:COL-3D-MON}(b) \& \ref{FIG:COL-3D-MON-TD}(b)).}

%%%%%%%%%%%%%%%
\begin{figure*}
\centerline{\psfig{figure=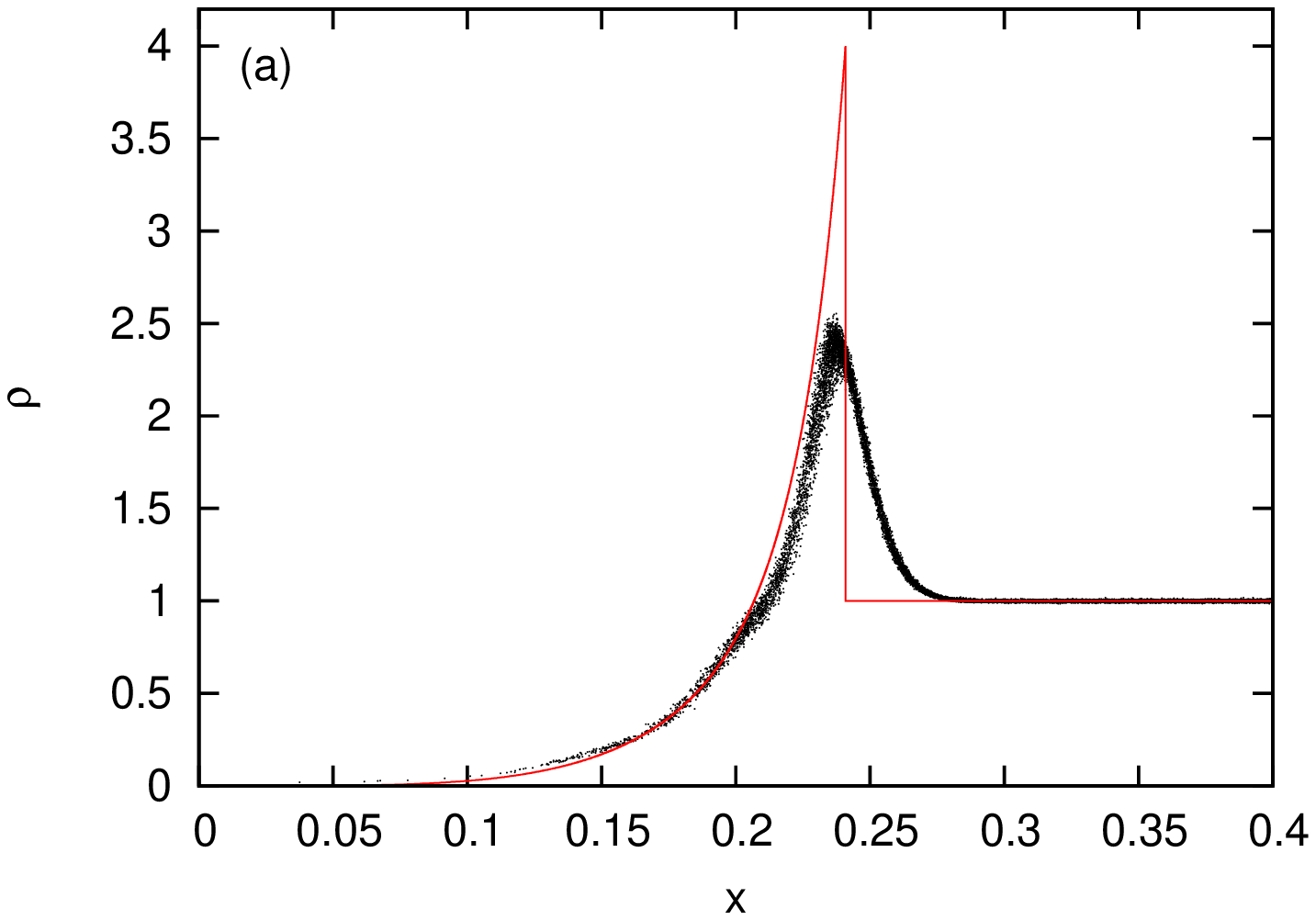,height=5cm,width=6.2cm,angle=270}
\hspace{0.05cm}\psfig{figure=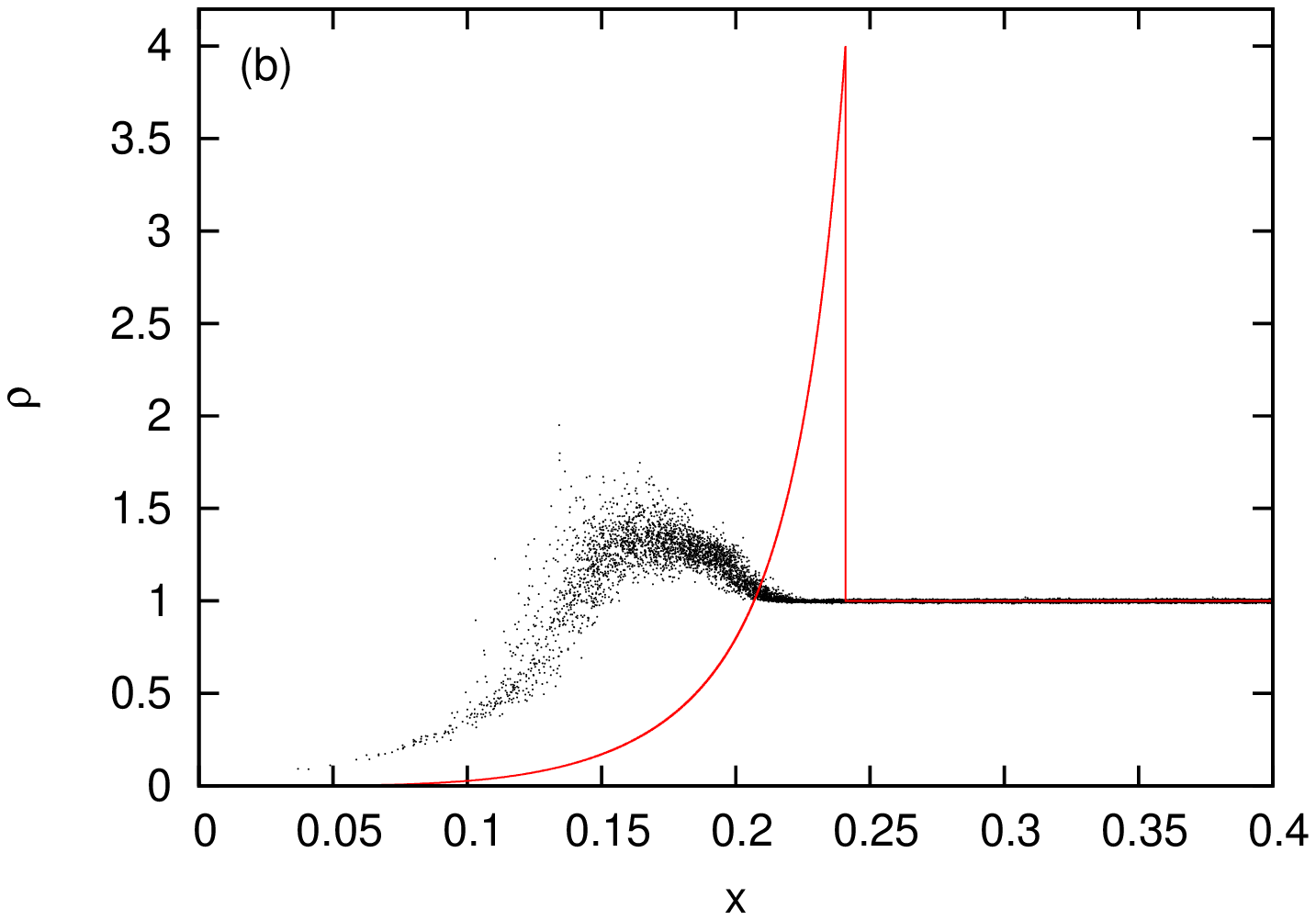,height=5cm,width=6.2cm,angle=270}
\hspace{0.05cm}\psfig{figure=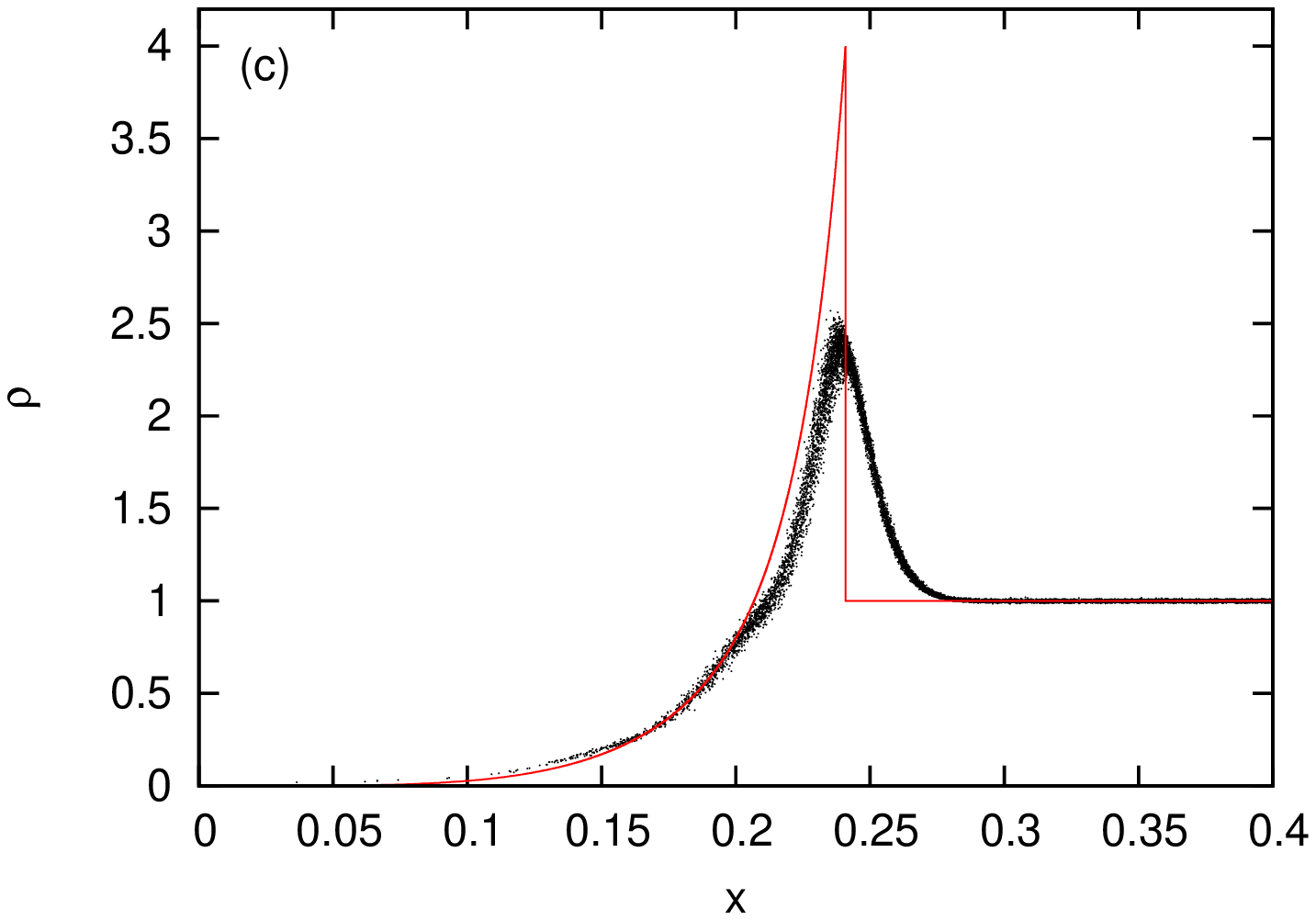,height=5cm,width=6.2cm,angle=270}}
\caption{Results of the Sedov blast wave test at a time $t = 0.02$ using (a) global timesteps; (b) individual block timesteps; (c) individual timesteps with the timestep-limiter described in Section \ref{SS:BLOCKTIMESTEPS} and in Saitoh \& Makino (2009).  The block dots represent the SPH results and the red line shows the semi-analytic solution provided by Sedov (1959).}
\label{FIG:SEDOVTEST}
\end{figure*}

%%%%%%%%%%%%%%%%%%%%%%%%%%%%%%%%%%%%%%%%%%%%%%%%%%%%%%%%%%%%%%%
\subsection{Sedov blast wave (Sedov 1959)} \label{SS:SEDOVTEST}

This test demonstrates that the code can handle the steep temperature and density gradients created by an explosion, and the consequent requirement for a timestep limiter (see Section \ref{SS:BLOCKTIMESTEPS} and Saitoh \& Makino 2009). A settled, uniform-density glass-like distribution of 200,000 SPH particles is created. Then the central particle and its ($\sim\! 50$) neighbours are given a net impulse of thermal energy $\Sigma U=1$, divided amongst them according to the smoothing kernel. The remaining particles have a total thermal energy $10^{-6}$ times smaller than the particle with the maximum internal energy (i.e. the particle closest to the centre). The impulse of thermal energy results in an outward propagating shock front which sweeps the surrounding gas into a dense layer. Sedov (1959) provides { an analytic similarity solution} for the subsequent evolution of this system (strictly speaking, one in which the surrounding particles start with zero thermal energy).

We perform three realisations of this test, with three different time-stepping schemes. The resulting density profiles at time $t = 0.02$ are shown in Fig. \ref{FIG:SEDOVTEST}, and compared with the semi-analytic solution. The SPH simulation with global timesteps (Fig. \ref{FIG:SEDOVTEST}a) shows good agreement with the semi-analytic solution; the maximum density in the shell is reduced by smoothing, but the position and width of the shock front are comparable with the analytic solution. The SPH simulation using hierarchical block time-steps (Fig. \ref{FIG:SEDOVTEST}b) fails to reproduce any of the features of the semi-analytic solution, because the cold particles have such a long timestep, compared with the hot ones, that they cannot respond to the pressure of the explosion and the hot particles penetrate through them. The SPH simulation using hierarchical block timesteps with a timestep limiter (i.e. not allowing any SPH particle to have a timestep more than four times longer than its neighbours; Fig. \ref{FIG:SEDOVTEST}c) produces results which are indistinguishable from the simulation using global timesteps, but uses $\,\sim\! 8 \%$ of the computing time.

%%%%%%%%%%%%%%%
\begin{figure*}
\centerline{\psfig{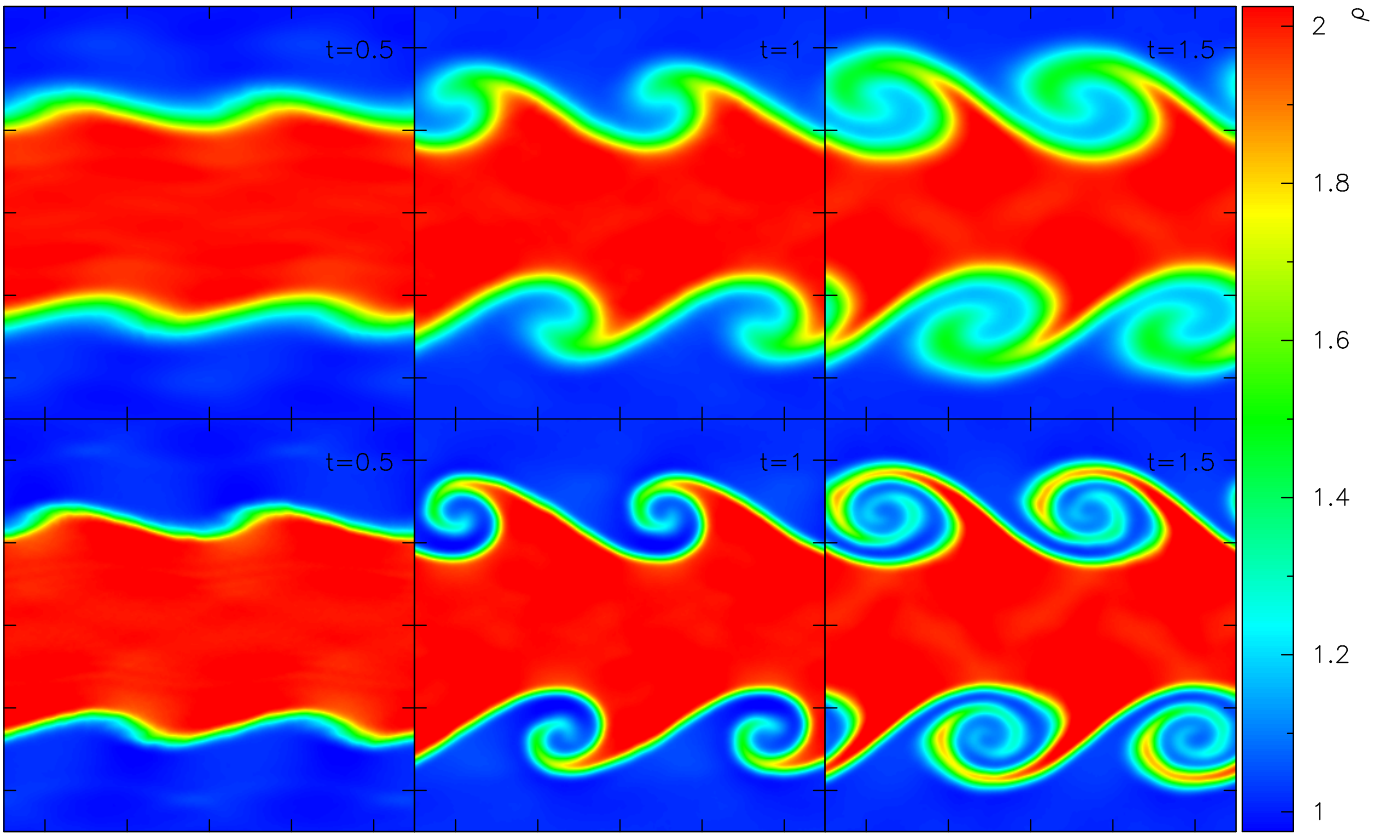}}
\caption{{ Development of the Kelvin-Helmholtz instability for the low-resolution case (top-row; 12,242 particles) and the high-resolution case (bottom-row; 98,290 particles).  The plots show the evolution of the density-field (colour bar on right-hand side) at times $t = 0.5$, $1.0$ and $1.5$ (left, middle and right columns respectively).}}
\label{FIG:KHI}
\end{figure*}

%%%%%%%%%%%%%%%%%%%%%%%%%%%%%%%%%%%%%%%%%%%%%%%%%%%%%%%%%%%%%%%%%
{
\subsection{Kelvin-Helmholtz instability} \label{SS:KHI}
The Kelvin-Helmholtz instability (hereafter KHI) is a classical hydrodynamical instability that occurs, in the simplest case, between two bulk flows that are shearing past one another.  It has been extensively studied in recent years as a diagnostic for comparing the ability of both SPH and grid codes to model mixing of interacting fluids (e.g. Agertz et al. 2006; Price 2008; Read et al. 2010).  In particular, this test has highlighted an intrinsic problem in the standard formulation of SPH and has led to several suggested modifications to SPH (e.g. Price 2008; Read et al. 2010).  

We use similar initial conditions to Springel (2010) where two fluids with densities $\rho_1 = 1$ and $\rho_2 = 2$ are in shear-flow along the $y = 0$ plane with relative velocity $|v_1 - v_2| = 1.0$.  The two fluids are in pressure-balance, $P = 2.5$, and have ratio of specific heats $\gamma = 5/3$.  Therefore, there is discontinuity in the specific internal energy, $u = P/(\gamma - 1)\,\rho$, and also in the specific entropy.  Both layers are contained within a periodic box of extent $-0.5 <x < 0.5$ and $-0.5 < y < 0.5$.  Springel (2010) adds a velocity perturbation of the form 
\begin{eqnarray} \label{EQN:KHIPERT}
v_y(x,y) &=& w_0\,\sin{ \left( \frac{2\,\pi}{\lambda} \right)}\,
\left\{ \exp{ \left[ -\frac{(y - y_{I1})^2}{2\sigma^2} \right]} + 
\exp{ \left[ -\frac{(y - y_{I2})^2}{2\sigma^2} \right]} \right\}
\end{eqnarray}x
where $\lambda = 0.5$ is the wavelength of the velocity perturbation between the two fluids, $w_0 = 0.1$ is its amplitude and $\sigma = 0.05/\sqrt{2}$ is the scale-height of the perturbation in the $y$-direction.  We invoke time-dependent artificial viscosity and artificial conductivity (see Section \ref{SS:ARTDISS}). We adopt the quintic kernel (see Appendix \ref{SS:QUINTIC}) for computing all SPH quantities, instead of the more common M4 kernel (see Appendix \ref{SS:M4}).  We follow the growth of the instability for a total dimensionless time of $t = 1.5$.  The linear growth-timescale of the instability is 
\begin{eqnarray} \label{EQN:KHITIME}
\tau_{\rm KH} &=& \frac{(\rho_1 + \rho_2)}
{\sqrt{\rho_1\,\rho_2}}\frac{\lambda}{|{\bf v}_2 - {\bf v}_1|}\,.
\end{eqnarray}
For our initial conditions, the growth timescale is $\tau_{\rm KH} = 1.06$. Therefore, by the end of the simulation the instability should have entered the non-linear phase where significant vorticity and mixing occur near the shearing interface.  We perform simulations using these initial conditions at both low (12,242 particles) and high (98,290 particles) resolutions.  

In Fig. \ref{FIG:KHI}, we show the evolution of the density field and the development of the instability at three different times, $t = 0.5$, $1.0$ and $1.5$.  For both the low and high resolution cases, the instability evolves at approximately the same rate through the linear-phase ($t = 0.5$), and subsequently during the non-linear phase ($t = 1.0$ and $1.5$) where significant vorticity develops.  The large-scale properties of the vortices formed are very similar in both the low and high resolutions cases.  The main difference between the two is the number of resolved spiral turns in a vortex.  The low resolution case has just enough spatial resolution to model the formation of one complete spiral loop by $t = 1.5$.  The high resolution case has enough resolution to model two complete spiral loops and can be seen to have less dispersion in the density field around the contact regions between the two fluids.

}

%%%%%%%%%%%%%%%%%%%%%%%%%%%%%%%%%%%%%%%%%%%%%%%%%%%%%%%%%%%%%%%%%
\subsection{Tree multipole expansion and scaling characteristics} \label{SS:MULTIPOLETEST}

{ We test the accuracy of the Barnes-Hut gravity tree and the multipole moment correction terms (Section \ref{SS:TREEGRAV}), by comparing the gravitational acceleration obtained by walking the tree, ${\bf a}_i^{^{\rm TREE}}$, with that obtained by a direct-summation over all particles, ${\bf a}_i^{^{\rm DIRECT}}$ (cf. McMillan \& Aarseth 1993). Specifically, we compute the root-mean-square fractional acceleration error,
\begin{eqnarray} \label{EQN:FORCEERROR2}
\epsilon&=&\left(\frac{1}{N}\sum\limits_{i=1}^{N}\left\{\frac{\left|{\bf a}_i^{^{\rm TREE}}-{\bf a}_i^{^{\rm DIRECT}}\right|^2}{\left|{\bf a}_i^{^{\rm DIRECT}}\right|^2}\right\}\right)^{1/2}\,.
\end{eqnarray}
The density field used in this test is a uniform-density, glass-like sphere (See Section \ref{SS:IC}) of 32,000 SPH particles; we note that this is actually a stiffer test of the tree than a highly structured density field. We compute $\epsilon$ using the Geometric MAC (Section \ref{SS:GEOMAC}) and the Eigenvalue MAC (Section \ref{SS:EIGENMAC}).  For the Geometric MAC, we compute $\epsilon$ using different values of $\theta_{_{\rm MAC}}$ in the range $0.1$ to $1.0$, and including terms up to monopole, quadrupole and octupole order.  For the Eigenvalue MAC, we compute $\epsilon$ using different values of $\alpha_{_{\rm MAC}}$ in the range $10^{-6}$ to $10^{-2}$, and including terms up to quadrupole and octupole order; we do not consider monopole-only since we must calculate the quadrupole moment terms anyway in order to formulate the Eigenvalue MAC. We do not include the effects of kernel-softening in this test and therefore we effectively set the smoothing lengths to zero for the purposes of using the SPH-neighbour opening criterion; Section \ref{SS:SPHNEIBMAC}.

\begin{figure*}
\centerline{\psfig{figure=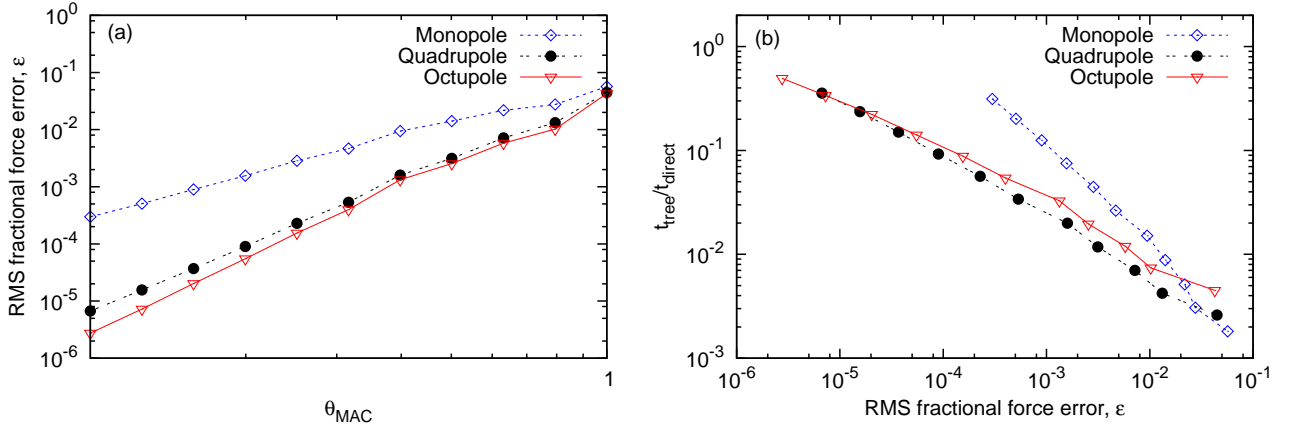,height=6cm,width=17.0cm,angle=270}}
\caption{(a) The root-mean-square fractional force error computing the gravitational forces for all particles in a uniform-density sphere with the Barnes-Hut tree using the Geometric MAC as a function of $\theta_{_{\rm MAC}}$, and (b) the ratio of CPU time for computing all gravitational forces with the tree to direct-summation as a function $\epsilon$.  The gravitational accelerations are calculated without kernel-softening, up to monopole (blue diamonds), quadrupole (solid black circles) and octupole (red triangles) order.}
\label{FIG:GEOMACERROR}
\end{figure*}

\begin{figure*}
\centerline{\psfig{figure=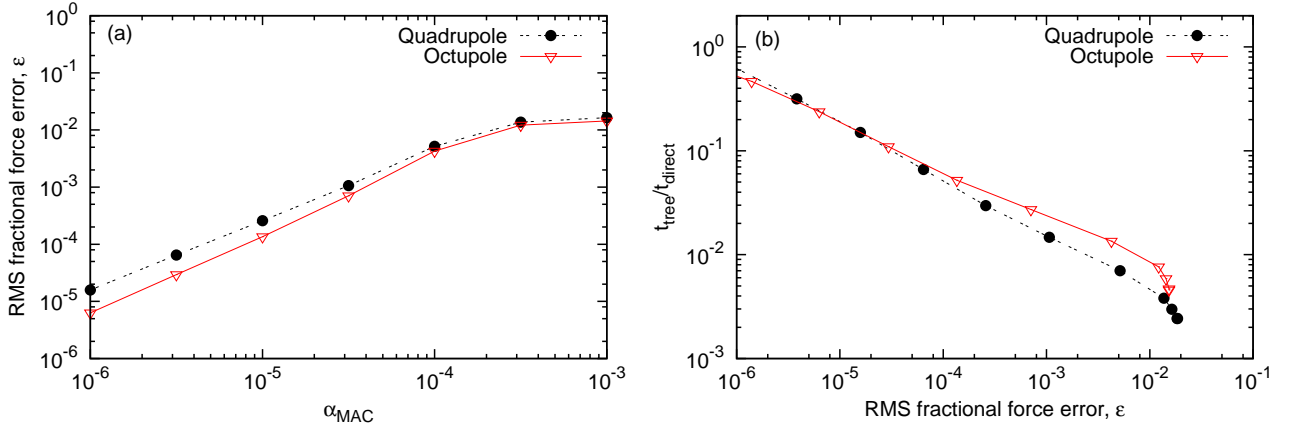,height=6cm,width=17.0cm,angle=270}}
\caption{(a) The root-mean-square fractional force error computing the gravitational forces for all particles in a uniform-density sphere with the Barnes-Hut tree using the Eigenvalue MAC as a function of $\alpha_{_{\rm MAC}}$, and (b) the ratio of CPU time for computing all gravitational forces with the tree to direct-summation as a function $\epsilon$.  The gravitational accelerations are calculated without kernel-softening, up to quadrupole (solid black circles) and octupole (red triangles) order.}
\label{FIG:EIGENMACERROR}
\end{figure*}

The resulting values of $\epsilon$ are plotted against $\theta_{_{\rm MAC}}$ and $\alpha_{_{\rm MAC}}$ in Figs. \ref{FIG:GEOMACERROR}(a) and \ref{FIG:EIGENMACERROR}(a). We see that for the Geometric MAC, $\epsilon$ decreases monotonically with decreasing $\theta_{_{\rm MAC}}$ and with the inclusion of higher-order multipole terms (cf. McMillan \& Aarseth 1993). Likewise the value of $\epsilon$ computed with the Eigenvalue MAC decreases monotonically for decreasing $\alpha_{_{\rm MAC}}$.

In Figs. \ref{FIG:GEOMACERROR}(b) and \ref{FIG:EIGENMACERROR}(b), we plot the CPU time required to compute all the gravitational accelerations using, respectively, the Geometric and Eigenvalue MACs, against the computed RMS fractional force error, $\epsilon$. For both MACS, and for all multipole-expansions, the CPU time increases as $\epsilon$ decreases. For the Geometric MAC, acceptably small values of $\epsilon$ are delivered much faster if the quadrupole terms are included. For both the Geometric and Eigenvalue MACs, the octupole terms do not deliver a big improvement in accuracy, and therefore -- in the interests of memory and CPU efficiency -- we normally evaluate only monopole and quadrupole terms.

The time required to calculate the gravitational accelerations using the tree is expected to scale as $N\,\log N$ (e.g. Pfalzner \& Gibbon 1996), compared with $N^2$ for a direct-summation.  Fig. \ref{FIG:TREESCALING} shows the average CPU time for calculating gravitational accelerations in a uniform density sphere using the tree with the Geometric MAC (red triangles) and using direct-summation (solid black circles). The two graphs scale as expected up to $10^5$ particles and beyond.}

%%%%%%%%%%%%%%%
\begin{figure*}
\centerline{\psfig{figure=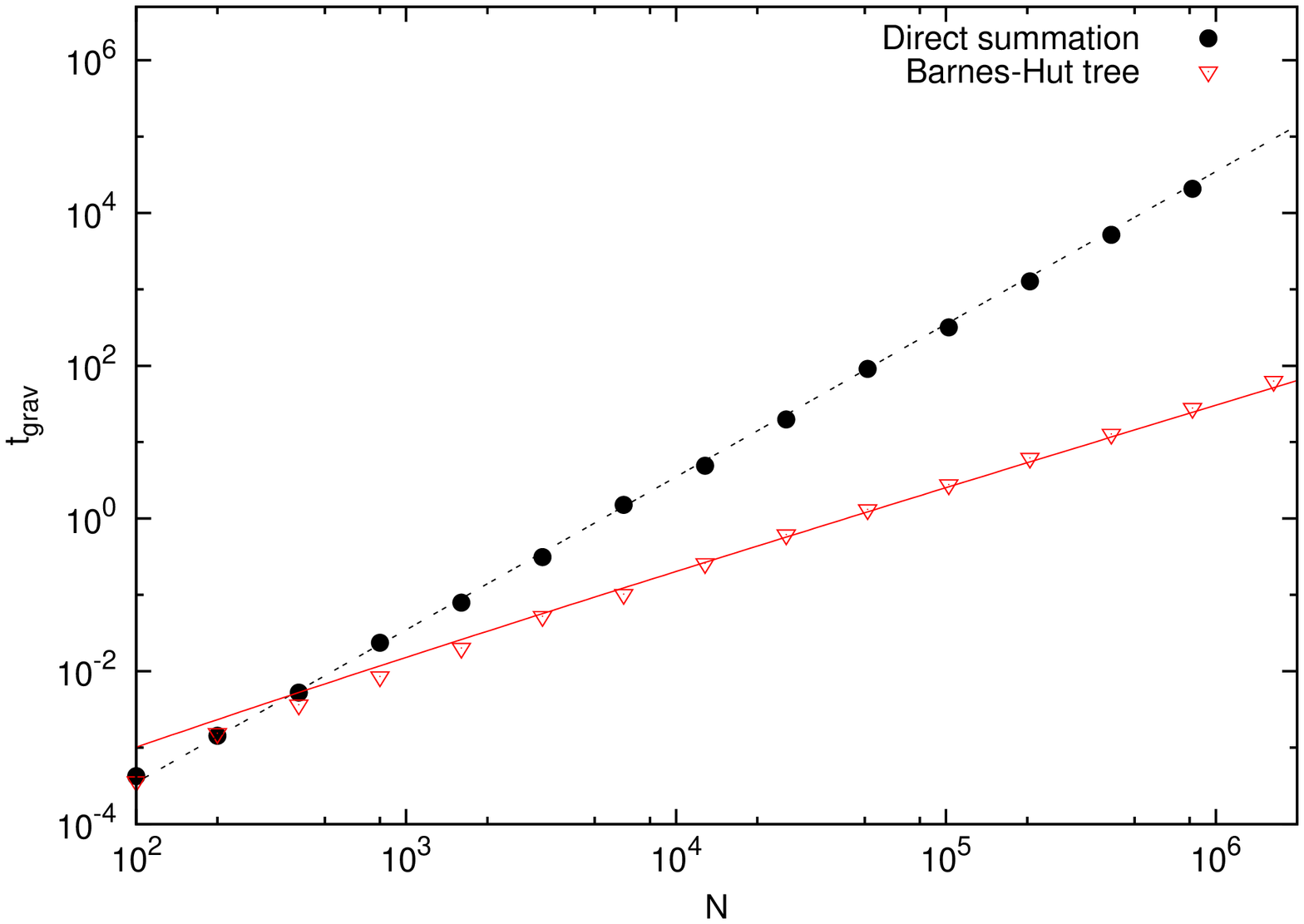,height=7.0cm,width=10.0cm,angle=270}}
\caption{Scaling characteristics of the Barnes-Hut tree code in SEREN.  The time taken to compute gravitational forces for all particles for direct-summation (solid black circles) and the Barnes-Hut tree (red triangles) as a function of particle number.  For reference, we show the expected scaling for $N^2$ (dashed line) and $N\,\log N$ (solid red line).}
\label{FIG:TREESCALING}
\end{figure*}
%%%%%%%%%%%%%

%%%%%%%%%%%%%%%
\begin{figure*}
\centerline{\psfig{figure=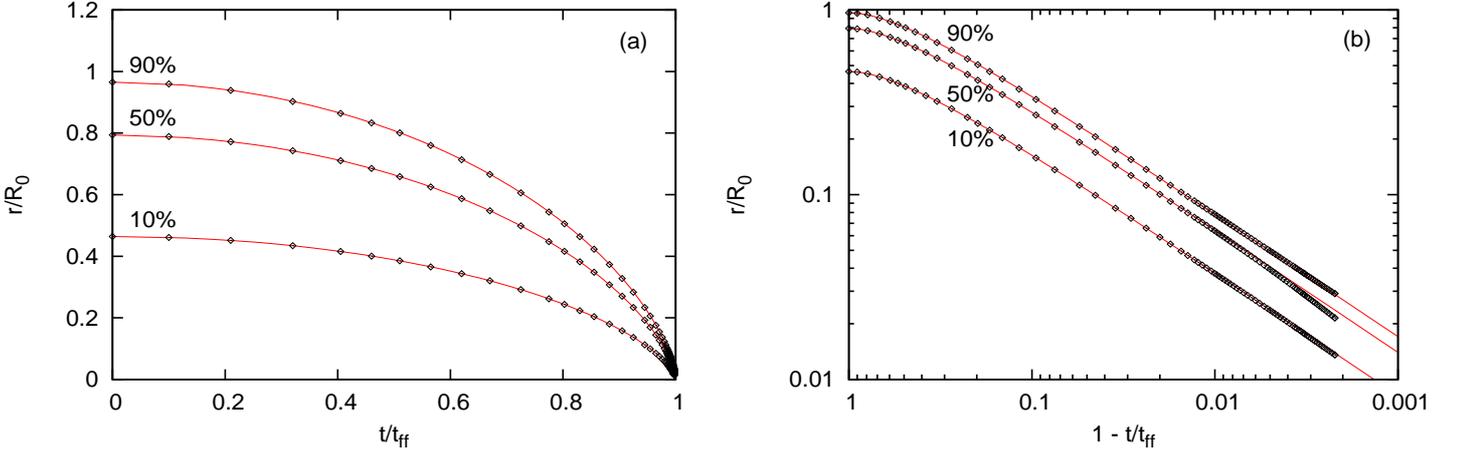,height=7.0cm,width=19.0cm,angle=270}}
\caption{(a) Freefall collapse of a pressure-less, uniform-density sphere. The figure shows the analytic solution (dashed lines), and the radial position of three representative particles at 90\%, 50\% and 10\% mass radii (filled circles). (b) Same as (a), but with time measured from the end of the collapse and using logarithmic axis.}
\label{FIG:FREEFALL}
\end{figure*}
%%%%%%%%%%%%%

%%%%%%%%%%%%%%%
\begin{figure*}
\centerline{\psfig{figure=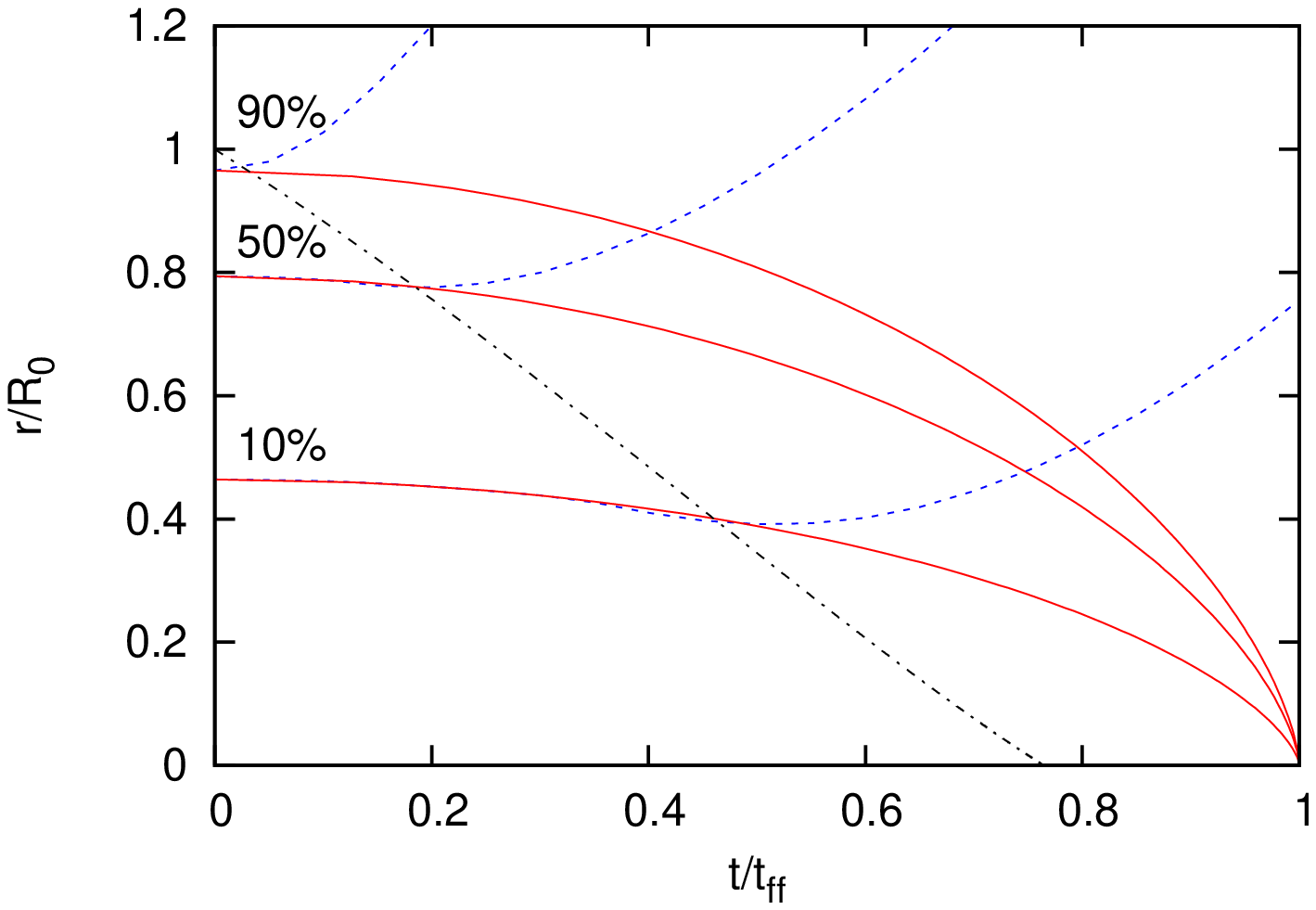,height=7.0cm,width=9.5cm,angle=270}}
\caption{Collapse of a uniform density sphere with an isothermal equation of state and dimensionless isothermal sound speed $c_s = 1$. The figure shows the analytic solution for pressure-less freefall collapse (solid red lines) and the radial position of the 90\%, 50\% and 10\% mass radii (blue dashed lines). The black dot-dashed line shows the analytic solution for the progression of the rarefaction wavefront.}
\label{FIG:ISOFREEFALL}
\end{figure*}
%%%%%%%%%%%%%

%%%%%%%%%%%%%%%%%%%%%%%%%%%%%%%%%%%%%%%%%%%%%%%%%%%%%%%%%%%%%%%%%%%%%%%%%%%%%%%%%%%%%%%%%%%%
\subsection{Freefall and isothermal collapse of a uniform density sphere}\label{SS:FREEFALL}

We test the accuracy of the gravitational acceleration evaluation in a dynamically-evolving system, by simulating the freefall collapse of a uniform density sphere. A static, uniform-density sphere with mass $M_{_{\rm O}}$, initial radius $R_{_{\rm O}}$ and initial density, $\rho_{_{\rm O}}=3M_{_{\rm O}}/4\pi R_{_{\rm O}}^3$, collapses to a singularity on a timescale $t_{_{\rm FF}}$; and a shell of the sphere which is initially ($t=0$) at radius $r_{_{\rm O}}$ is at subsequent times ($0<t\leq t_{_{\rm FF}}$) at radius $r$, given by
\begin{eqnarray} \label{EQN:FREEFALLRADIUS}
\frac{t}{t_{_{\rm FF}}}&=&\frac{2}{\pi}\left\{\cos^{-1}\left(\frac{r}{r_{_{\rm O}}}\right)^{1/2}+\left(\frac{r}{r_{_{\rm O}}}\right)^{1/2}\left(1-\frac{r}{r_{_{\rm O}}}\right)^{1/2}\right\}\,,\hspace{1.0cm}t_{_{\rm FF}}\;\,=\;\,\frac{\pi}{2}\left(\frac{R_{_{\rm O}}^3}{2\,G\,M_{_{\rm O}}}\right)^{1/2}\;\,=\;\,\left(\frac{3\,\pi}{32\,G\,\rho_{_{\rm O}}}\right)^{1/2}\,.
\end{eqnarray}
{ We set up the initial conditions by constructing a glass-like uniform-density sphere containing 100,000 SPH particles (as described in Section \ref{SS:IC})}.  The subsequent evolution of the particles is then followed invoking gravitational accelerations only. Fig. \ref{FIG:FREEFALL}(a) compares the 90\%, 50\% and 10\% mass radii as a function of time (dots) with the analytic solution (dashed lines). Significant divergence between the numerical results and the analytic solution -- due to gravitational softening, particle noise, and integration error -- occurs only after the density has increased by more than $10^7$ (See Fig. \ref{FIG:FREEFALL}(b)).

This test has been repeated, but now imposing an isothermal equation of state and invoking both gravitational and hydrostatic accelerations. The collapse is no longer homologous, since there is a pressure gradient at the edge of the sphere, and this drives a rarefaction wave into the cloud. Ahead of the rarefaction wave, the gas collapses in freefall, as before, but behind it the gas decelerates and then expands. Fig. \ref{FIG:ISOFREEFALL} compares the 90\%, 50\% and 10\% mass radii as a function of time (solid lines) with the analytic solution for the pressure-less collapse (Eqn. \ref{EQN:FREEFALLRADIUS}; dashed lines) and the position of the rarefaction wave as a function of time (Truelove et al. 1998; dot-dashed line). We use a dimensionless isothermal sound speed, $c_{_{\rm S}}=1$, so that the rarefaction wave reaches the centre of the sphere in less than a freefall time, preventing collapse to a singularity. Fig. \ref{FIG:ISOFREEFALL} shows that the gas motion diverges from freefall collapse just after the rarefaction wave passes, as it should. Slight deviations before this juncture are due to smoothing, gravitational softening, particle noise, and integration error.

\begin{figure*}
\centerline{\psfig{figure=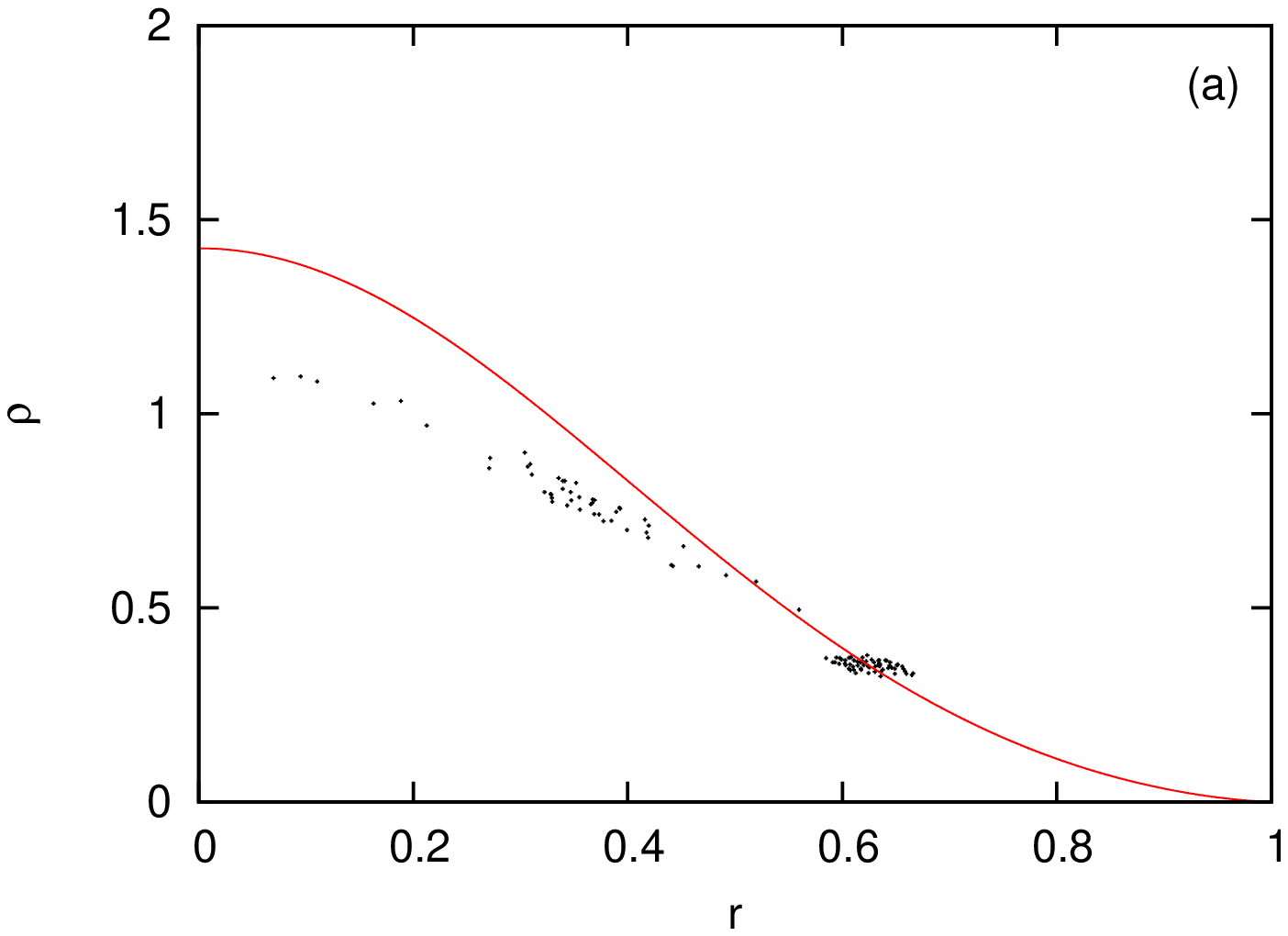,height=5cm,width=6.25cm,angle=270}
\hspace{0.05cm}\psfig{figure=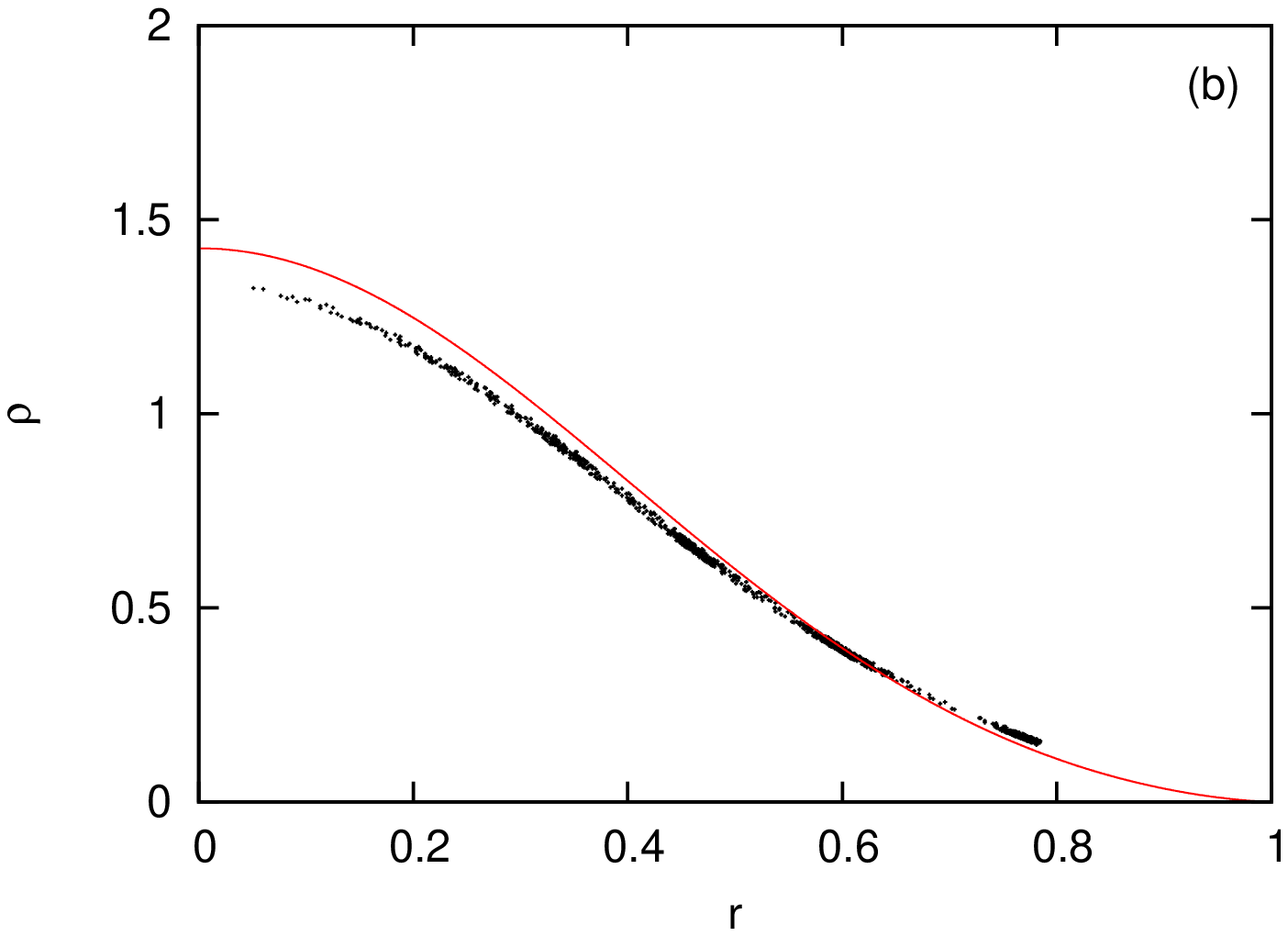,height=5cm,width=6.25cm,angle=270}\hspace{0.05cm}\psfig{figure=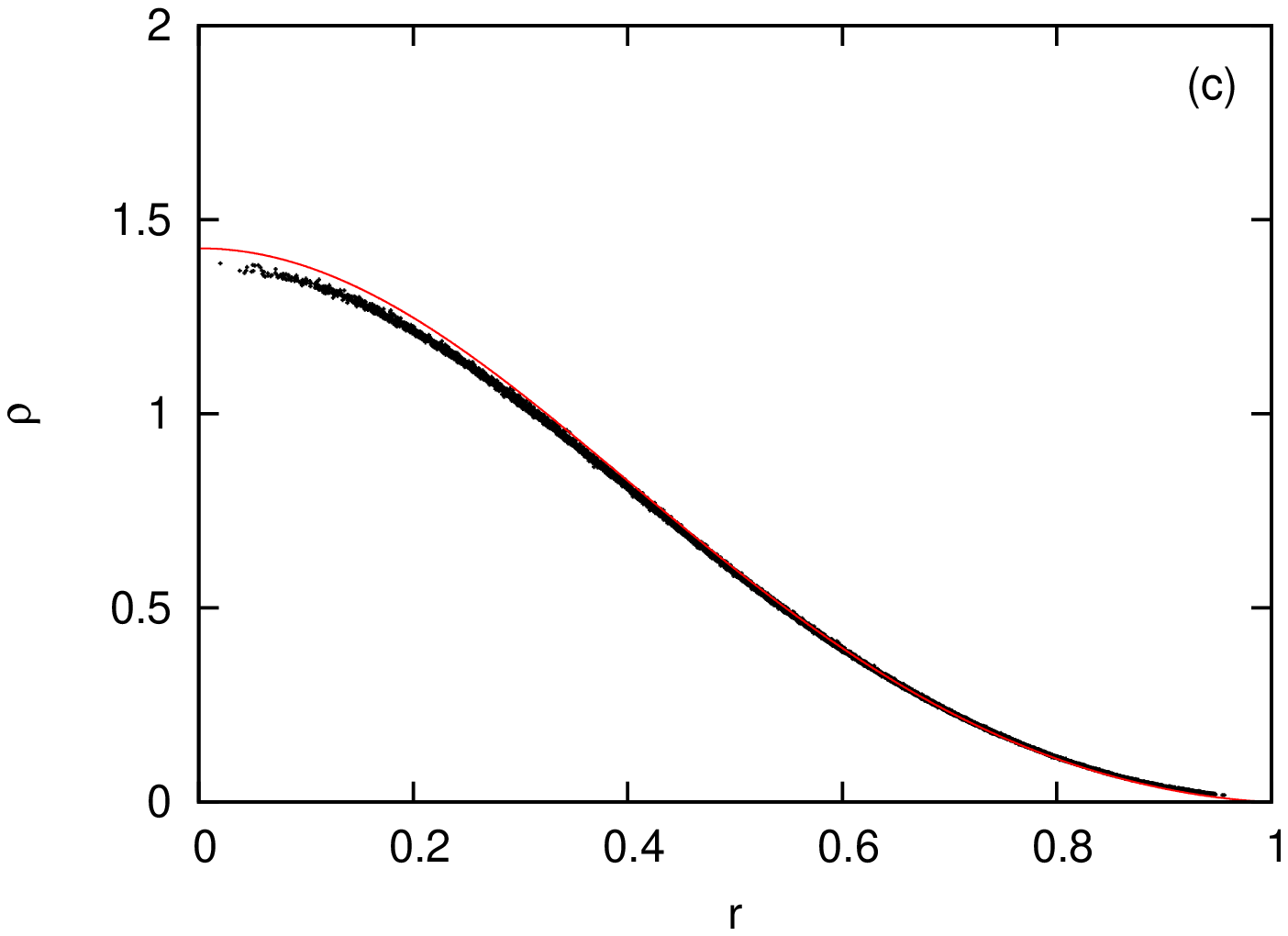,height=5cm,width=6.25cm,angle=270}}
\caption{Results of the polytrope test for an $\eta=5/3$ polytrope, using (a) 114, (b) 1086, and (c) $10^5$ SPH particles.}
\label{FIG:POLYTROPE}
\end{figure*}

%%%%%%%%%%%%%%%%%%%%%%%%%%%%%%%%%%%%%%%%%%%%
\subsection{Polytropes} \label{SS:POLYTROPE}
\begin{figure*}
\centerline{\psfig{figure=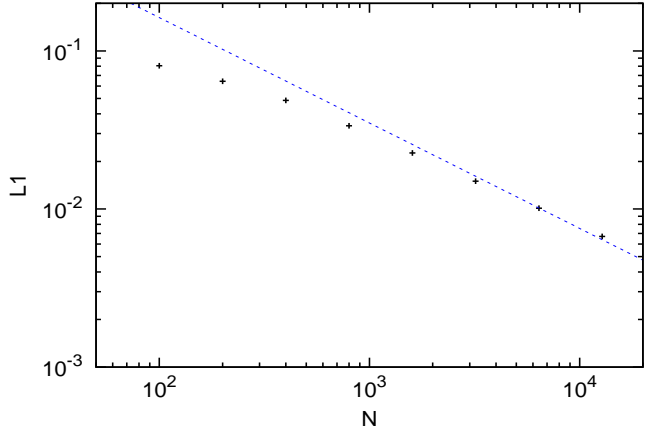,height=6cm,width=9cm,angle=270}}
\caption{{L1 error norm as a function of particle number for the static polytrope test.  The expected behaviour for an ideal 2nd-order numerical hydrodynamics scheme, $L1 \propto N^{-2/3}$,  is also plotted for reference (blue dashed line).}}
\label{FIG:L1ERRORPOLY}
\end{figure*}

This test demonstrates that SEREN can model the structure of an $\eta=5/3$ polytrope, and therefore should be able to handle general self-gravitating equilibria. The density profile of a polytrope is obtained by solving the Lane-Emden equation (Chandrasekhar 1939, Ch. IV). An $\eta=5/3$ polytrope with mass $M=1$ and radius $R=1$ (in dimensionless code units) has polytropic constant $K=0.4246$ (cf. Price \& Monaghan 2007). The initial conditions are generated by cutting a unit-mass, unit-radius sphere from a cube of settled particles ({ see Section \ref{SS:IC}}), and then stretching the particles radially so that, in spherical polar co-ordinates, the new radius of particle $i$, $r_i'$, is related to its old radius, $r_i$, by $M_{_{\rm POLY}}(r'_i)=r_i^3$, where $M_{_{\rm POLY}}(r')$ is the mass interior to radius $r'$ in the polytropic configuration; the angular co-ordinates of particle $i$ are not changed. Stretching distorts the local arrangement of individual particles, and so the new configuration is not in detailed equilibrium. We therefore evolve it with $\eta_{_{\rm SPH}}=1.2$, using artificial viscosity, until the system reaches equilibrium. This test has been performed with $N=114,\;1086\;\,{\rm and}\;\,10^5$ SPH particles. For $\eta_{_{\rm SPH}}=1.2$, $\bar{N}_{_{\rm NEIB}}\simeq 57$. Bate \& Burkert (1997) suggest that in SPH only condensations with $N \geq 2\bar{N}_{_{\rm NEIB}}$ particles are resolved. Therefore our very low-resolution test with $N=114=2\bar{N}_{_{\rm NEIB}}$ particles (Fig. \ref{FIG:POLYTROPE}(a)) demonstrates that SEREN can indeed { crudely model such a condensation, albeit with only approximately the correct radius and central density}; the grouping of particles near the boundary (at $r\sim 0.6$) in Fig. \ref{FIG:POLYTROPE}(a) reflects the tendency for well relaxed distributions of SPH particles to adopt a glass-like arrangement. Fig. \ref{FIG:POLYTROPE}(b) shows that with $N=1086\simeq 20\bar{N}_{_{\rm NEIB}}$ the polytrope is much better resolved, and Fig. \ref{FIG:POLYTROPE}(c) shows that with $N=10^5$ the density profile { almost exactly matches the Lane-Emden solution.  

We test convergence with the exact solution by calculating the L1 error norm as a function of particle number for varying resolutions. SPH is formally second-order accurate in space (e.g. Monaghan 1992) and therefore the L1 error should scale as $L1 \propto h^2 \propto N^{-2/D}$ where $D$ is the dimensionality (cf. Springel 2010b).  However, the discretization of the gas into particles introduces additional errors, so the error scales less well than second-order (e.g. $L1 \propto N^{-1}\,\log{N}$; e.g. Monaghan 1991).  Figure \ref{FIG:L1ERRORPOLY} demonstrates the L1 error norm as a function of total particle number.  We see that the L1 error norm decreases with increasing particle number, and therefore converges with increasing resolution.  Also plotted on Fig. \ref{FIG:L1ERRORPOLY} is the expected scaling for an ideal second-order scheme.  It can be seen that the convergence rate is similar to the ideal case, but a little shallower suggesting discretization errors are reducing the effective order of the scheme.}

%%%%%%%%%%%%%%%%%%%%%%%%%%%%%%%%%%%%%%%%%%%%

%%%%%%%%%%%%%%%%
\begin{table}[b]
\centering
\begin{tabular}{rccccccrccccc}
\hline\hline
{\sc figure-eight}\;\;\;ID & $m$ &         $x$ &        $y$ &        $v_x$ &       $v_y$ &\;& {\sc Burrau}\;\;\;ID & $m$ &  $x$ &  $y$ & $v_x$ & $v_y$ \\\hline
                       1\, & 1.0 &  0.97000436 & -0.2430875 &  0.466203685 &  0.43236573 &  &                  1\, & 3.0 &  1.0 &  3.0 &   0.0 &   0.0 \\
                       2\, & 1.0 & -0.97000436 &  0.2430875 &  0.466203685 &  0.43236573 &  &                  2\, & 4.0 & -2.0 & -1.0 &   0.0 &   0.0 \\
                       3\, & 1.0 &  0.0        &  0.0       & -0.93240737  & -0.86473146 &  &                  3\, & 5.0 &  1.0 & -1.0 &   0.0 &   0.0 \\
\end{tabular}
\caption{Initial conditions for the figure-eight 3-body problem (Chencier \& Montgomery 2000; Columns 1 to 6) and the Burrau 3-body problem (Burrau 1913; Columns 7 to 12). In both problems, the centre of mass is at the origin, the net linear and angular momenta are zero, and dimensionless units are used, such that $G=1$.}
\label{FIG:ICFIG8}
\end{table}
%%%%%%%%%%%

%%%%%%%%%%%%%%%%%%%%%%%%%%%%%%%%%%%%%%%%%%%%%%%%%%%%%%%%%%%%%%%%%
{
\subsection{Boss-Bodenheimer test} \label{SS:BBSIT}

%%%%%%%%%%%%%%%
\begin{figure*}
\centerline{\psfig{figure=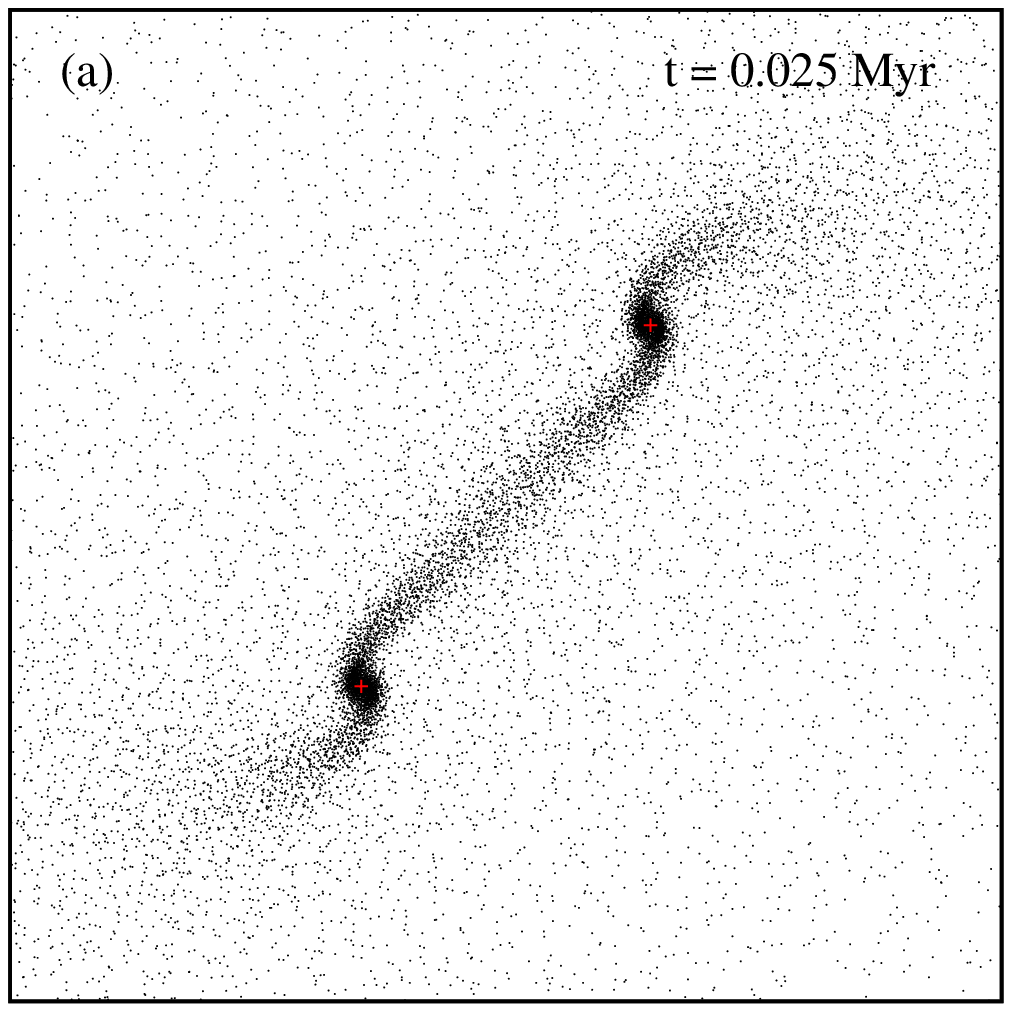,height=5.5cm,width=6.5cm,angle=270}
\psfig{figure=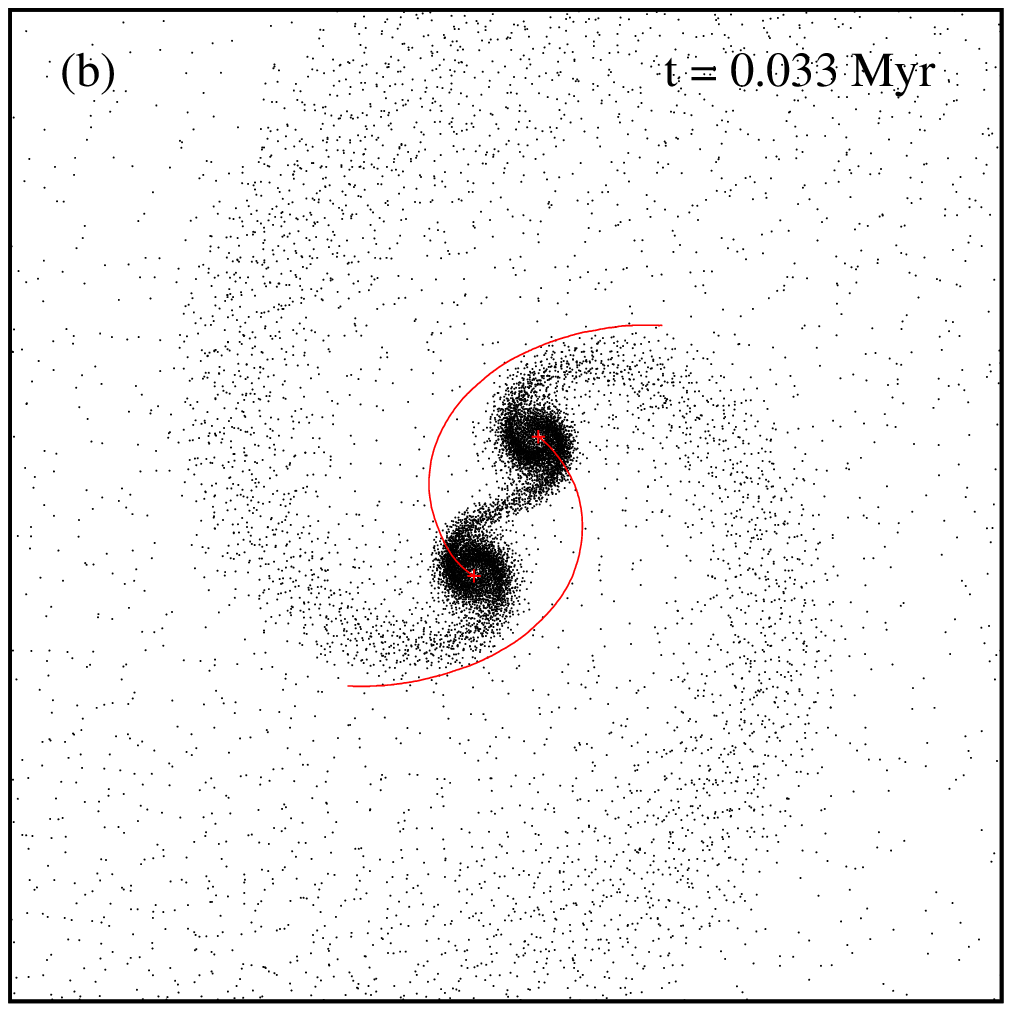,height=5.5cm,width=6.5cm,angle=270}
\psfig{figure=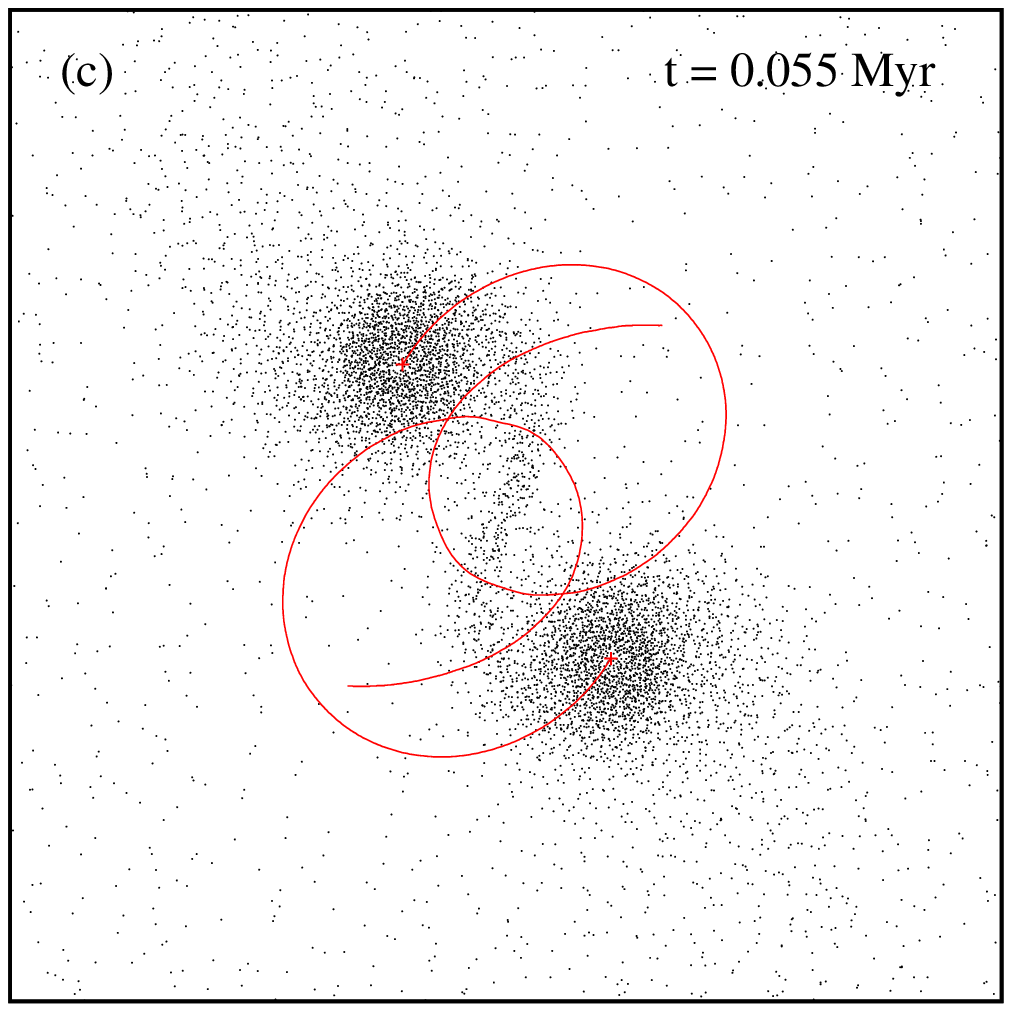,height=5.5cm,width=6.5cm,angle=270}}
\caption{SPH and sink particle plots of Boss-Bodenheimer test at times (a) $t = 0.025\,{\rm Myr}$, (b) $t = 0.033\,{\rm Myr}$ and (c) $t = 0.055\,{\rm Myr}$.  SPH particles are represented by black dots (only one in every three plotted for clarity) and the position and motion of the sink particles are represented by the red lines.  The first tile ($t = 0.025\,{\rm Myr}$) shows the particle distribution just after the formation of the two sinks in the condensations that form either end of the bar.  The subsequent times shows the motion of the sink particles as they move with the gas and the small disks that form around each sink.  All figures show the region $-0.005 < x < 0.005$, $-0.005 < y < 0.005$.}
\label{FIG:BBSIT}
\end{figure*}

The Boss-Bodenheimer test (Boss \& Bodenheimer 1979) is a standard test of star formation codes designed to investigate the non-axisymmetric collapse and fragmentation of a rotating, self-gravitating gas cloud.  The rotating cloud is seeded with an $m=2$ azimuthal perturbation and therefore collapses under self-gravity and forms a bar-like structure.  At the ends of the bar, dense condensations are formed.  

The initial conditions are set up as follows.  A relaxed (i.e. glass-like) uniform density sphere (see Section \ref{SS:IC} for details) is rescaled to produce the correct total mass, $M = 1\,{\rm M}_{\odot}$, radius, $R = 3.2 \times 10^{16}\,{\rm cm}$, and density $\rho_0 = 1.44 \times 10^{-17}\,{\rm g\,cm^{-3}}$.  We then add a sinusoidal, azimuthal density perturbation of the form 
\begin{eqnarray} \label{EQN:BBSIT}
\rho &=& \rho_0\,\left[ 1 + A\,\sin{\left(m\,\phi \right)} \right]
\end{eqnarray}
where $\phi$ is the azimuthal angle about the z-axis, $A = 0.5$ is the magnitude of the perturbation, and $m = 2$ is the order of the azimuthal perturbation.  The density perturbation is achieved by altering the particle positions rather than changing the masses of the particles.  The cloud is initially set in solid-body rotation with an angular velocity of $\Omega = 1.6 \times 10^{-12}\,{\rm rad\,s^{-1}}$.  
In our simulation, we use a barotropic EOS (Eqn. \ref{EQN:BAROTROPIC} with $T_{_0} = 10\,{\rm K}$ and $\rho_{_{\rm CRIT}} = 10^{-14}\,{\rm g\,cm^{-3}}$) in order to set a minimum scale for fragmentation of the cloud.  We use sink particles with a sink formation density of $\rho_{_{\rm SINK}} = 2 \times 10^{-12}\,{\rm g\,cm^{-3}}$ and sink radius $r_{_{\rm SINK}} = 2\,h_{_{\rm FORM}}$ where $h_{_{\rm FORM}}$ is the smoothing length of the SPH particle that triggers sink formation.  The freefall collapse timescale of the original unperturbed cloud is $t_{_{\rm FF}} = 17.4\,{\rm yr}$.  We use $50,000$ SPH particles in the original cloud in order to adequately resolve gravitational fragmentation with our choice of EOS (Bate \& Burkert 1997; Hubber, Goodwin \& Whitworth 2006).  The simulation is run until a time of $t = 100\,{\rm kyr}$.

The gas initially collapses under self-gravity to form a thin, dense `bar' with two denser condensations at either end.  The barotropic EOS (along with the relatively low resolution of the bar) prevents the bar collapsing to high densities once its density exceeds $\rho_{_{\rm CRIT}}$.  The denser condensations at either end of the bar are able to collapse to higher densities.  Eventually, the two condensations form sinks.  Fig. \ref{FIG:BBSIT}(a) shows the particle positions just after the formation of the two sinks.  The gas surrounding the sinks has some angular momentum relative to the sinks (from the original rotational field of the cloud) and assembles into two small disks which are connected together by the bar.  Subsequently, the two sinks follow eccentric orbits with a series of close approaches, during which the increased compression loads more mass into the disks, from both the surrounding gas and the bar. This leads to a period of rapid accretion, followed by a relatively quiet period as the sinks move towards apastron and the accretion rate drops off.  We note that in the presence of a reservoir of gas that constantly feeds accretion, the orbital properties of the system change continuously until the gas supply becomes negligible.

}

%%%%%%%%%%%%%%%%%%%%%%%%%%%%%%%%%%%%%%%%%%
\subsection{3-body tests} \label{SS:3BODY}

Since the N-body integrator in SEREN is intended to follow small-N systems, it is appropriate to perform 3-body tests for which accurate solutions are known (rather then large-N tests to which only statistical constraints can be applied). We limit ourselves to two such tests. 

\begin{figure*}
\centerline{\psfig{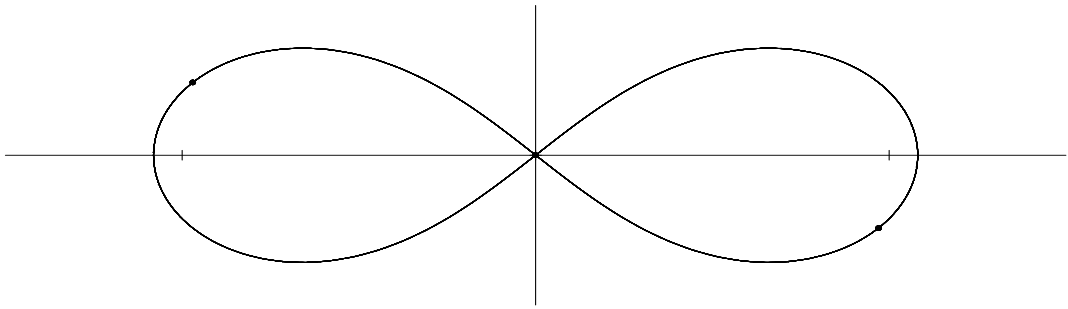}}
\caption{Tracks for the first 20 periods of the figure-eight 3-body problem. The positions of the stars are plotted every period, with solid black dots.}
\label{FIG:FIGURE8}
\end{figure*}

The first test is the figure-eight 3-body problem defined by Chenciner \& Montgomery (2000). The initial conditions for this test are summarised in Table \ref{FIG:ICFIG8} (left side). With these initial conditions, all three particles follow the same figure-eight trajectory, with period ${\cal P} = 6.32591398$. We have evolved this system using SEREN's Hermite N-body integrator with a timestep multiplier of $\gamma=0.05$, without the trajectory being corrupted. Fig. \ref{FIG:FIGURE8} shows the trajectory, and the positions of the three stars at $t={\cal N}{\cal P}$ for ${\cal N}=0,\,1,\,2,\,...\,20$, demonstrating that the stars return to the same positions every period. After 100 orbits, energy is conserved to better than one part in $10^6$, and the errors in the net linear and angular momenta are of order machine rounding error. The Hermite integrator therefore appears to be very stable.

\begin{figure*}
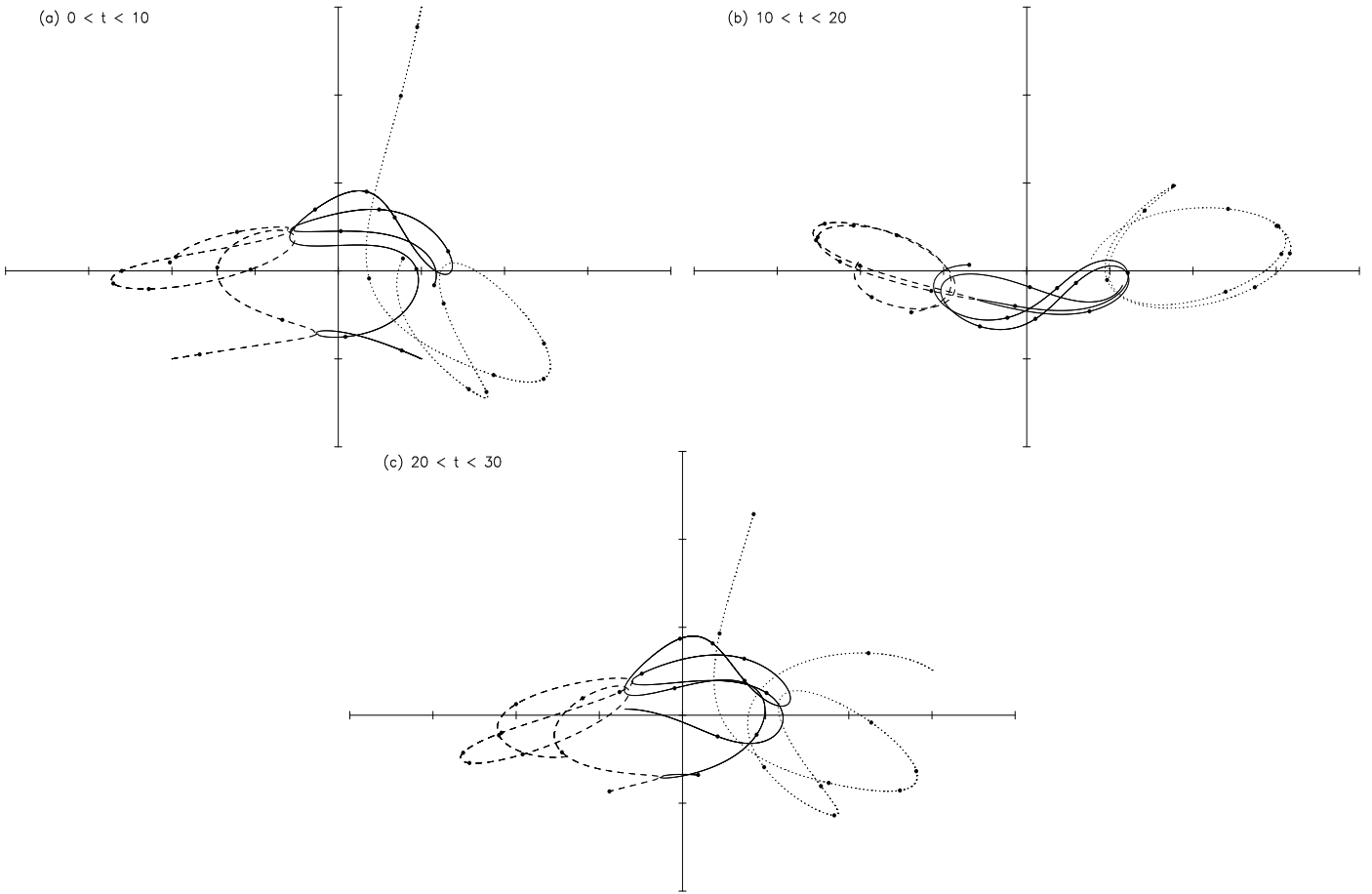

\centerline{\psfig{figure=14949-fg21.ps,height=6cm,width=9cm,angle=270}
\hspace{0.2cm}\psfig{figure=14949-fg22.ps,height=6cm,width=9cm,angle=270}}
\centerline{\psfig{figure=14949-fg23.ps,height=6cm,width=9cm,angle=270}} 
\caption{Tracks for the Burrau problem in the time intervals (a) $0 < t < 10$, (b) $10 < t < 20$, and (c) $20 < t < 30$.  The dotted lines track star 1, the dashed lines star 2, and the solid lines star 3. Each track includes solid dots at intervals of one time unit (i.e. at $t=1,\,2,\,3,\,\;{\rm etc.}$). These tracks should be compared with those presented by Szebehely \& Peters (1967; their Figures 2, 3, 4).}
\label{FIG:BURRAUTRACKS1}
\end{figure*}

\begin{figure*}
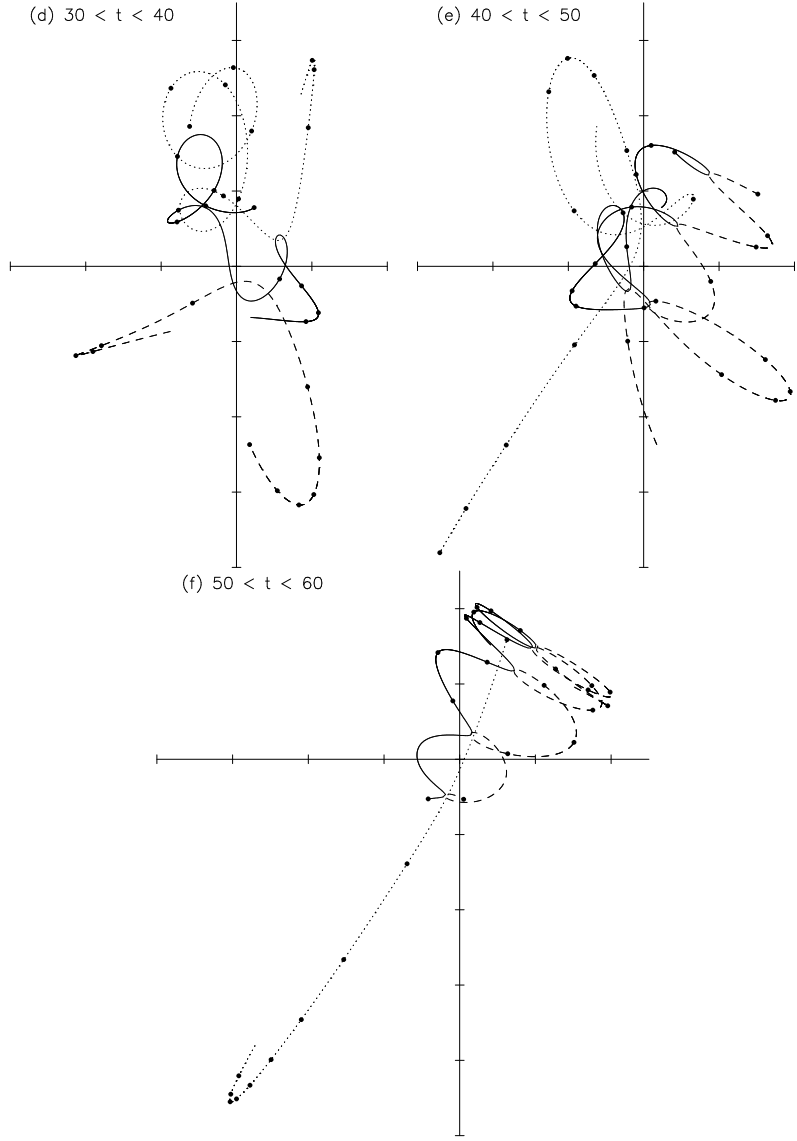

\centerline{\psfig{figure=14949-fg24.ps,height=7.5cm,width=5cm,angle=270}
\hspace{0.2cm} \psfig{figure=14949-fg25.ps,height=7.5cm,width=5cm,angle=270}}
\centerline{\psfig{figure=14949-fg26.ps,height=7.5cm,width=6.5cm,angle=270}}
\caption{As Fig. \ref{FIG:BURRAUTRACKS1}, but for the time intervals (d) $30 < t < 40$, (e) $40 < t < 50$, and (f) $50 < t < 60$. These tracks should be compared with Figs. 5, 6 and 7 in Szebehely \& Peters (1967).}
\label{FIG:BURRAUTRACKS2}
\end{figure*}

The second test is the 3-body problem devised by Burrau (1913), in which three particles are placed at the vertices of a right-angled triangle with sides 5, 4 and 3, and each particle has a mass equal to the length of the side opposite it. The initial conditions for this test are summarised in Table \ref{FIG:ICFIG8} (right side). The subsequent evolution involves close encounters (separations $|\Delta{\bf r}_{ij}|\la 10^{-3}$), and is therefore highly chaotic. The Burrau problem was first integrated numerically, to the point where one star is ejected permanently, by Szebehely \& Peters (1967), using the two-dimensional Levi-Civita (1904) regularisation method. We have evolved this system up to $t=70$, using SEREN's Hermite integrator with a low timestep multiplier of $\gamma=0.02$ and a smoothing length of $h=10^{-4}$. Energy is conserved to one part in $10^7$, and the errors in the net linear and angular momenta are of order machine rounding error. In Figs. \ref{FIG:BURRAUTRACKS1} and \ref{FIG:BURRAUTRACKS2} we plot orbital tracks for the same time intervals as Szebehely \& Peters (1967) and using the same line styles. The close agreement between our tracks and those of Szebehely \& Peters (1967) demonstrates the accuracy and robustness of SEREN's Hermite N-body integrator. 

%%%%%%%%%%%%%%%
%%%%%%%%%%%%%

%%%%%%%%%%%%%%%%
%%%%%%%%%%%%%

%%%%%%%%%%%%%%%
%%%%%%%%%%%%%

%%%%%%%%%%%%%%%%%%%%%%%%%%%%%%%%%%%%%%%%%%%%%%%%%%%%%%%%%%%%%%%
\section{Memory and cache optimisations} \label{S:OPTIMISATION}

Here we describe the features of SEREN which are designed to improve cache performance and reduce overall memory usage.

%%%%%%%%%%%%%%%%%%%%%%%%%%%%%%%%%%%%%%%%%%%%%%%%%%%%%%%%%%%%%%
\subsection{Particle re-ordering} \label{SS:SPACEFILLINGCURVE}

Particle data arrays are arranged in tree-walk order, i.e. the order in which individual particles are interrogated during a tree-walk. This ensures that all particles in the same leaf cell are contiguous in memory, and particles in nearby branches are likely to be in a nearby (if not contiguous) part of the memory, i.e. they are likely to be within the same cache block. This requires that the data arrays are repeatedly re-ordered, but the computational cost of re-ordering is relatively low compared with the run time saved by optimising the cache usage. For large numbers of particles run times are more than halved.  We do not use a more sophisticated space-filling curve, such as the Hilbert space-filling curve used in GADGET 2 (Springel 2005), which is optimal for distributed-memory architectures requiring large amounts of communication between nodes.

%%%%%%%%%%%%%%%%%%%%%%%%%%%%%%%%%%%
\subsection{Grouping particle data}

Since the run times of most SEREN simulations are dominated by the routines that compute gravitational accelerations, we group together in a single array all the data required for particle-particle gravitational interactions (i.e. position, mass, smoothing length). This optimises cache usage by ensuring that all the data required for calculating the gravitational interaction due to a particle is loaded in the same cache block, and therefore avoids thrashing the memory while loading the required variables.

%%%%%%%%%%%%%%%%%%%%%%%%%%%%%%%%%%%%%%%%%
\subsection{Minimising memory allocation}

For subroutines that compute SPH sums, we first walk the neighbour tree to obtain a potential-neighbour list. In the first instance, the code only allocates a small amount of memory to store the potential neighbour list ($N_{_{\rm MAX}}$ elements, where $N_{_{\rm MAX}} \ll N$), in order to reduce memory fragmentation. For example, in a 3D simulation, the expected mean number of neighbours might be $\bar{N}_{_{\rm NEIB}}=50$ (in the grad-$h$ formulation, $\bar{N}_{_{\rm NEIB}}=32\pi\eta_{_{\rm SPH}}^3/3$), and in this case an appropriate choice would be $N_{_{\rm MAX}} = 200$). Then, in the rare instances where more than $N_{_{\rm MAX}}$ potential neighbours are found (e.g. an isolated particle with a very large smoothing length), the memory is deallocated and reallocated to $N$ elements.

%%%%%%%%%%%%%%%%%%%%%%%%%%%%%%%%%%%%%%%%%%%%%%%%%%%
\section{Parallelisation} \label{S:PARALLELISATION}

SEREN is parallelised using OpenMP, for use on shared-memory architectures (for example, symmetric multiprocessing (SMP) and non-uniform memory access (NUMA) machines). { OpenMP requires that each processor can see all the data, and so there is no need for any explicit transfer of data between each processor's RAM, although there is some overhead associated with transferring data from the shared RAM to the local caches of individual processors}. OpenMP works by parallelising do-loops. If the operations executed in each cycle of a loop are independent of those executed in the other cycles, this can be achieved simply by adding OpenMP directives at the beginning and end of the loop. The cycles of the loop are then farmed out to the available processors; if there are $N$ cycles (corresponding to $N$ particles) and $N_p$ processors, each processor receives $N_{_{\rm BATCH}}=N/N_p$ cycles to execute. 

The scaling of a parallel code, ${\cal S}(N_p)$, is defined as the wall-clock time, $t(1)$, the code takes to perform a reference simulation on one processor, divided by the time, $t(N_p)$, it takes to perform the same simulation on $N_p$ processors, i.e. ${\cal S}(N_p)=t(1)/t(N_p)$. A perfectly scaling code has ${\cal S}(N_p)=N_p$, but normally scaling is less than perfect, because (i) some fraction of the code is inherently serial and cannot be parallelised (Amdahl's law), (ii) the code is not perfectly load-balanced at all times (i.e. not all processors are equally busy at all times), and (iii) { there is latency, for example due to communication of data between the OpenMP master node and the slave nodes}. In SEREN, these difficulties are compounded by the implementation of hierarchical block timesteps (See Section \ref{SS:BLOCKTIMESTEPS}). 

The main routines in SEREN are those that (a) construct and stock the tree, (b) determine the SPH smoothing lengths, densities and other local functions of state, (c) compute the hydrodynamic accelerations and heating terms, (d) compute the gravitational accelerations, and (e) advance the particle positions, velocities and internal energies. Of these the last four can be parallelised quite straightforwardly and effectively, but the first (tree-building) can not. In particular, the assigning of new tree cells, the construction of linked lists, and the re-ordering of particles (See Section \ref{SS:SPACEFILLINGCURVE}) can not be parallelised efficiently, and it is these elements, along with the other smaller serial sections of code, that ultimately limit the scalability of SEREN.

Even if the operations executed in each cycle of a do-loop are independent, na{\"i}ve application of OpenMP directives to the beginning and end of the do-loop will not guarantee load balancing, because the individual cycles do not necessarily entail similar amounts of computation. For example, walking the gravity tree is a much more compute-intensive operation for a particle in the densest regions of a fragmenting prestellar core than for a particle in the diffuse outer envelope of the same core. Therefore, in order to improve load balancing, the code delivers to each processor a small batch of cycles, and when the processor is finished executing these cycles it requests another batch. We find, empirically, that SEREN runs most efficiently with $N_{_{\rm BATCH}}\sim 10^{-3}N/N_p$.

When hierarchical block timesteps are used, SEREN maintains a list of the IDs of all the active particles (i.e. the $N_{_{\rm ACTIVE}}$ particles whose accelerations and heating terms are being computed on the current timestep). Load balancing is then achieved by only looping over this active list, and farming out batches of size $N_{_{\rm BATCH}}\sim 10^{-3}N_{_{\rm ACTIVE}}/N_p$ to the individual processors.

%%%%%%%%%%%%%%%%
\begin{figure*} 
\centerline{\psfig{figure=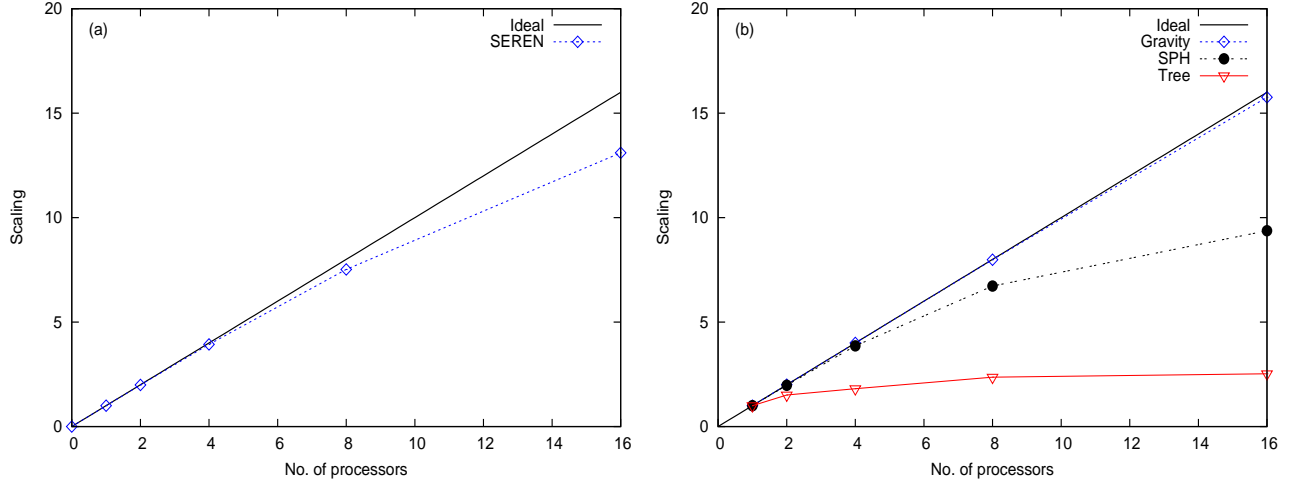,height=6.5cm,width=17cm,angle=270}}
\caption{(a) Net scaling of SEREN, using OpenMP on a 16-core SMP machine, as a function of the number of processors. (b) Scaling of the individual gravity, SPH and tree-building routines, as a function of the number of processors.}
\label{FIG:OMPSCALING1}
\end{figure*}
%%%%%%%%%%%%%

%%%%%%%%%%%%%%%%%%%%%%%%%%%%%%%%%%%%%%%%%%%%%
\subsection{Scaling tests} \label{SS:SCALING}

To test the scaling of SEREN we revisit the collapse of a spherical isothermal cloud which initially is at rest with uniform density (see Section \ref{SS:FREEFALL}). We model the cloud with $10^6$ particles, and follow the evolution to dimensionless time $t=0.6$, using global timesteps. Since the cloud does not develop any complicated internal structure, this is a relatively undemanding test. Fig. \ref{FIG:OMPSCALING1}a shows the net scaling obtained on a 16-core SMP node of the Cardiff University Merlin cluster, using 1, 2, 4, 8 and 16 processors. The scaling is good up to 8 processors, but for 16 processors is starting to deteriorate (${\cal S}(16)\sim 13$). This indicates that SEREN is not likely to be able to exploit SMP machines with 100+ cores. Fig. \ref{FIG:OMPSCALING1}b shows how the main routines in SEREN scale individually. Evidently the gravity routines scale very well, almost perfectly; the SPH routines start to deteriorate at 8 processors (${\cal S}_{_{\rm SPH}}(8)\sim 7$); and the tree-building routines scale very poorly. It is the tree-building routines that limit the net scaling.

%%%%%%%%%%%%%%%%%%%%%%%%%%%%%%%%%%%%%%%%%%%%%%%%%%
\section{Future development} \label{S:FUTUREPLANS}

We are continuing to develop SEREN and add new features.  Some of these features have already been implemented in our development code and will be released to the main working version of SEREN once they are fully tested and debugged. The most important of these additions are (i) an MPI-parallelised version of the code (McLeod et al. in preparation); (ii)  an hybrid flux-limited diffusion and radiative cooling scheme (Forgan et al. 2009); (iii) the use of different timesteps for hydrodynamical and gravitational accelerations (Saitoh \& Makino 2010); { (iv) improved sink particles with feedback; (v) an integrated N-body and SPH integrator to model cluster dynamics with a live gas potential;} (vi) ideal MHD using divergence cleaning and/or Euler potentials (Price \& Monaghan 2004a, 2004b, 2005, Rosswog \& Price 2007) and (vii) non-ideal MHD (Hosking \& Whitworth 2004).  
The MPI version of SEREN will be a hybrid MPI/OpenMP code that can parallelise a group of shared memory nodes using MPI communication to link them together.  This will reduce the amount of communication between nodes, which is often the bottleneck to good scalability over a large number of processors. The remaining additions to SEREN are implementations of existing algorithms. We will provide up-to-date information on the development status of SEREN, and any further tests we have performed, at the web address {\rm http://www.astro.group.shef.ac.uk/seren}.

%%%%%%%%%%%%%%%%%%%%%%%%
\begin{acknowledgements}
We would like to thank Simon Goodwin for providing the DRAGON SPH code to the authors on which some of the initial development of SEREN was based upon.  We would also like to thank Sumedh Anathpindika, Thomas Bisbas, Steinar B{\o}rve, Murat Kaplan, Dimitrios Stamatellos, Jan Trulsen, Stephanie Walch \& Richard W\"{u}nsch for helpful suggestions and comments during the development of the code { and for donating some routines to SEREN}.  We also thank the anonymous referee for useful suggestions that have improved the paper.  DAH is funded by a Leverhulme Trust Research Project Grant (F/00 118/BJ). CPB is funded by an STFC studentship.  AM is funded by an STFC studentship and was funded as an Early-Stage Researcher by the EC-funded CONSTELLATION Marie Curie Training Network MRTN-CT-2006-035890. APW gratefully acknowledges the support of the STFC rolling grant PP/E000967/1, and the CONSTELLATION network.  We thank Daniel Price for the use SPLASH (Price 2007) for creating some images.
\end{acknowledgements}

%%%%%%%%%
\appendix

%%%%%%%%%%%%%%%%%%%%%%%%%%%%%%%%%%%%%%%%%%%
\section{Kernel functions} \label{A:KERNEL}

%%%%%%%%%%%%%%%%%%%%%%%%%%%%%%%%%%%%%%%%%%%%%%%%%
\subsection{M4 cubic spline kernel} \label{SS:M4}

The M4 cubic spline kernel (Monaghan \& Lattanzio 1985) is used in many implementations of SPH, due to its simple form and its compact support.  The M4 kernel is a function of $s\equiv r/h$ only. For $D=1,\;2,\;{\rm and}\;3\,$ dimensions, it takes the form
\begin{eqnarray} \label{EQN:M4} 
W (s) &=& \frac{\sigma_{\!_D}}{h^D} 
\left\{ \;\; \begin{array}{ll} 
1 - \frac{3}{2}s^2 + \frac{3}{4}s^3 \;\;\;\; & \;\;0 \leq s \leq 
1\,; \\ \frac{1}{4}(2-s)^3 \;\;\;\; & \;\;1 \leq s \leq 2\,; \\ 
0 \;\;\;\; & \;\;s > 2 \,.
\end{array} \right . 
\end{eqnarray}
where $\sigma_{_1}=2/3$, $\sigma_{_2}=10/7\pi$, and $\sigma_{_3}=1/\pi$. The first spatial derivative is
\begin{eqnarray} \label{EQN:M4DERIV}
\frac{dW}{dr}(s) &=& - \frac{\sigma_{\!_D}}{h^{D+1}}
\left\{ \;\; \begin{array}{ll}
3s - \frac{9}{4}s^2 \;\;\;\; & \;\;0 \leq s \leq 1\,; \\ 
\frac{3}{4}(2-s)^2 \;\;\;\; & \;\;1 \leq s \leq 2\,; \\ 
0 \;\;\;\; & \;\;s > 2 \,.
\end{array} \right .
\end{eqnarray} 
SEREN also allows the modified derivative proposed by Thomas \& Couchman (1992) to prevent the clumping instability,
\begin{eqnarray} \label{EQN:M4DERIV2}
\left.\frac{dW}{dr}\right|_{_{\rm TC}}(s) &=& - \frac{\sigma_{\!_D}}{h^{D+1}}
\left\{ \;\; \begin{array}{ll}
1 \;\;\;\; & \;\; 0 \leq s \leq \frac{2}{3}\,; \\
3s - \frac{9}{4}s^2 \;\;\;\; & \;\;\frac{2}{3} \leq s \leq 1\,; \\
\frac{3}{4}(2-s)^2 \;\;\;\; & \;\;1 \leq s \leq 2\,; \\
0 \;\;\;\; & \;\;s > 2 \,.
\end{array} \right. 
\end{eqnarray}
For `grad-h' SPH, the $\Omega$ correction kernel function is given by 
\begin{eqnarray} \label{EQN:M4PMH} 
\frac{\partial W}{\partial h}(s) = \frac{\sigma_{\!_D}}{h^{D+1}} 
\left\{ \;\; \begin{array}{ll} 
-D + \frac{3}{2}(D + 2)\,s^2 - \frac{3}{4}(D + 3)\,s^3 
\;\;\;\; & \;\;0 \leq s \leq 1\,; \\ 
-2\,D + 3\,(D + 1)\,s - \frac{3}{2}(D + 2)\,s^2 + \frac{1}{4}(D + 3)\,s^3 
 \;\;\;\; & \;\;1 \leq s \leq 2\,; \\ 
0 \;\;\;\; & \;\;s > 2 \,.
\end{array} \right . 
\end{eqnarray}
For kernel-softened gravity (three dimensions only), the kernel function $\phi'$ is
\begin{eqnarray} \label{EQN:M4KS} 
\phi'(s) &=& \frac{1}{h^2} \left\{ \;\; \begin{array}{ll} 
\frac{4}{3}s - \frac{6}{5}s^3 + \frac{1}{2}s^4 \;\;\;\; & 
\;\;0 \leq s \leq 1\,; \\ 
\frac{8}{3}s - 3s^2 + \frac{6}{5}s^3 - \frac{1}{6}s^4 - \frac{1}{15}\frac{1}{s^2} \;\;\;\; & \;\;1 \leq s \leq 2\,; \\ 
1/s^2 \;\;\;\; & \;\;s > 2 \,.
\end{array} \right . 
\end{eqnarray}
For calculating the gravitational potential, the kernel function $\phi$ is 
\begin{eqnarray} \label{EQN:M4POT} 
\phi (s) &=& -\frac{1}{h} \left\{ \;\; \begin{array}{ll} 
\frac{7}{5} - \frac{2}{3}s^2 + \frac{3}{10}s^4 - \frac{1}{10}s^5 \;\;\;\; & 
\;\;0 \leq s \leq 1\,; \\ 
\frac{8}{5} - \frac{4}{3}s^2 + s^3 - \frac{3}{10}s^4 + \frac{1}{30}s^5 - \frac{1}{15}\frac{1}{s} \;\;\;\; & \;\;1 \leq s \leq 2\,; \\ 
1/s \;\;\;\; & \;\;s > 2 \,.
\end{array} \right . 
\end{eqnarray}
For `grad-h' gravity (Price \& Monaghan 2007), the kernel function $\zeta$ is calculated using 
\begin{eqnarray} \label{EQN:M4PMGRAV} 
\frac{\partial \phi}{\partial h} (s) &=& \frac{1}{h^2} 
\left\{ \;\; \begin{array}{ll} 
\frac{7}{5} - 2\,s^2 + \frac{3}{2}\,s^4 - \frac{3}{5}\,s^5
\;\;\;\; & \;\;0 \leq s \leq 1\,; \\ 
\frac{8}{5} - 4\,s^2 + 4\,s^3 - \frac{3}{2}\,s^4 + \frac{1}{5}\,s^5
 \;\;\;\; & \;\;1 \leq s \leq 2\,; \\ 
0 \;\;\;\; & \;\;s > 2 \,.
\end{array} \right . 
\end{eqnarray}

%%%%%%%%%%%%%%%%%%%%%%%%%%%%%%%%%%%%%%%%%%%%%%%%%%%%%
\subsection{Quintic spline kernel} \label{SS:QUINTIC}

The quintic spline kernel (Morris 1996) is a fifth-order polynomial function with compact support. It was originally presented in the form of a factorised polynomial.  However, to facilitate the processes of differentiation and integration that are required to compute the other kernel functions, we expand the brackets into a simple power series:
\begin{eqnarray} \label{EQN:QUINTIC} 
W (s) &=& \frac{\sigma_{\!_D}}{h^D} 
\left\{ \;\; \begin{array}{ll} 
66 - 60s^2 + 30s^4 - 10s^5 \;\;\;\; & \;\;0 \leq s \leq 1\,; \\ 
51 + 75s - 210s^2 + 150s^3 - 45s^4 + 5s^5 \;\;\;\; & \;\;1 \leq s \leq 2\,; \\ 
243 - 405s + 270s^2 - 90s^3 + 15s^4 - s^5 \;\;\;\; & \;\;2 \leq s \leq 3\,; \\ 
0 \;\;\;\; & \;\;s > 3 \,.
\end{array} \right . 
\end{eqnarray}
where $s \equiv r/h$ and $\sigma_{_1}=120$, $\sigma_{_2}=7/478\pi$, and $\sigma_{_3}=3/359\pi$. The first spatial derivative is 
\begin{eqnarray} \label{EQN:QUINTICDERIV} 
\frac{dW}{dr}(s) &=& \frac{\sigma_{\!_D}}{h^{D+1}} 
\left\{ \;\; \begin{array}{ll} 
- 120s + 120s^3 - 50s^4 \;\;\;\; & \;\;0 \leq s \leq 1\,; \\ 
75 - 420s + 450s^2 - 180s^3 + 25s^4 \;\;\;\; & \;\;1 \leq s \leq 2\,; \\ 
- 405 + 540s - 270s^2 + 60s^3 - 5s^4 \;\;\;\; & \;\;2 \leq s \leq 3\,; \\ 
0 \;\;\;\; & \;\;s > 3 \,.
\end{array} \right . 
\end{eqnarray}
For `grad-h' SPH, 
\begin{eqnarray} \label{EQN:QUINTICGRADH} 
\frac{\partial W}{\partial h}(s) &=& - \frac{\sigma_{\!_D}}{h^{D+1}} 
\left\{ \;\; \begin{array}{ll} 
66D - 60(D + 2)s^2 + 30(D + 4)s^4 - 10(D + 5)s^5 \;\;\;\; & \;\;0 \leq s \leq 1\,; \\ 
51D + 75(D + 1)s - 210(D + 2)s^2 + 150(D + 3)s^3 - 45(D + 4)s^4 + 5(D + 5)s^5 \;\;\;\; & \;\;1 \leq s \leq 2\,; \\ 
243D - 405(D + 1)s + 270(D + 2)s^2 - 90(D + 3)s^3 + 15(D + 4)s^4 - (D + 5)s^5 \;\;\;\; & \;\;2 \leq s \leq 3\,; \\ 
0 \;\;\;\; & \;\;s > 3 \,.
\end{array} \right . 
\end{eqnarray}
For kernel-softened gravity (three dimensions only), the kernel function $\phi'$ is
\begin{eqnarray} \label{EQN:QUINTICKS} 
\phi'(s) &=& \frac{4\,\pi\,\sigma}{h^2} \left\{ \;\; \begin{array}{ll} 
22s - 12s^3 + \frac{30}{7}s^5 - \frac{5}{4}s^6 \;\;\;\; & 
\;\;0 \leq s \leq 1\,; \\ 
17s + \frac{75}{4}s^2 - 42s^3 + 25s^4 - \frac{45}{7}s^5 + \frac{5}{8}s^6 + \frac{5}{56}\frac{1}{s^2} \;\;\;\; & \;\;1 \leq s \leq 2\,; \\ 
81s - \frac{405}{4}s^2 + 54s^3 - 15s^4 + \frac{15}{7}s^5 - \frac{1}{8}s^6 + \frac{507}{56}\frac{1}{s^2} \;\;\;\; & \;\;2 \leq s \leq 3\,; \\ 
1/s^2 \;\;\;\; & \;\;s > 3 \,.
\end{array} \right . 
\end{eqnarray}
The gravitational potential kernel $\phi$ is (three dimensions only) 
\begin{eqnarray} \label{EQN:QUINTICPOT} 
\phi (s) &=& -\frac{4\,\pi\,\sigma}{h} \left\{ \;\; \begin{array}{ll} 
\frac{478}{14} - 11s^2 + 3s^4 - \frac{5}{7}s^6 + \frac{5}{28}s^7 \;\;\;\; & 
\;\;0 \leq s \leq 1\,; \\ 
\frac{473}{14} - \frac{17}{2}s^2 - \frac{25}{4}s^3 + \frac{21}{2}s^4 - 5s^5 + \frac{15}{14}s^6 - \frac{5}{56}s^7 + \frac{5}{56}\frac{1}{s} \;\;\;\; & \;\;1 \leq s \leq 2\,; \\ 
\frac{729}{14} - \frac{81}{2}s^2 + \frac{135}{4}s^3 - \frac{27}{2}s^4 + 3s^5 - \frac{5}{14}s^6 + \frac{1}{56}s^7 - \frac{507}{56}\frac{1}{s} \;\;\;\; & \;\;1 \leq s \leq 2\,; \\
1/s \;\;\;\; & \;\;s > 3 \,.
\end{array} \right . 
\end{eqnarray}
For `grad-h' gravity (three dimensions only), the kernel function $\zeta$ is calculated using 
\begin{eqnarray} \label{EQN:QUINTICPMGRAV} 
\frac{\partial \phi}{\partial h} (s) &=& \frac{4\,\pi\,\sigma}{h} \left\{ \;\; \begin{array}{ll} 
\frac{478}{14} - 33s^2 + 15s^4 - 5s^6 + \frac{10}{7}s^7 \;\;\;\; & 
\;\;0 \leq s \leq 1\,; \\ 
\frac{473}{14} - \frac{51}{2}s^2 - 25s^3 + \frac{105}{2}s^4 - 30s^5 + \frac{15}{2}s^6 - \frac{5}{7}s^7 \;\;\;\; & \;\;1 \leq s \leq 2\,; \\ 
\frac{729}{14} - \frac{243}{2}s^2 + 135s^3 - \frac{135}{2}s^4 + 18s^5 - \frac{5}{2}s^6 + \frac{1}{7}s^7 \;\;\;\; & \;\;1 \leq s \leq 2\,; \\
0 \;\;\;\; & \;\;s > 3 \,.
\end{array} \right . 
\end{eqnarray}

%%%%%%%%%%%%%%%%%%%%%%%%%%%%%%%%%%%%%%%%%%%%%%%
\section{Multipole moments} \label{A:MULTIPOLE}

When calculating gravitational accelerations, using the Barnes-Hut gravity tree, SEREN calculate the contribution from a cell up to octupole order, if requested. The multipole moments of each cell are computed relative to the centre of mass of the cell; this means that the dipole term is always zero. The components of the quadrupole moment tensor, $Q$, for a leaf cell, $c$, are given by 
\begin{eqnarray} 
Q_{ab,c} &=& \sum \limits_{i} m\ssi \left( 3\,x_{a,i} x_{b,i} - 
r_i^2 \delta_{ab} \right)\,, \label{EQN:QUADMOM1}
\end{eqnarray}
where the summation is over all the particles $i$ in the leaf cell. If the cell is not a leaf cell, the quadrupole moment tensor is given by
\begin{eqnarray} 
Q_{ab,c} &=& \sum \limits_{d} m_d \left( 3\,x_{a,d} x_{b,d} - 
r_d^2 \delta_{ab} \right) + \sum \limits_d Q_{ab,d}\,, \label{EQN:QUADMOM2}
\end{eqnarray}
where the summation is over all the daughter cells $d$. The octupole moment tensor, $S$, for a leaf cell, $c$, is given by 
\begin{eqnarray}
S_{ab,c} &=& \sum \limits_{i}  m\ssi\left[ 5 \left( 3 - 2\delta_{ab} \right) 
x_{a,i}^2 - 3 r_i^2 \right] x_{b,i}\,, \label{OCTMOM1} \\
S_{123,c} &=& 15 \sum \limits_{i} m\ssi x_{1,i} x_{2,i} x_{3,i}\,; \label{OCTMOM2}
\end{eqnarray}
and for a non-leaf cell by 
\begin{eqnarray}
S_{ab,c} &=& \sum \limits_{d} m\ssi\left[ 5 \left( 3 - 2\delta_{ab} \right) 
x_{a,d}^2 - 3 r_d^2 \right] x_{b,d} + \sum \limits_{d} \left[ 
5 \left( 1 - \delta_{ab} \right) x_{a,d}Q_{ab,d} + \tfrac{5}{2}x_{b,d}Q_{aa,d} 
 - x_{l,d}Q_{bl,d} + S_{ab,d} \right]\,, \label{OCTMOM3} \\
S_{123,c} &=& 15 \sum \limits_{d} m\ssi x_{1,d} x_{2,d} x_{3,d} + \sum \limits_{d} \left[ \tfrac{5}{3} \left( x_{1,d}Q_{23,d} + x_{2,d}Q_{31,d} 
+ x_{3,d}Q_{12,d} \right) + S_{123,d} \right]\,. \label{OCTMOM4}
\end{eqnarray}

%%%%%%%%%%%%%%%%%%%%%%%%%%


\begin{thebibliography}{}%
%%%%%%%%%%%%%%%%%%%%%%%%%%

  \bibitem[2001]{Aarseth} Aarseth, S.J., 2001, NewA, 6, 277
  \bibitem[2003]{AarsethBook} Aarseth, S.J., 2003, `Gravitational N-Body Simulations: Tools and Algorithms' (Cambridge University Press)
  \bibitem[2002]{ENZO} Abel, T., Bryan, G.I. \& Norman, M.L., Sci, 2002, Vol. 295, Issue 5552, 93
  \bibitem[2007]{Agertz} Agertz, O., Moore, B., Stadel, J., et. al. 2007, MNRAS, 380, 963
  \bibitem[1995]{Bals} Balsara, D.S., 1995, JCoPh, 121, 357
  \bibitem[1986]{Tree} Barnes, J. \& Hut, P., 1986, Nat, 324, 446
  \bibitem[1997]{SPHres} Bate, M.R. \& Burkert, A., 1997, MNRAS, 288, 1060
  \bibitem[1995]{Sink} Bate, M.R., Bonnell, I.A., \& Price, N.M., 1995, MNRAS, 277, 362
  \bibitem[1990]{BenzSPH} Benz, W., 1990, `Numerical Modelling of Nonlinear Stellar Pulsations: Problems and Prospects' (Kluwer Academic Publishers), ed. Buchler, J.R., 269
  \bibitem[2009]{HEALPix} Bisbas, T. G., W\"{u}nsch, R., Whitworth, A. P. \& Hubber, D. A., 2009, A\&A, 497, 649
  \bibitem[1979]{SIT} Boss, A. P. \& Bodenheimer, P., 1979, ApJ, 234, 289
  \bibitem[1913]{Burrau} Burrau, C., 1913, AN, 195, 113
  \bibitem[2002]{GPH2} Cha, S. H., \& Whitworth, A., 2003, MNRAS, 340, 73
  \bibitem[1939]{Chandra} Chandrasekhar, S., 1939,  `An Introduction to the Study of Stellar Structure' (Dover Publs. Inc.; New York)
  \bibitem[2000]{C&M} Chenciner, A. \& Montgomery, R., 2000, AnMat, 152, 881
  \bibitem[2003]{DD03} Delgado Donate, E.J., Clarke, C.J. \& Bate, M.R., 2003, MNRAS, 342, 1926
  \bibitem[2009]{HybridRT} Forgan, D., Rice, K., Stamatellos, D. \& Whitworth, A.P., 2009, MNRAS, 394, 882
  \bibitem[2000]{FLASH} { Fryxell, B. et al., 2000, ApJS, 131, 273}
  \bibitem[1977]{G&M} Gingold, R.A., \& Monaghan, J.J., 1977, MNRAS, 181, 375
  \bibitem[1987]{Hern} Hernquist, L., 1987, ApJS, 64, 715
  \bibitem[1989]{TreeSPH} Hernquist, L. \& Katz, N., 1989, ApJS, 70, 419
  \bibitem[1991]{Ewald} Hernquist, L., Bouchet, F.R. \& Suto, Y., 1991, ApJS, 75, 231
  \bibitem[2004]{SPNIMHD} Hosking, J.G. \& Whitworth, A.P., 2004, MNRAS, 347, 994
  \bibitem[2006]{SPHres2} Hubber, D.A., Goodwin, S.P. \& Whitworth, A.P., 2006, 450, 881
  \bibitem[2002]{GPH1} Inutsuka, S., 2002, JCoPh, 179, 238
  \bibitem[1997]{Kless} Klessen, R.S., 1997, MNRAS, 292, 11
  \bibitem[2000]{TurbMC} Klessen, R.S., Heitsch, F. \& Mac Low, M-M, 2000, ApJ, 535, 887
  \bibitem[1904]{LCReg} Levi-Civita, T., 1904, Ann. Mat. Ser., 9, 1
  \bibitem[1977]{lucy} Lucy, L., 1977, AJ, 82, 1013
  \bibitem[1998]{Maths2} Martin, G.E., 1998, `Geometric constructions' (Springer-Verlag)
  \bibitem[1991]{Herm1} Makino, J., 1991, ApJ, 369, 200
  \bibitem[1992]{Herm2} Makino, J. \& Aarseth, S.J., 1992, PASJ, 44, 141
  \bibitem[2003]{GRAPE} { Makino, J., Fukushige, T., Koga, M., Namura, K., 2003, PASJ, 55, 1163}
  \bibitem[1993]{M&A} McMillan, S.L.W. \& Aarseth, S.J., 1993, AJ, 414, 200
  \bibitem[2010]{EvoL} { Merlin, E., Buonomo, U., Grassi, T., Piovan, L., Chiosi, C., 2010, A\&A, 513, 36}
  \bibitem[1992]{SPH} Monaghan, J.J., 1992, ARA\&A, 30, 543
  \bibitem[1997]{ACond} Monaghan, J.J., 1997, JCoPh, 136, 298
  \bibitem[2002]{SPHTurb} Monaghan, J.J., 2002, MNRAS, 335, 843
  \bibitem[2005]{SPH2} Monaghan, J.J., 2005, RPPh, 68, 1703
  \bibitem[2006]{MonK} Monaghan, J.J., 2006, MNRAS, 365, 199
  \bibitem[1983]{ABvisc} Monaghan, J.J. \& Gingold, R.A., 1983, J. Comp. Phys, 52, 374
  \bibitem[1985]{M4kernel} Monaghan, J.J. \& Lattanzio, J.C., 1985, A\&A, 149, 135
  \bibitem[1996]{MorrisPhD} Morris, J.P., 1996, PhD Thesis - `Analysis of Smoothed Particle Hydrodynamics with Applications', Monash University
  \bibitem[1997]{M&M} Morris, J.P. \& Monaghan, J.J. 1997, JCoPh, 136, 41
  \bibitem[1994]{N&P} Nelson, R.P. \& Papaloizou, J.C.B., 1994, MNRAS, 270, 1
  \bibitem[2008]{8thHerm} Nitadori, K. \& Makino, J., 2008, NewA, 13, 498
  \bibitem[1996]{P&G} Pfalzner, S. \& Gibbon, S., 1996, `Many-body tree methods in Physics' (Cambridge University Press)
  \bibitem[2001]{KIRA} Portegies Zwart, S.F., McMillan, S.L.W, Hut, P. \& Makino, J., 2001, MNRAS, 321, 199
  \bibitem[2007]{SPLASH} { Price, D.J., 2007, PASA, 24, 159}
  \bibitem[2008]{Vsig} Price, D.J., 2008, JCoPh, 227, 10040
  \bibitem[2004]{SPMHD1} Price, D.J., \& Monaghan, J.J., 2004, MNRAS, 348, 123
  \bibitem[2004]{SPMHD2} Price, D.J., \& Monaghan, J.J., 2004, MNRAS, 348, 139
  \bibitem[2005]{SPMHD3} Price, D.J., \& Monaghan, J.J., 2005, MNRAS, 364, 384
  \bibitem[2007]{SPHgrav} Price, D.J., \& Monaghan, J.J., 2007, MNRAS, 374, 1347
  \bibitem[2010]{OSPH} { Read, J.I., Hayfield, T., \& Agertz, O., 2010, MNRAS, 405, 1513}
  \bibitem[1998]{Maths} Riley, K.F., Hobson, M.P. \& Bence, S.J., 1998, `Mathematical methods for physics and engineering' (Cambridge University Press)
  \bibitem[2007]{EulerPot} Rosswog, S. \& Price, D.J., 2007, MNRAS, 379, 915
  \bibitem[2006]{2ndFrag} Saigo, K. \& Tomisaka, K., 2006, ApJ, 645, 381
  \bibitem[2009]{Timestep} Saitoh, T.R. \& Makino, J., 2009, ApJ, 697, 99
  \bibitem[2010]{FAST} Saitoh, T.R., \& Makino, J., 2010, PASJ, 62, 301
  \bibitem[1959]{Sedov} Sedov, L.I., 1959, `Similarity and Dimensional Methods in Mechanics' (New York, Academic Press)
  \bibitem[1978]{Sod} Sod, G.A., 1978, JCoPh., 27, 1
  \bibitem[2005]{GADGET2} Springel, V., 2005, MNRAS, 364, 1105
  \bibitem[2009]{AREPO} {Springel, V., 2010, MNRAS, 401, 791}
  \bibitem[2005]{MilSim} Springel, V. et al., 2005, Nat, 435, 629
  \bibitem[2002]{S&H} Springel, V. \& Hernquist L., 2002, MNRAS, 333, 649
  \bibitem[2005]{GADGET1} Springel, V., Yoshida, N. \& White, S.D.M., 2001, NewA, 6, 79
  \bibitem[2007]{RadT} Stamatellos, D., Whitworth, A.P., Bisbas, T. \& Goodwin, S.P., 2007, MNRAS, 475, 37
  \bibitem[1971]{SS} Stiefel, E. L. \& Scheifele, G., 1971, `Linear and regular celestial mechanics' (Springer-Verlag)
  \bibitem[1992]{ZEUS} Stone, J.M. \& Norman, M.L., 1992, AJS, 80, 753
  \bibitem[1967]{SP} Szebehely, V. \& Peters, C. F., 1967, AJ, 72, 876
  \bibitem[2002]{RAMSES} Teyssier, R., 2002, A\&A, 385, 337
  \bibitem[1992]{T&C} Thomas, P.A. \& Couchman, H.M.P., 1992, MNRAS, 257, 11
  \bibitem[1998]{TL} Truelove, J.K., Klein R.I., McKee, C.F., Holliman II, J.H., Howell, L.H., Greenough, J.A. \& Woods, D.T., 1998, ApJ, 495, 821
  \bibitem[1968]{VA} Van Albada, T.S., 1968, BAN, 19, 479
  \bibitem[2009]{VINE1} Wetzstein, M., Nelson, A.F., Naab, T. \& Burkert, A., 2009, ApJS, 184, 298

\end{thebibliography}
\end{document}